\documentstyle[general_cite,graphicx,subfigure,graphics]{mn}

\title[The extended 12--micron Seyferts at 8.4~GHz]{High--resolution
radio observations of Seyfert galaxies in the extended 12--micron
sample -- I. The observations.}

\author[A.Thean, A.Pedlar, M.Kukula, S.Baum and C.O'Dea]
{Andy~Thean$^{1,2}$, Alan~Pedlar$^{2}$, Marek~J. Kukula$^{3}$, Stefi
A. Baum$^{4}$ \cr
and Christopher P. O'Dea$^{4}$\\ 
$^{1}$~Istituto di Radioastronomia del CNR, Via P. Gobetti 101,
I--40129 Bologna, Italy\\ 
$^{2}$~Nuffield Radio Astronomy Laboratories, University of
Manchester, Jodrell~Bank,  
Macclesfield, Cheshire SK11~9DL, U.K.\\ 
$^{3}$~Institute for Astronomy, University of Edinburgh, Royal Observatory, Blackford Hill, Edinburgh EH9 3HJ\\
$^{4}$~Space Telescope Science Institute, 3700 San Martin Drive,
Baltimore, Maryland 21218, USA}

\date{26th Jan 2000}

\begin{document}

\maketitle

\begin{abstract}

We present 8.4 GHz VLA A--configuration observations of 87 sources from the 
mid--infrared--selected AGN sample of \scite{RMS93}.
These 0.25 arcsec resolution observations allow elongated radio
structures tens of parsecs in size to be resolved and enable 
radio components smaller than 3.5 arcsec to be isolated from
diffuse galactic disc emission.
When combined with previous data, matched radio observations covering 
ninety percent of the sample have been collected and these represent
the largest sub--arcsecond--resolution 
radio imaging survey of a homogeneously--selected sample of Seyfert
galaxies to date.  

We use our observations to identify 5 radio--loud AGN in the sample.
The nature of the
radio emission from Seyfert nuclei will be discussed in subsequent papers.

\end{abstract}


\section{Introduction}

Exceptional amounts of energy are being released at the centres of a
few percent of all galaxies. 
These Active Galactic Nuclei (AGN) are among the most luminous objects
in the Universe and may emit more radiation than an entire galaxy from a region
thought to be around ten thousand times as small.
Seyfert nuclei, which are usually found in nearby spiral galaxies, are
an important class of AGN because of the high quality and wide 
variety of information we may obtain about them and their host
galaxies: they are sufficiently close and sufficiently luminous to be observed
with good linear resolution using a variety of important techniques. 
For these reasons, samples of Seyfert galaxies may be defined more selectively 
than other classes of AGN and permit detailed comparisons of a wider
range of properties. 

Radio studies of several Seyferts show highly--collimated structures 
similar to those found in radio galaxies e.g. Markarian 3
\cite{KukulaMkn3}, Markarian 6 \cite{KukulaMkn6}, Markarian 463
\cite{Mazzarella91}, NGC 1068 \cite{Ulvestad87},  and NGC 4151
\cite{Pedlar4151cont}.  
Hubble Space Telescope images have confirmed that small--scale radio
structures are often associated with individual narrow--line--region
features (\pcite{Falcke98}; \pcite{Capetti99}), as previously suspected from 
ground--based spectroscopy (\pcite{Whittle88}; \pcite{Haniff88}) and
it is now clear that the outflows which cause collimated radio
structures have a direct influence on the narrow--line 
emission we observe (see models by \pcite{Steffen97c_Ib} and
\pcite{Bicknell98}).
Despite this, the importance of collimated outflows from Seyfert
nuclei is often overlooked.
In order to incorporate the radio properties of Seyfert nuclei into
models of their activity and provide the context for  
studies of well--known individual Seyfert galaxies,
it is necessary to document their generic radio properties.

Early statistical studies of the radio properties of 
Seyfert galaxies were carried out by \scite{Wade68} and \scite{deBruyn+W76}.
The radio properties of the following samples of Seyferts have been
studied subsequently; Seyferts from the lists of Markarian 
(\pcite{Meurs+W84}; \pcite{Ulvestad+W84a}), distance--limited samples
(\pcite{Ulvestad+W84b}; \pcite{Ulvestad+W89}; \pcite{Nagar99}), an X--ray
flux--limited sample \cite{Unger87}, the CfA
Seyfert sample (\pcite{Edelson87}; \pcite{Kukula95};
\pcite{RushM+E96}), samples selected from the literature
(\pcite{Giuricin90}; \pcite{Whittle92}), far--infrared--selected
samples (\pcite{Roy94}; \pcite{Giuricin96}; \pcite{Roy98}) and the
mid--infrared--selected extended 12 $\mu$m sample \cite{RushM+E96}. 
Large, unbiased samples are required to compare
the properties of the two types of Seyfert galaxy, 
but the definition of such samples is problematic. 
Ideally, all physically--similar objects within a
certain volume of space should be selected, however
this is extremely difficult to achieve. In practice 
the selection criteria together with observational limitations 
tend to bias samples towards certain classes of object. 

In this paper we present 
observations of the extended 12 $\mu$m AGN sample of \scite{RMS93}
using an observing technique which has been chosen to optimize 
sensitivity to small--scale radio structures i.e. 
the VLA in A--configuration at 8.4 GHz.
This sample is one of the largest homogeneously--selected samples of
Seyfert galaxies available and  
contains well--matched populations of type 1 and type 2 sources.
The paper is organized as follows; in Section \ref{12um.sec} we briefly
describe the sample, in Section \ref{obs.sec} we describe
the observations and data reduction, in
Section \ref{results.sec} we present the results and  
in Section \ref{radio-loud.sec} we explain how 5 
radio--loud objects from the sample may be identified.
An analysis of the radio properties of the sample will be carried out
in subsequent papers.

A value of H$_{\circ}$ = 75 km$\,$s$^{-1}$Mpc$^{-1}$ is assumed throughout.

\section{THE EXTENDED 12--MICRON SAMPLE}
\label{12um.sec}

The extended 12 $\mu$m AGN sample of \scite{RMS93}
is an extension of the original 12 $\mu$m AGN sample of
\scite{SM89} to fainter flux levels using the 
{\it IRAS Faint Source Catalogue Version
2} \cite{Moshir91}. 
From an initial sample of 893 mid--infrared--bright sources,
AGN catalogues were used to define a 
subsample of active galaxies which contains 118 objects, the majority of
which are Seyfert galaxies.
The sample was selected at 12 $\mu$m in order
to minimize wavelength--dependent 
selection effects. \scite{SM89} proposed that this 
wavelength carries an approximately constant fraction, around 20\%,
of the bolometric flux for quasars and both types of Seyfert.

Previous radio observations of part of the extended 12 $\mu$m AGN
sample have been made by \scite{Nagar99} who observed 16 sources  
with the same resolution as the observations presented in this paper
and \scite{RushM+E96} who presented lower resolution observations of
the original 12 $\mu$m sample which includes 51 sources from the
extended sample. 
The hard X--ray properties of the sample have been studied by
\scite{Barcons95} and the host galaxies of the sample have been
studied by \scite{Hunt99}. 

\section{THE OBSERVATIONS}
\label{obs.sec}

We have made new observations of 87 sources and when 
combined with matched observations of 19 sources from the CfA Seyfert
sample \cite{Kukula95} these observations cover 91\% of the AGN sample
of \scite{RMS93}. 
Of the 12 sources not observed, 10 are unobservable at the VLA due to
their low declination and two are well--studied radio sources (3C 120
and 3C 273). 

An almost identical observing strategy to that used by \scite{Kukula95}
was followed for all observations.
The Very Large Array\footnote{Operated by Associated Universities Inc.
under contract with the National Science
Foundation. \scite{Thompson80} give instrumental details.} (VLA) was used
in A--configuration at 8.4 GHz in snapshot mode with approximately 16 minutes 
on each source and approximately 6 minutes on a corresponding phase calibrator.
J2000 co--ordinates were used, with phase calibrators
selected from the A--category NRAO list and the list of
\scite{Patnaik92}. 
In theory the positional accuracy of the final images is 0.005 arcsec, but  
to allow for atmospheric fluctuations we adopt a conservative
estimate of 0.05 arcsec.
Due to an on--line computer failure during the initial observing run,
the 87 new sources observed were split between two separate observing
runs. 
The final radio database will therefore contain 
data from 3 separate epochs; 15th July 1995 (63 sources), 
25th November 1996 (24 sources) and June 1991 (19 sources from
the CfA sample). 
The epoch of each new observation is indicated in Table \ref{beams.tab}.

Between the 1995 run and the 1996 run, the two default VLA observing
frequencies changed from 8.415 and 8.464 GHz to 8.435 and 
8.485 GHz to avoid interference. 
The mean 1--$\sigma$ noise level for all maps was 53$\pm$20 $\mu$Jy;
56$\pm$16 $\mu$Jy for the 1995 maps compared with 44$\pm$26 $\mu$Jy
beam$^{-1}$ for the 1996 maps. Table \ref{beams.tab} gives the noise
levels for individual maps. Mean noise values have been calculated
using only those thermal--noise--limited sources (peak flux $<$ 100 mJy
beam$^{-1}$) observed at high elevation (axial ratio of beam $<$ 5), 
errors represent the standard deviation of the mean.

All data processing, including calibration and mapping, was performed using the
Astronomical Image Processing System (AIPS) in the standard way. 
The data were Fourier--transformed using a natural weighting scheme in order to
maximize sensitivity.  
After CLEAN deconvolution, the maps were restored with a 0.25 arcsec FWHM 
Gaussian beam. 
The largest detectable angular size in this configuration is 3.5 arcsec.
Seventeen strong sources, whose peak flux density
exceeded 10 mJy, were subjected to several cycles of
self--calibration. 
In the final maps the sensitivity approached thermal noise levels 
(1--$\sigma$ $<$ 120 $\mu$Jy~beam$^{-1}$) for all except 5
sources, these were either bright ($>$ 100 mJy beam$^{-1}$) or observed at low
elevation (axial ratio of beam $>$ 5).

\section{RESULTS AND ANALYSIS}
\label{results.sec}

\subsection{Observational results}

\begin{figure*}
\centering
      \subfigure[) Markarian 938]{
        \includegraphics[width=5.1cm,clip,trim=0 47 0 35]{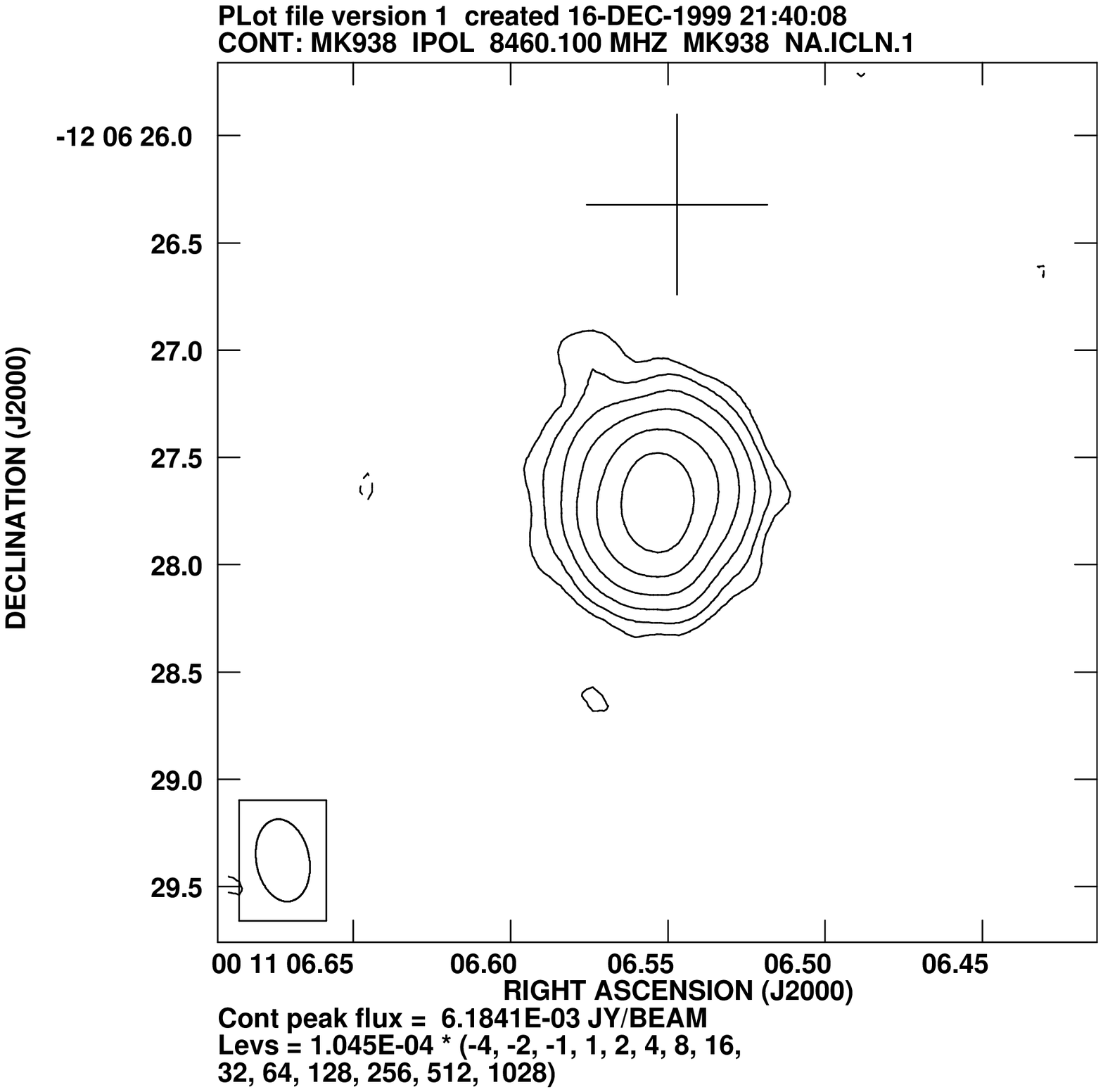}}
      \subfigure[) NGC 262 (Markarian 348)]{
       \includegraphics[width=5.0cm,clip,trim=0 47 0 35]{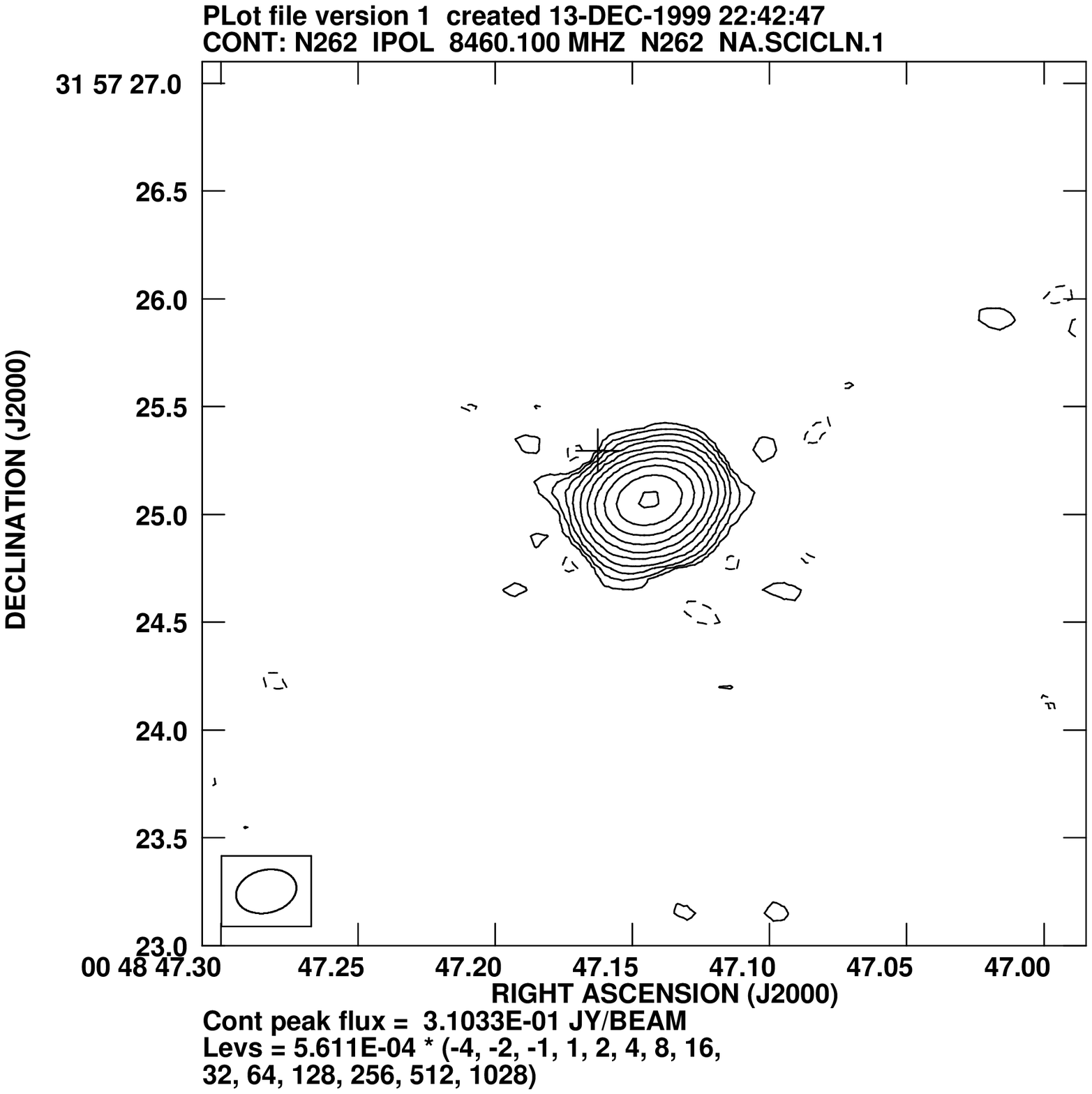}}
      \subfigure[) E541-IG12]{
        \includegraphics[width=4.8cm,clip,trim=0 62 0 35]{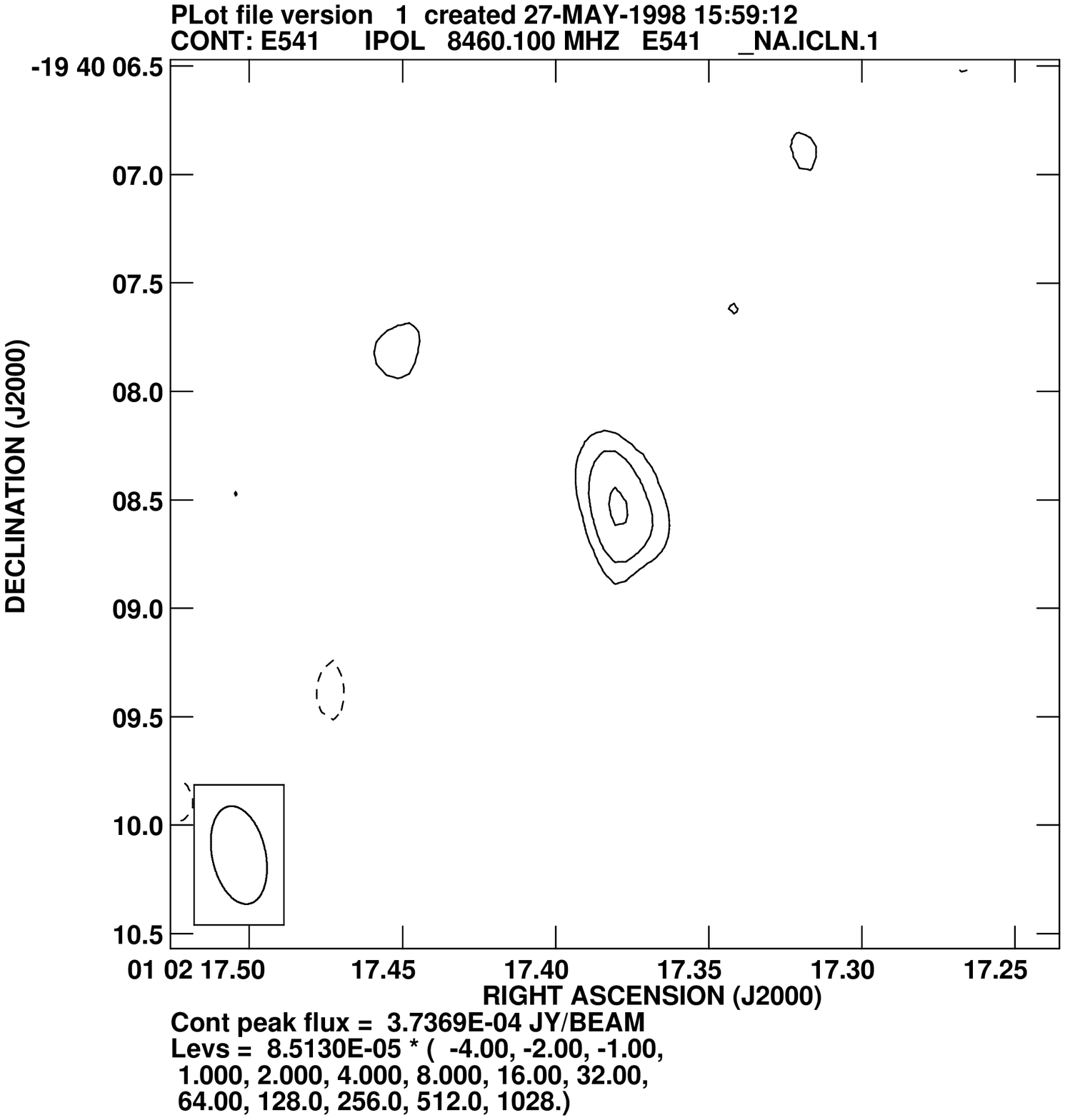}}
      \subfigure[) NGC 424]{
         \includegraphics[width=4.8cm,clip,trim=0 62 0 35]{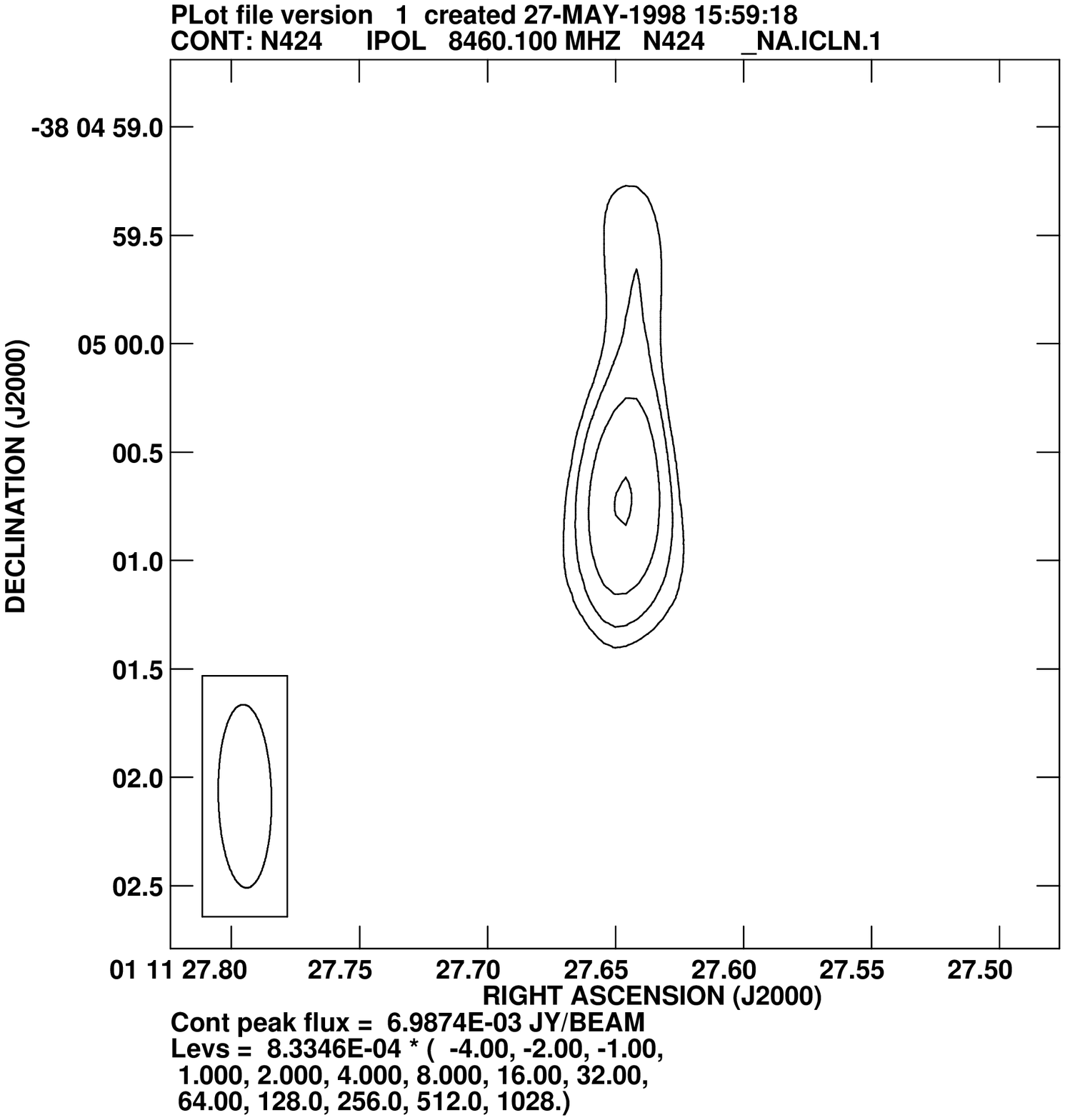}}
      \subfigure[) NGC 526A]{
        \includegraphics[width=4.8cm,clip,trim=0 62 0 35]{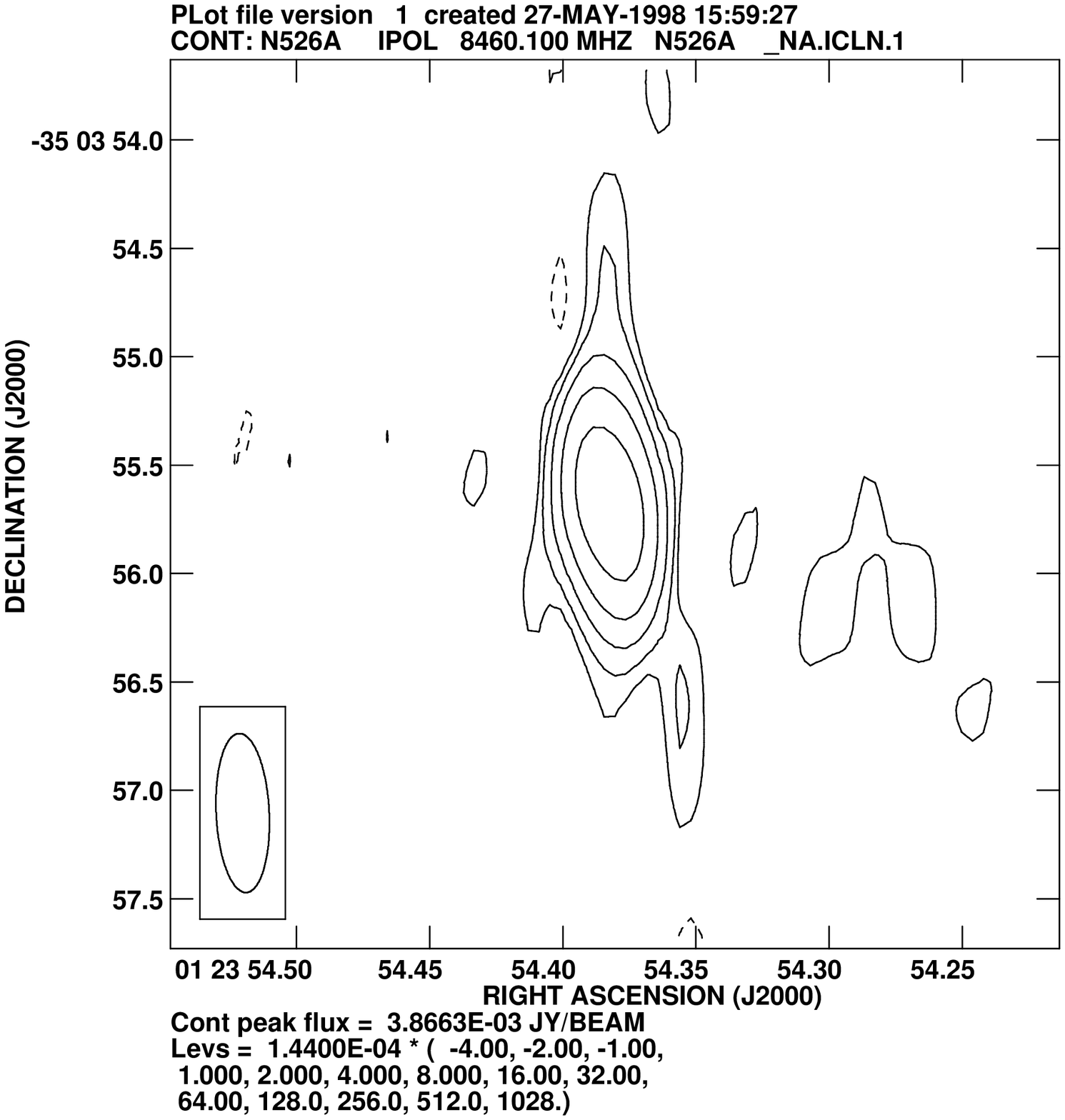}}
      \subfigure[) NGC 513]{
        \includegraphics[width=4.9cm,clip,trim=0 47 0 35]{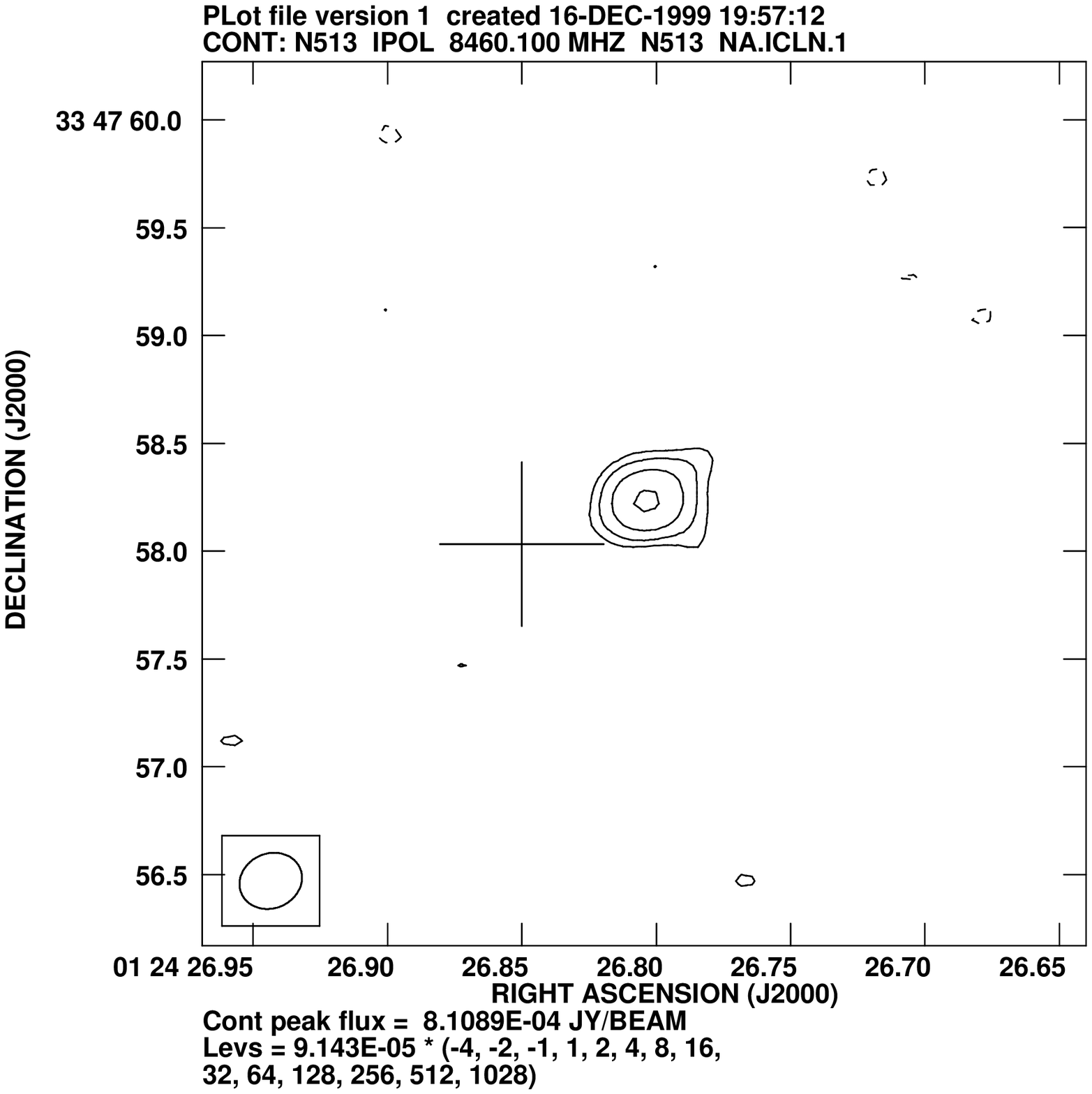}}
      \subfigure[) F01475-0740]{
        \includegraphics[width=4.9cm,clip,trim=0 62 0 35]{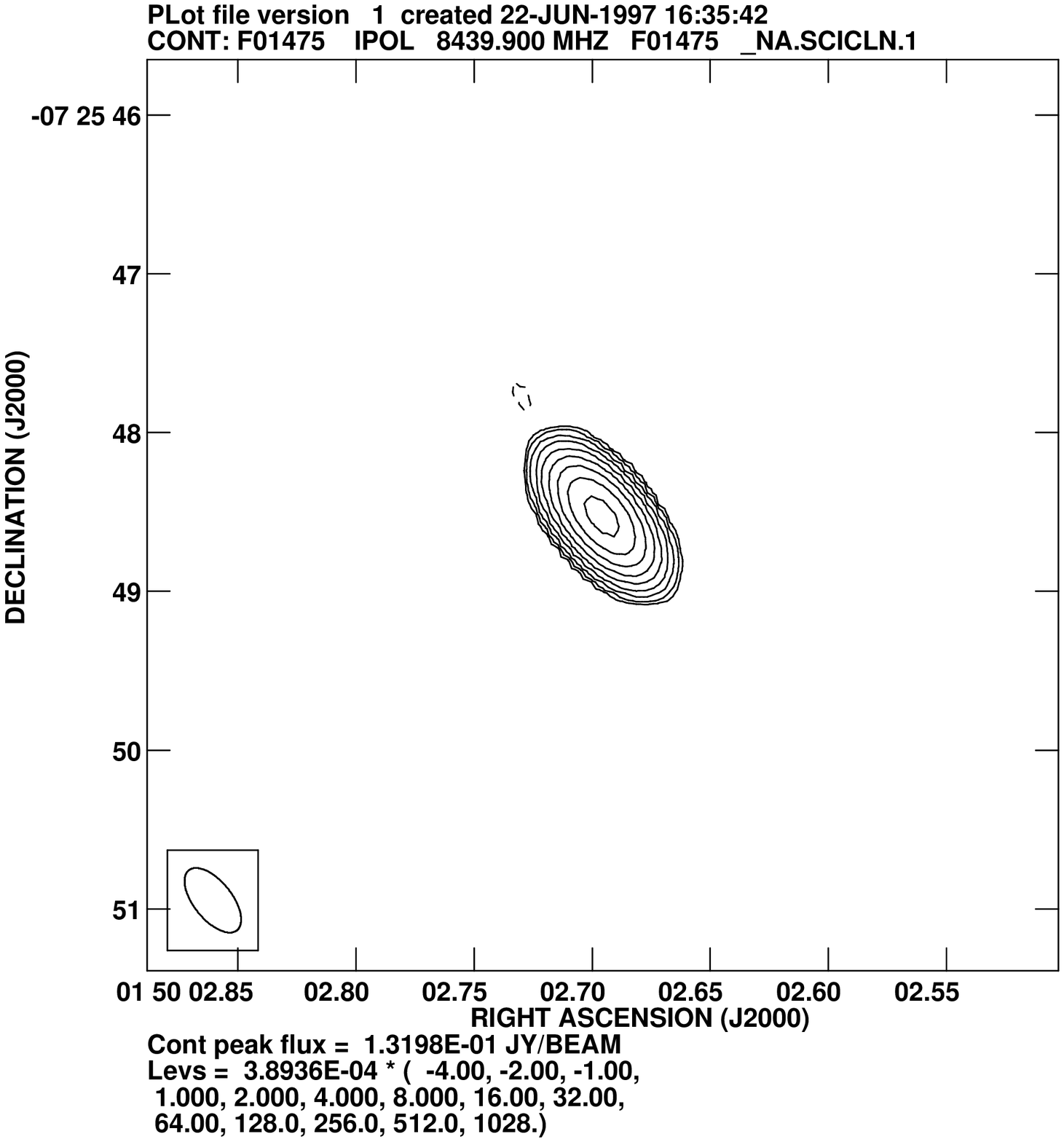}}
       \subfigure[) Markarian 1034]{
       \includegraphics[width=4.8cm,clip,trim=0 62 0 35]{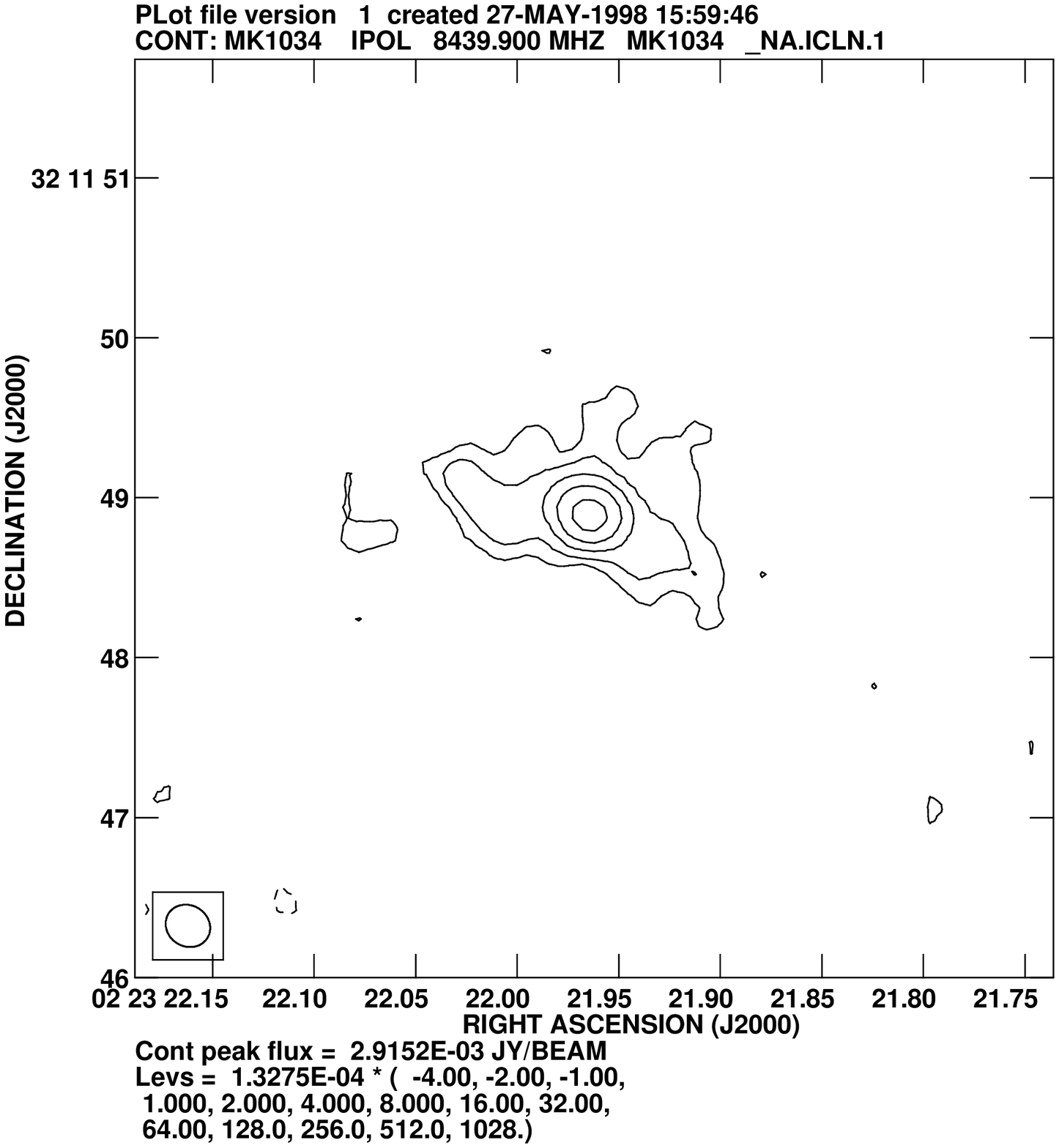}}
      \subfigure[) MCG-3-7-11]{
        \includegraphics[width=4.9cm,clip,trim=0 62 0 35]{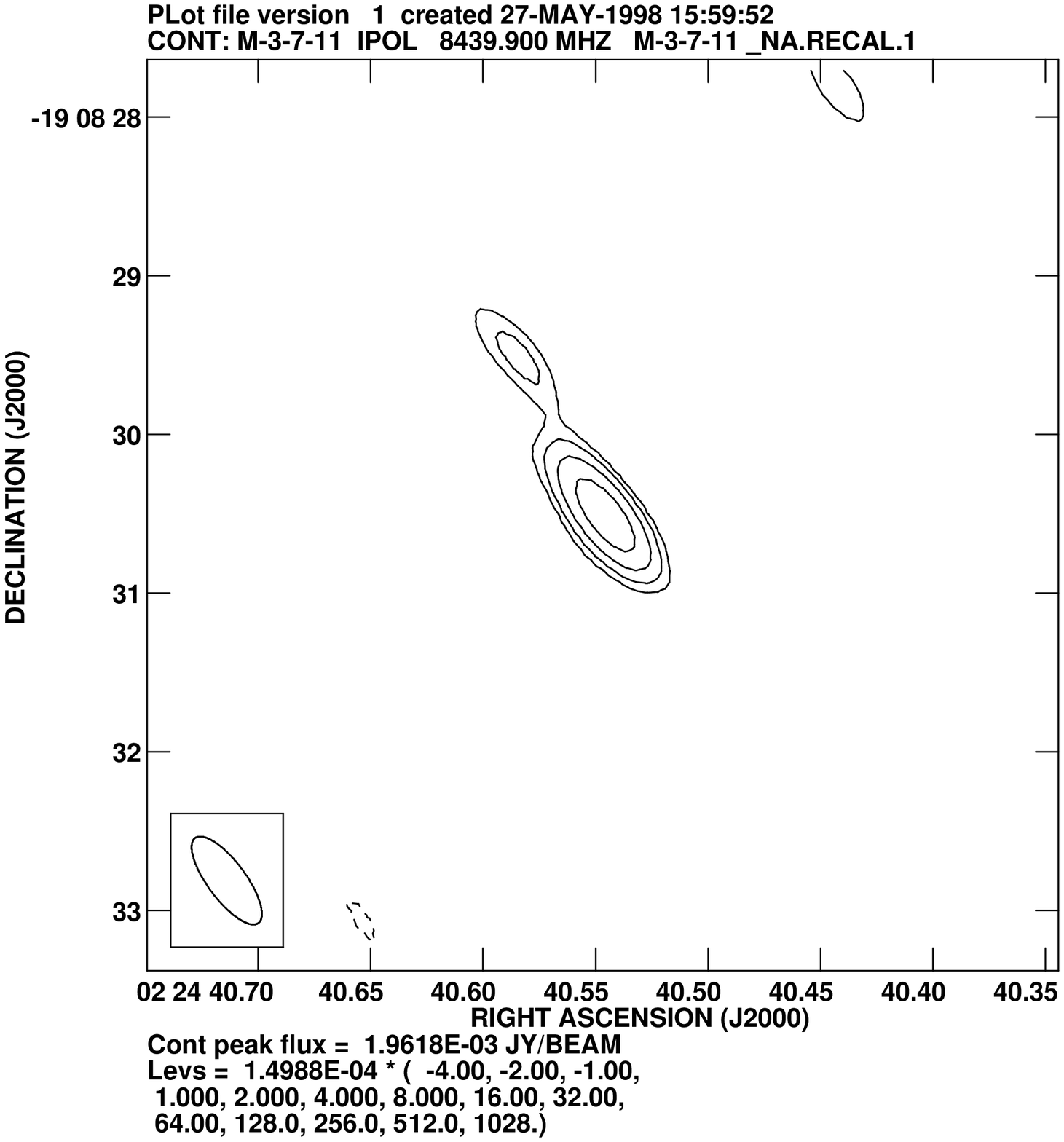}}
      \subfigure[) NGC 1056]{
        \includegraphics[width=4.8cm,clip,trim=0 62 0 35]{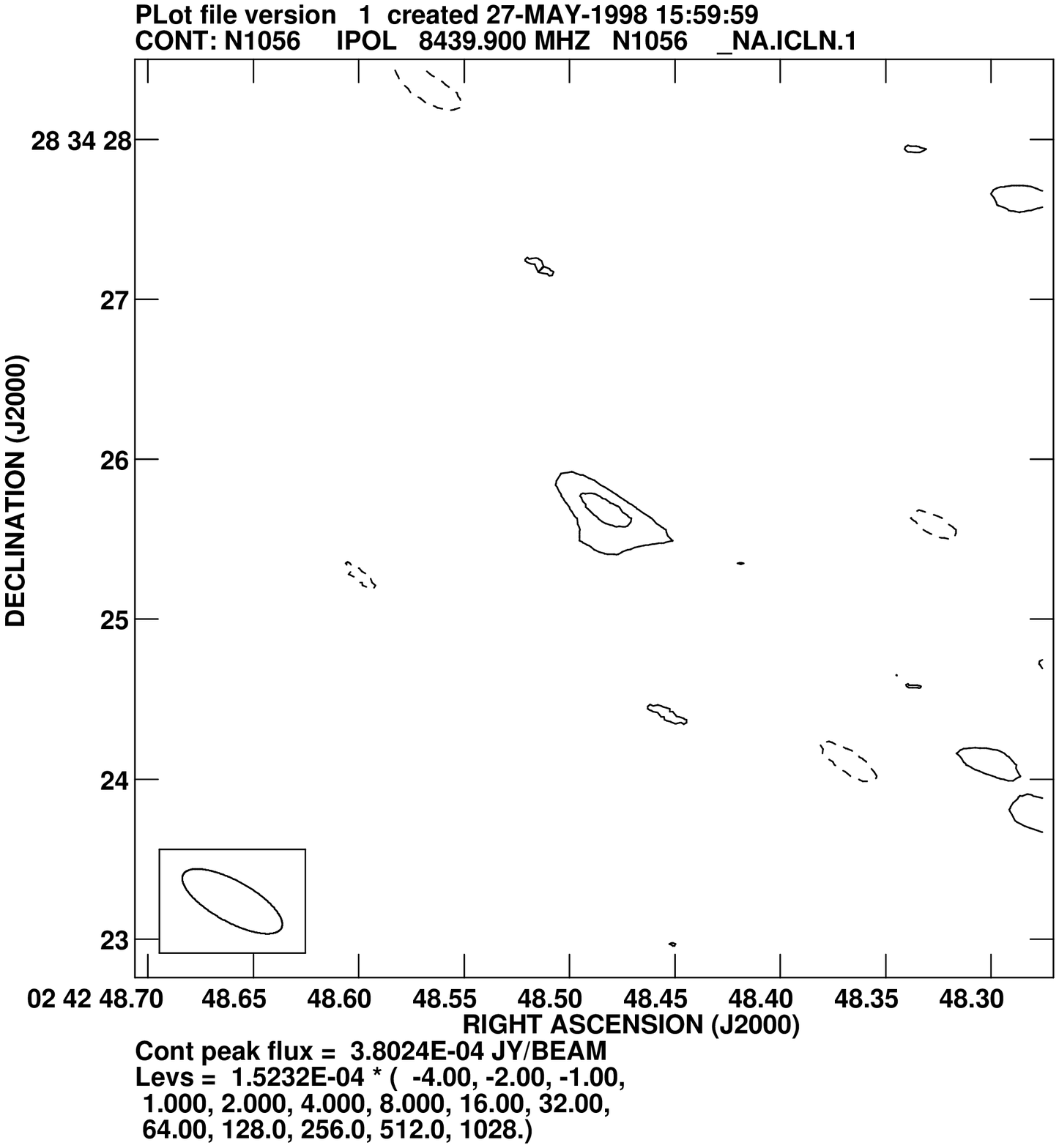}}
      \subfigure[) NGC 1097]{
        \includegraphics[width=4.9cm,clip,trim=0 62 0 35]{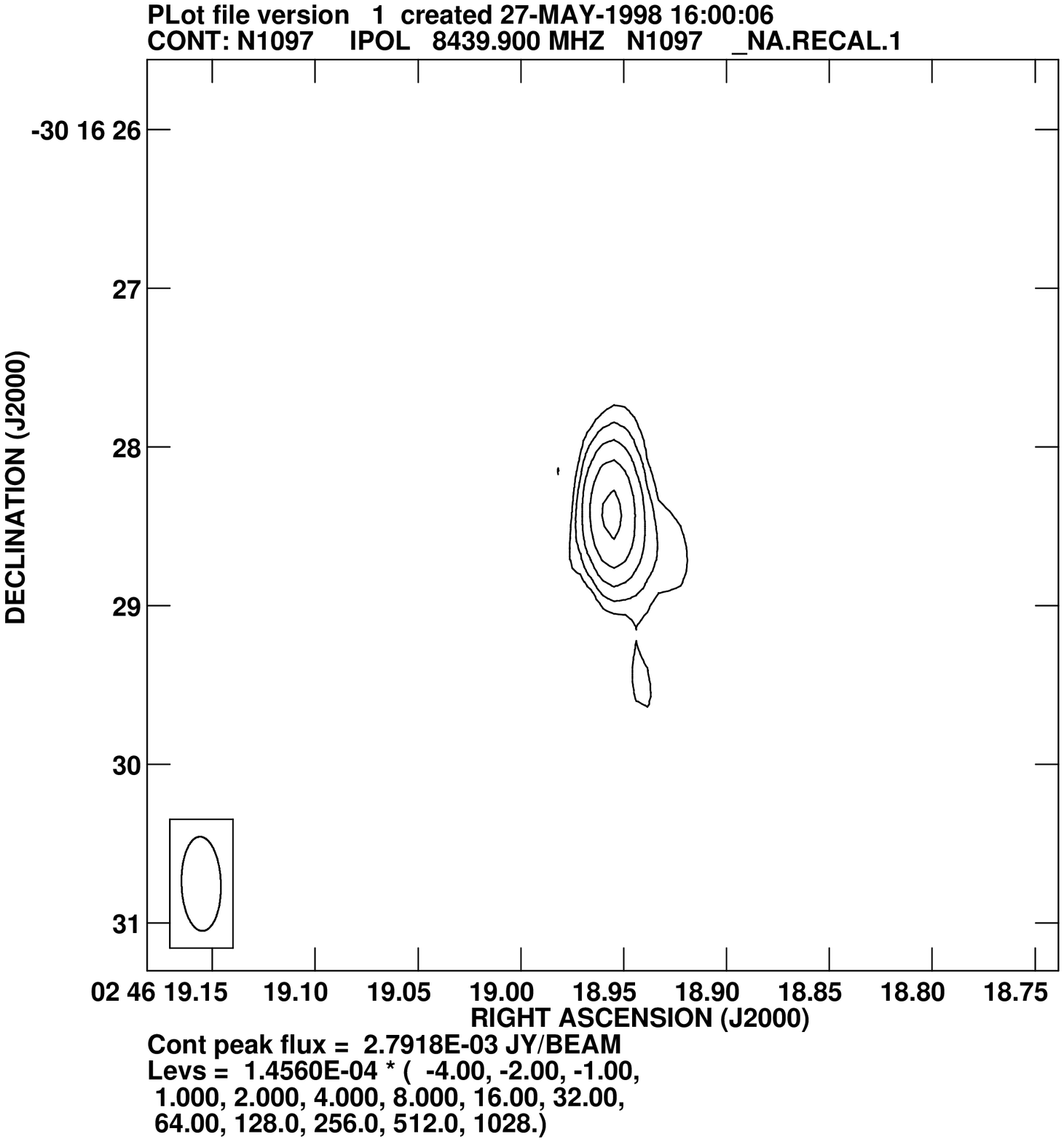}}
      \subfigure[) NGC 1125]{
        \includegraphics[width=5.0cm,clip,trim=0 62 0 35]{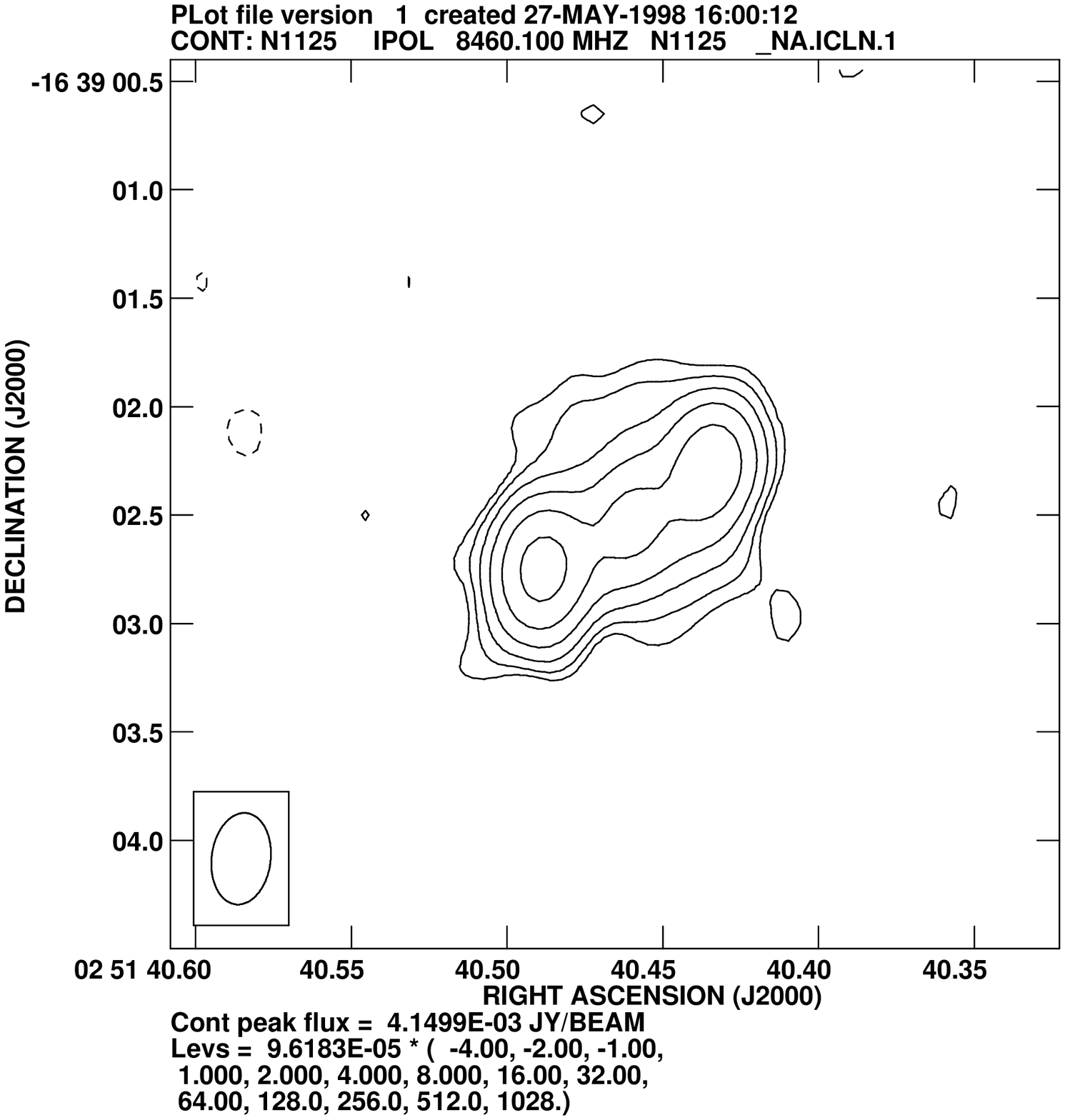}}
\caption{A-configuration 8.4 GHz images.}
\label{contours.fig}
\end{figure*}

Contour maps of all detected sources are
shown in Figure \ref{contours.fig}.
The ellipse in the lower left--hand corner shows
the shape of the restoring beam at half power. 
Optical nuclear positions are marked by a cross where they
are available from \scite{Clements81}, \scite{Clements83} 
and \scite{Argyle90}; 
the diameter of the cross shows the 2--$\sigma$ positional uncertainty.

Descriptions of the radio maps are shown in Table 
\ref{beams.tab} which is arranged as follows;
{\it Column 1:} Galaxy name, a dagger ($\dagger$) indicates 
data from the 
July 1995 observing run and an asterisk ($\ast$) 
indicates data from the 
November 1996 observing run.
{\it Column 2:} Beam major axis, $\theta _{maj}$.
{\it Column 3:} Beam minor axis, $\theta _{min}$.
{\it Column 4:} Beam position angle in degrees, PA.
{\it Column 5:} Peak flux density, S$_{peak}$ (mJy/beam).
{\it Column 6:} The root mean square noise, $\sigma$, measured at the
edge of the field where no deconvolution techniques were applied
($\mu$Jy/beam). 
{\it Column 7:} Contour levels for each map ($\mu$Jy/beam).
Where possible the base contour level was set at
3--$\sigma$ however in 28\% of the maps
instrumental `side--lobes' were present and the base contour level was
set at a level which excluded obvious spurious features. 
Side--lobes result from the incomplete coverage of the uv--plane of
the interferometer. 
They are usually 
removed by deconvolution techniques such as CLEAN but can remain strong 
when the original data are of poor quality or badly calibrated, 
or when the target source is particularly bright or observed at
low elevation.

\begin{table*}
\scriptsize
\begin{center}
\begin{tabular}{|l|c|c|c|c|c|l|} \hline 
{\bf Galaxy}  & {\bf $\theta _{maj}$($''$)}  & {\bf $\theta _{min}$($''$)} & {\bf PA($^{\circ}$)} & 
{\bf S$_{peak}$(mJy/B)}&{\bf 1--$\sigma$($\mu$Jy/B)} & {\bf Contour levels ($\mu$Jy/B)}\\ \hline
$^{\ast}$Mrk 938         & 0.39 &  0.25 & ~11 &  ~~~6.2 & ~~35 & ~3--$\sigma$$\times$-1,1,2,4,8,16,32  \\
$^{\ast}$NGC 262         & 0.28 &  0.20 & -76 &  ~310.3 & ~187 & ~3--$\sigma$$\times$-1,1,2,4,8,16,32,64,128,256,512  \\
$^{\ast}$E541            & 0.46 &  0.25 & ~12 &  ~~~0.4 & ~~28 & ~3--$\sigma$$\times$-1,1,2,4  \\
$^{\ast}$NGC 424         & 0.85 &  0.25 & ~~1 &  ~~~7.0 & ~~69 & $\!$12--$\sigma$$\times$-1,1,2,4,8  \\
$^{\ast}$NGC 526A        & 0.73 &  0.24 & ~~3 &  ~~~3.9 & ~~48 & ~3--$\sigma$$\times$-1,1,2,4,8,16  \\
$^{\ast}$NGC 513         & 0.30 &  0.25 & -67 &  ~~~0.8 & ~~30 & ~3--$\sigma$$\times$-1,1,2,4,8  \\
$^{\dagger}$F01475-0740  & 0.48 &  0.24 & ~39 &  ~132.0 & ~~65 & ~6--$\sigma$$\times$-1,1,2,4,8,16,32,64,128,256  \\
$^{\dagger}$Mrk 1034     & 0.29 &  0.26 & ~57 &  ~~~2.9 & ~~44 & ~3--$\sigma$$\times$-1,1,2,4,8,16  \\
$^{\dagger}$MCG-3-7-11   & 0.67 &  0.24 & ~37 &  ~~~2.0 & ~~50 & ~3--$\sigma$$\times$-1,1,2,4,8  \\
$^{\dagger}$NGC 1056     & 0.70 &  0.26 & ~~1 &  ~~~0.4 & ~~51 & ~3--$\sigma$$\times$-1,1,2  \\
$^{\dagger}$NGC 1097     & 0.59 &  0.25 & ~~2 &  ~~~2.8 & ~~49 & ~3--$\sigma$$\times$-1,1,2,4,8,16  \\
$^{\ast}$NGC 1125        & 0.42 &  0.27 & ~-7 &  ~~~4.2 & ~~32 & ~3--$\sigma$$\times$-1,1,2,4,8,16,32 \\
$^{\dagger}$NGC 1194     & 0.58 &  0.25 & ~51 &  ~~~0.7 & ~~51 & ~3--$\sigma$$\times$-1,1,2,4  \\
$^{\ast}$NGC 1241        & 0.36 &  0.27 & -13 &  ~~~5.3 & ~~46 & ~4--$\sigma$$\times$-1,1,2,4,8,16,32 \\
$^{\dagger}$NGC 1320     & 0.60 &  0.25 & ~51 &  ~~~0.7 & ~~55 & ~3--$\sigma$$\times$-1,1,2,4  \\
$^{\dagger}$NGC 1365     & 0.76 &  0.25 & ~-1 &  ~~~1.5 & ~~59 & ~3--$\sigma$$\times$-1,1,2,4,8  \\
$^{\dagger}$NGC 1386     & 0.78 &  0.25 & ~-7 &  ~~~5.6 & ~~52 & $\!$15--$\sigma$$\times$-1,1,2,4  \\
$^{\dagger}$F03362-1642  & 1.01 &  0.26 & -48 &  ~~~0.7 & ~~46 & ~3--$\sigma$$\times$-1,1,2,4  \\
$^{\ast}$F03450+0055     & 0.32 &  0.27 & -17 &  ~~~4.2 & ~~31 & ~3--$\sigma$$\times$-1,1,2,4,8,16,32  \\
$^{\ast}$Mrk 618         & 0.38 &  0.27 & -14 &  ~~~1.7 & ~~26 & ~3--$\sigma$$\times$-1,1,2,4,8,16  \\
$^{\dagger}$F04385-0828  & 0.75 &  0.26 & -50 &  ~~~4.7 & ~~43 & ~6--$\sigma$$\times$-1,1,2,4,8,16  \\
$^{\dagger}$NGC 1667     & 0.64 &  0.26 & -50 &  ~~~0.4 & ~~48 & ~3--$\sigma$$\times$-1,1,2  \\
$^{\dagger}$MCG-5-13-17  & 0.65 &  0.25 & ~~8 &  ~~~1.4 & ~~70 & ~3--$\sigma$$\times$-1,1,2,4  \\
$^{\ast}$F05189-2524     & 0.53 &  0.27 & ~-8 &  ~~~5.0 & ~~57 & ~6--$\sigma$$\times$-1,1,2,4,8  \\
$^{\dagger}$E253-G3      & 1.49 &  0.24 & ~~2 &  ~~~5.7 & ~485 & ~3--$\sigma$$\times$-1,1  \\
$^{\dagger}$F05563-3820  & 0.85 &  0.26 & ~~4 &  ~~~5.1 & ~107 & $\!$15--$\sigma$$\times$-1,1,2 \\
$^{\dagger}$Mrk 6        & 0.46 &  0.24 & -85 &  ~~12.8 & ~~36 & ~3--$\sigma$$\times$-1,1,2,4,8,16,32  \\
$^{\dagger}$Mrk 9        & 0.55 &  0.24 & -71 &  ~~~0.4 & ~~40 & ~3--$\sigma$$\times$-1,1,2  \\
$^{\dagger}$Mrk 79       & 0.27 &  0.26 & -41 &  ~~~0.8 & ~~49 & ~3--$\sigma$$\times$-1,1,2,4  \\
$^{\ast}$F07599+6508     & 0.34 &  0.29 & ~61 &  ~~~5.0 & ~~35 & ~3--$\sigma$$\times$-1,1,2,4,8,16,32  \\
$^{\dagger}$NGC 2639     & 0.26 &  0.26 & ~~0 &  ~~90.6 & ~~67 & ~9--$\sigma$$\times$-1,1,2,4,8,16,32,64,128  \\
$^{\dagger}$OJ 287       & 0.31 &  0.26 & ~-1 &  $\!$1515.3 & $\!$1175 & ~9--$\sigma$$\times$-1,1,2,4,8,16,32,64,128  \\
$^{\dagger}$F08572+3915  & 0.26 &  0.25 & ~80 &  ~~~2.6 & ~~54 & ~6--$\sigma$$\times$-1,1,2,4  \\
$^{\dagger}$Mrk 704      & 0.27 &  0.26 & ~21 &  ~~~0.7 & ~~62 & ~3--$\sigma$$\times$-1,1,2  \\
$^{\ast}$UGC 5101        & 0.30 &  0.21 & ~63 &  ~~29.0 & ~~97 & ~3--$\sigma$$\times$-1,1,2,4,8,16,32,64  \\
$^{\dagger}$NGC 2992     & 0.37 &  0.26 & ~-8 &  ~~~3.9 & ~~57 & ~3--$\sigma$$\times$-1,1,2,4,8,16  \\
$^{\dagger}$Mrk 1239     & 0.31 &  0.26 & ~~7 &  ~~~6.8 & ~~90 & ~6--$\sigma$$\times$-1,1,2,4,8,16  \\
$^{\ast}$NGC 3031        & 0.32 &  0.20 & ~60 &  ~221.2 & ~133 & ~3--$\sigma$$\times$-1,1,2,4,8,16,32,64,128,256,512  \\
$^{\dagger}$3C 234       & 0.26 &  0.25 & ~68 &  ~~33.5 & ~~67 & ~4--$\sigma$$\times$-1,1,2,4,8,16,32,64  \\
$^{\dagger}$NGC 4579     & 1.32 &  0.26 & ~56 &  ~~33.9 & ~~59 & $\!$15--$\sigma$$\times$-1,1,2,4,8,16,32  \\
$^{\dagger}$NGC 4593     & 0.65 &  0.25 & ~49 &  ~~~1.0 & ~~45 & ~3--$\sigma$$\times$-1,1,2,4  \\
$^{\dagger}$NGC 4594     & 0.94 &  0.25 & ~50 &  ~~84.7 & ~~53 & ~3--$\sigma$$\times$-1,1,2,4,8,16,32,64,128,256,512  \\
$^{\dagger}$TOL1238      & 0.90 &  0.30 & -17 &  ~~~1.4 & ~102 & ~3--$\sigma$$\times$-1,1,2  \\
$^{\dagger}$MCG-2-33-34  & 0.38 &  0.26 & ~13 &  ~~~0.6 & ~~55 & ~3--$\sigma$$\times$-1,1,2  \\
$^{\dagger}$PGC 044896   & 0.45 &  0.26 & ~65 &  ~~~5.0 & ~~69 & ~6--$\sigma$$\times$-1,1,2,4,8,16  \\
$^{\dagger}$NGC 4941     & 0.45 &  0.26 & ~41 &  ~~~1.9 & ~~48 & ~3--$\sigma$$\times$-1,1,2,4,8  \\
$^{\dagger}$NGC 4968     & 0.47 &  0.26 & ~~1 &  ~~~2.8 & ~~47 & ~6--$\sigma$$\times$-1,1,2,4,8\\
$^{\dagger}$NGC 5005     & 0.56 &  0.26 & ~64 &  ~~~1.2 & ~~46 & ~3--$\sigma$$\times$-1,1,2,4,8 \\
$^{\dagger}$MCG-3-34-63  & 0.41 &  0.26 & ~-8 &  ~~24.5 & ~~64 & $\!$12--$\sigma$$\times$-1,1,2,4,8,16,32  \\
$^{\dagger}$NGC 5194     & 0.54 &  0.26 & ~68 &  ~~~0.3 & ~~47 & ~3--$\sigma$$\times$-1,1,2 \\
$^{\dagger}$MCG-6-30-15  & 0.82 &  0.24 & -22 &  ~~~0.4 & ~~68 & ~3--$\sigma$$\times$-1,1  \\
$^{\dagger}$F13349+2438  & 0.38 &  0.25 & ~66 &  ~~~4.0 & ~~52 & ~3--$\sigma$$\times$-1,1,2,4,8,16  \\
$^{\dagger}$NGC 5256     & 0.57 &  0.26 & ~67 &  ~~~3.5 & ~~39 & ~3--$\sigma$$\times$-1,1,2,4,8,16   \\
$^{\dagger}$Mrk 273      & 0.55 &  0.26 & ~69 &  ~~22.2 & ~~38 & ~6--$\sigma$$\times$-1,1,2,4,8,16,32,64 \\
$^{\dagger}$IC 4329A     & 0.70 &  0.26 & -21 &  ~~~4.9 & ~~91 & ~3--$\sigma$$\times$-1,1,2,4,8,16  \\
$^{\dagger}$NGC 5347     & 1.00 &  0.26 & ~58 &  ~~~0.7 & ~~59 & ~3--$\sigma$$\times$-1,1,2 \\
$^{\dagger}$Mrk 463      & 0.63 &  0.25 & ~58 &  ~~38.2 & ~~43 & ~3--$\sigma$$\times$-1,1,2,4,8,16,32,64,128,256  \\
$^{\dagger}$NGC 5506     & 0.35 &  0.26 & -31 &  ~~50.8 & ~~77 & ~3--$\sigma$$\times$-1,1,2,4,8,16,32,64,128  \\
$^{\dagger}$Mrk 817      & 0.67 &  0.26 & ~61 &  ~~~2.9 & ~~58 & ~3--$\sigma$$\times$-1,1,2,4,8,16  \\
$^{\dagger}$F15091-2107  & 0.45 &  0.26 & ~-6 &  ~~~6.0 & ~~68 & ~3--$\sigma$$\times$-1,1,2,4,8,16  \\
$^{\dagger}$NGC 5953     & 0.89 &  0.26 & ~57 &  ~~~0.4 & ~~55 & ~3--$\sigma$$\times$-1,1,2  \\
$^{\dagger}$Arp 220      & 0.86 &  0.26 & ~58 &  ~~54.6 & ~~41 & $\!$24--$\sigma$$\times$-1,1,2,4,8,16,32  \\
$^{\dagger}$MCG-2-40-4   & 0.38 &  0.26 & -13 &  ~~~1.6 & ~~42 & ~3--$\sigma$$\times$-1,1,2,4,8  \\
$^{\dagger}$F15480-0344  & 0.33 &  0.26 & ~-6 &  ~~~9.5 & ~~85 & ~3--$\sigma$$\times$-1,1,2,4,8,16,32  \\
$^{\dagger}$NGC 6890     & 1.35 &  0.24 & ~~5 &  ~~~0.6 & ~~53 & ~3--$\sigma$$\times$-1,1,2 \\
$^{\ast}$Mrk 509         & 0.40 &  0.25 & ~20 &  ~~~1.8 & ~~32 & ~3--$\sigma$$\times$-1,1,2,4,8,16  \\
$^{\ast}$Mrk 897         & 0.42 &  0.26 & ~48 &  ~~~1.7 & ~~30 & ~3--$\sigma$$\times$-1,1,2,4,8,16  \\
$^{\dagger}$NGC 7130     & 0.74 &  0.24 & ~~8 &  ~~~8.9 & ~~98 & ~6--$\sigma$$\times$-1,1,2,4,8 \\
$^{\dagger}$NGC 7172     & 0.65 &  0.24 & ~10 &  ~~~2.0 & ~~59 & ~3--$\sigma$$\times$-1,1,2,4,8  \\
$^{\ast}$F22017+0319     & 0.37 &  0.26 & ~41 &  ~~~0.7 & ~~28 & ~3--$\sigma$$\times$-1,1,2,4  \\
$^{\dagger}$NGC 7213     & 1.71 &  0.24 & ~-1 &  ~132.2 & ~183 & ~3--$\sigma$$\times$-1,1,2,4,8,16,32,64,128,256  \\
$^{\ast}$3C 445          & 0.35 &  0.25 & ~24 &  ~~56.3 & ~~36 & ~3--$\sigma$$\times$-1,1,2,4,8,16,32,64,128,256,512  \\
$^{\ast}$NGC 7314        & 0.53 &  0.25 & ~-2 &  ~~~0.8 & ~~41 & ~3--$\sigma$$\times$-1,1,2,4  \\
$^{\ast}$MCG-3-58-7      & 0.44 &  0.24 & ~-1 &  ~~~0.3 & ~~31 & ~3--$\sigma$$\times$-1,1,2  \\
$^{\ast}$NGC 7496        & 1.17 &  0.24 & ~-1 &  ~~~3.2 & ~~56 & ~3--$\sigma$$\times$-1,1,2,4,8,16  \\
$^{\ast}$NGC 7582        & 1.09 &  0.25 & ~~2 &  ~~~2.4 & ~~80 & ~3--$\sigma$$\times$-1,1,2,4,8  \\
$^{\ast}$CG 381          & 0.37 &  0.26 & ~40 &  ~~~0.3 & ~~29 & ~3--$\sigma$$\times$-1,1,2  \\  \hline
\end{tabular}
\end{center}
\caption{Key to radio maps presented in Figure \ref{contours.fig}.}
\label{beams.tab}
\end{table*}

\renewcommand{\thesubfigure}{\roman{subfigure}}

\setcounter{figure}{0}
\setcounter{subfigure}{12}
\begin{figure*}
\centering 
      \subfigure[) NGC 1194]{
        \includegraphics[width=4.8cm,clip,trim=0 62 0 35]{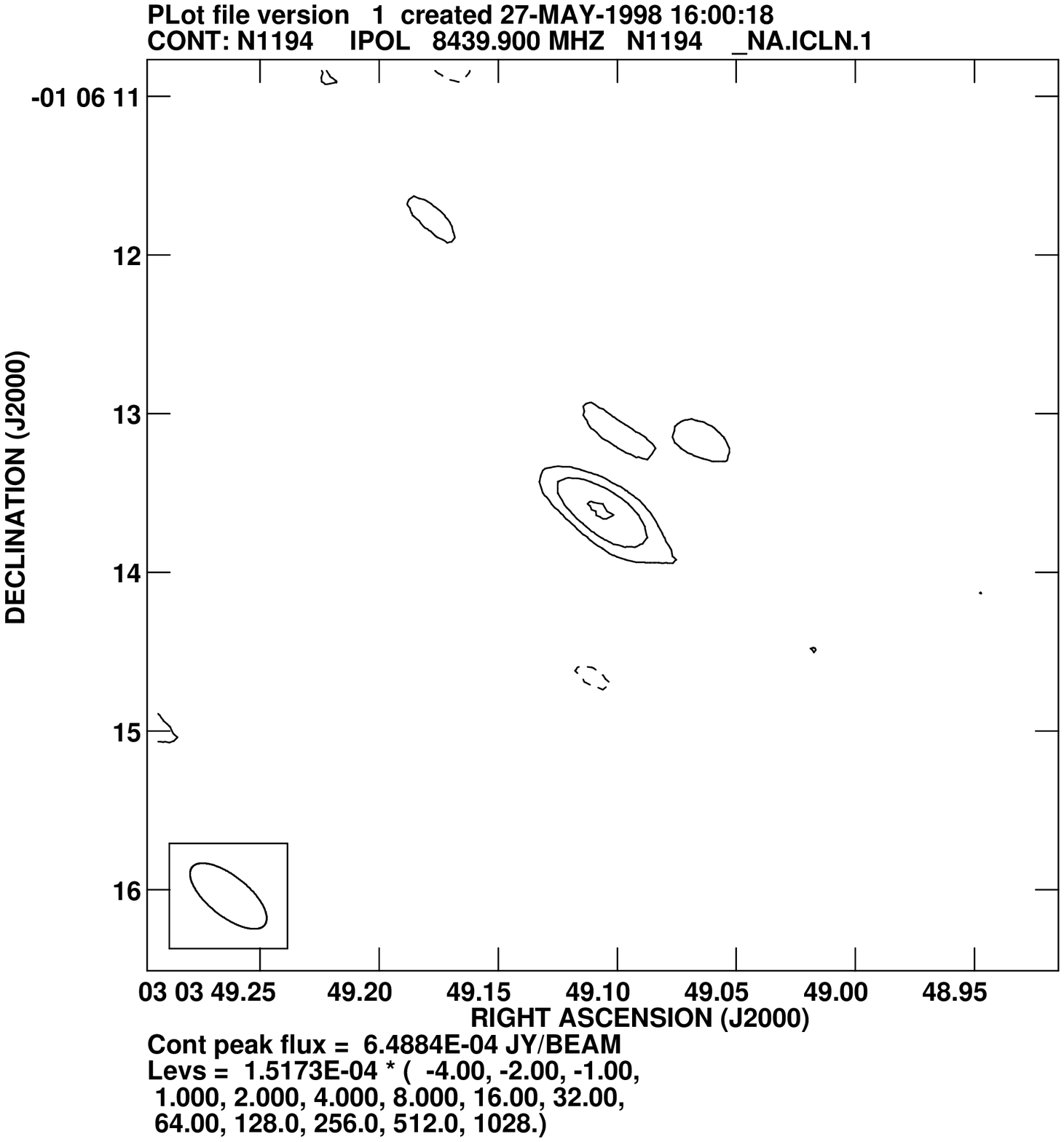}}
      \subfigure[) NGC 1241]{
        \includegraphics[width=4.9cm,clip,trim=0 62 0 35]{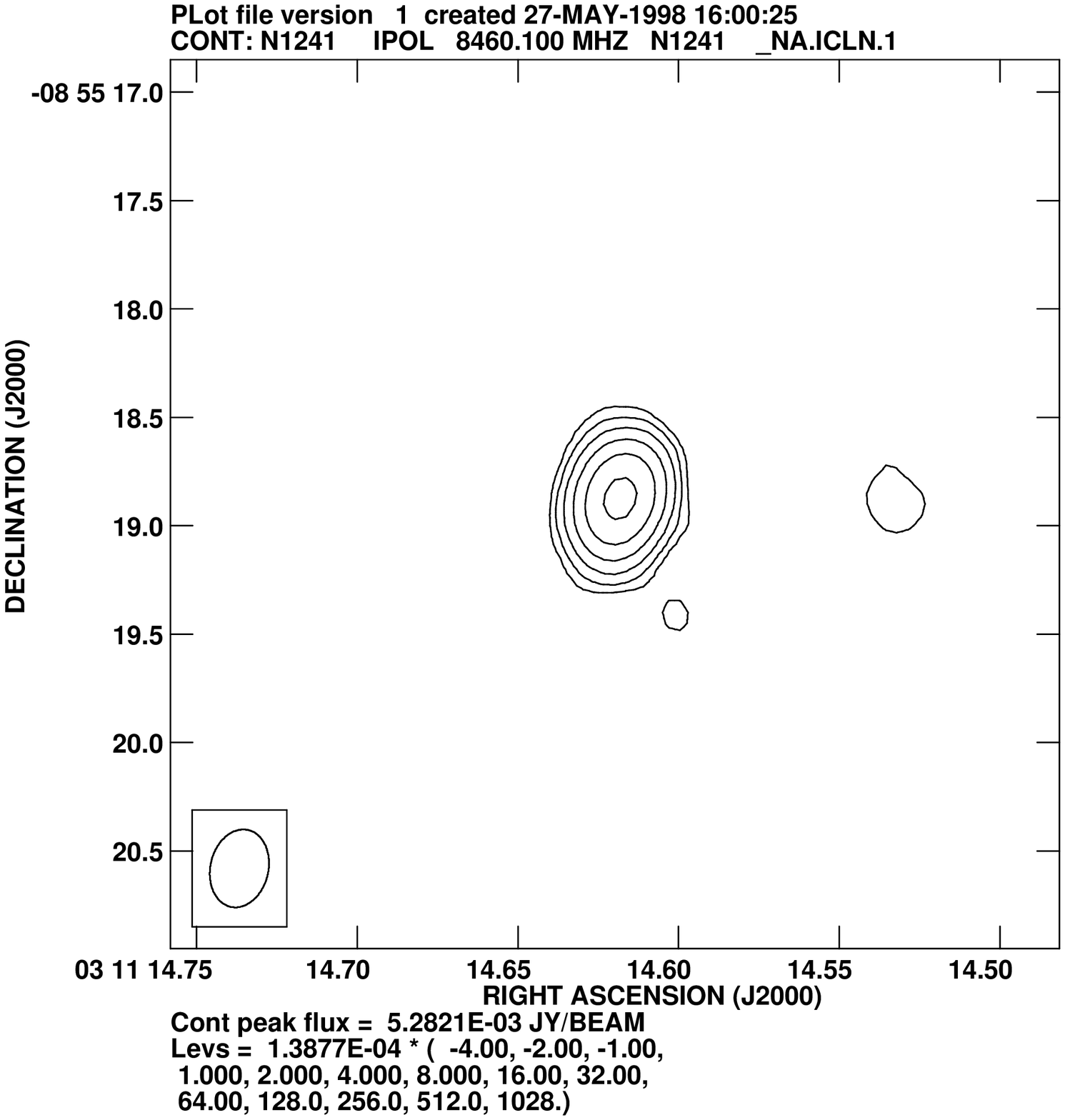}}
      \subfigure[) NGC 1320]{
        \includegraphics[width=4.8cm,clip,trim=0 62 0 35]{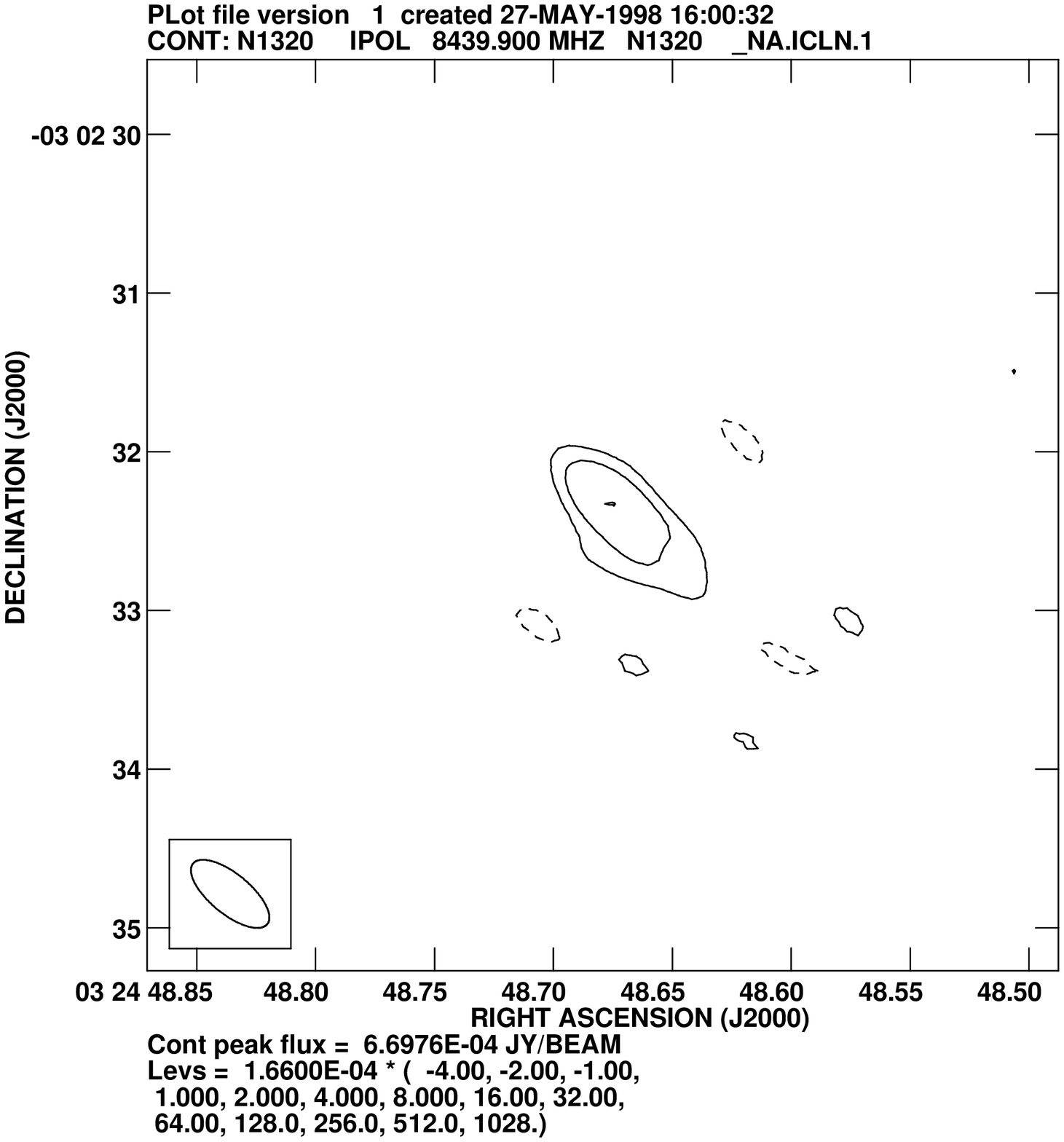}}      \subfigure[) 
NGC 1365]{
        \includegraphics[width=4.8cm,clip,trim=0 62 0 35]{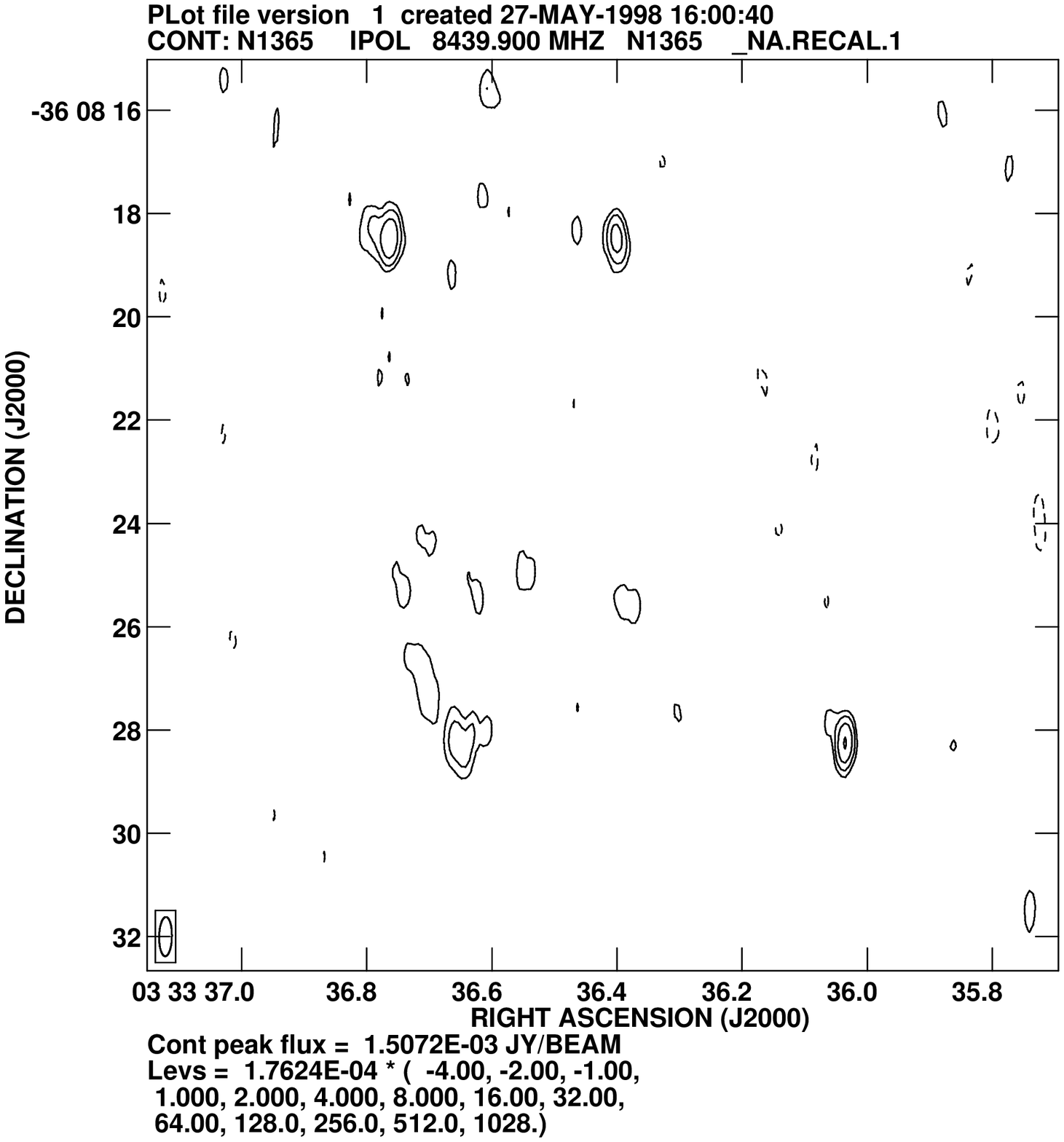}}
      \subfigure[) NGC 1386]{
        \includegraphics[width=4.8cm,clip,trim=0 62 0 35]{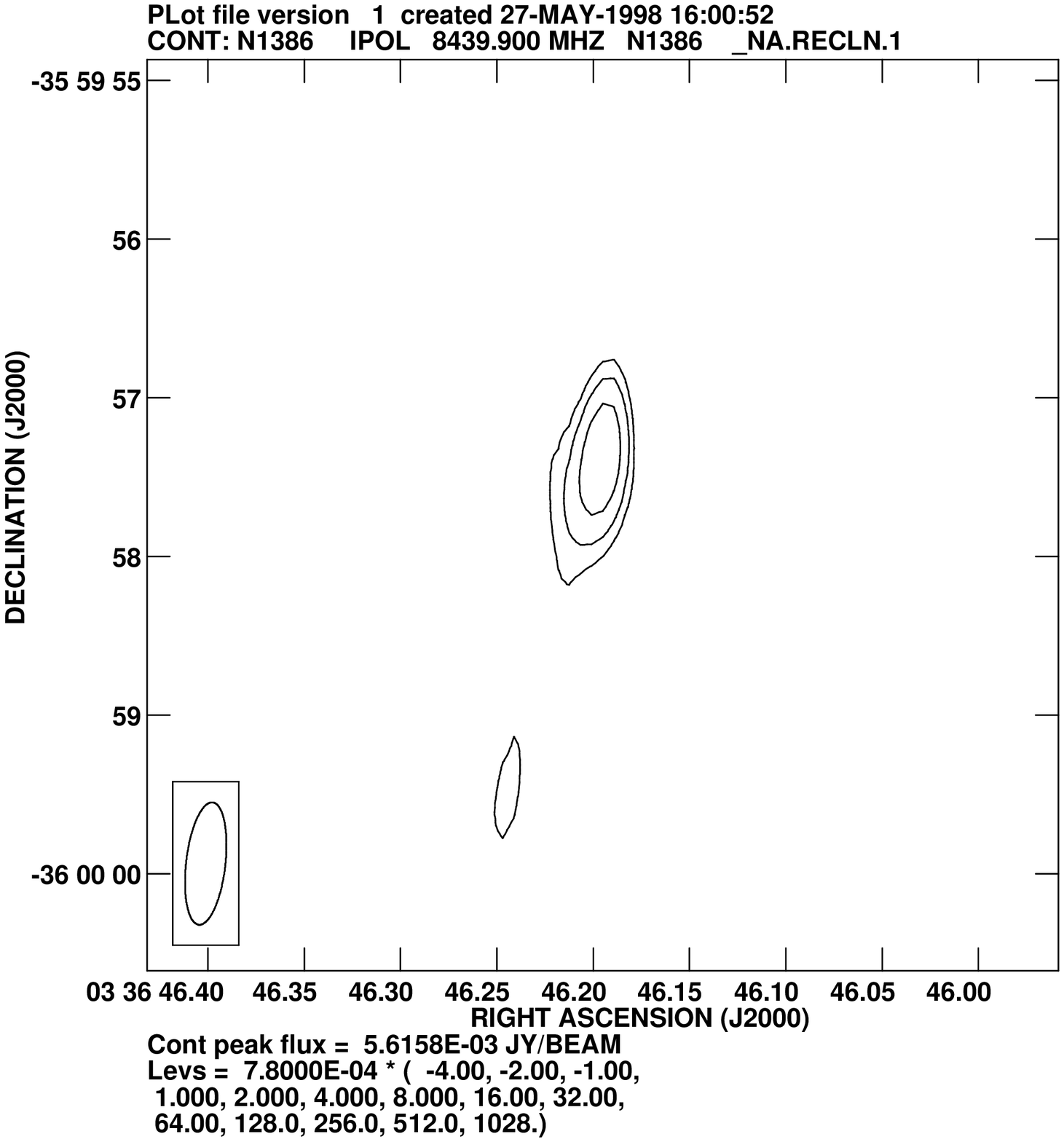}}
      \subfigure[) F03362-1642]{
        \includegraphics[width=4.8cm,clip,trim=0 62 0 35]{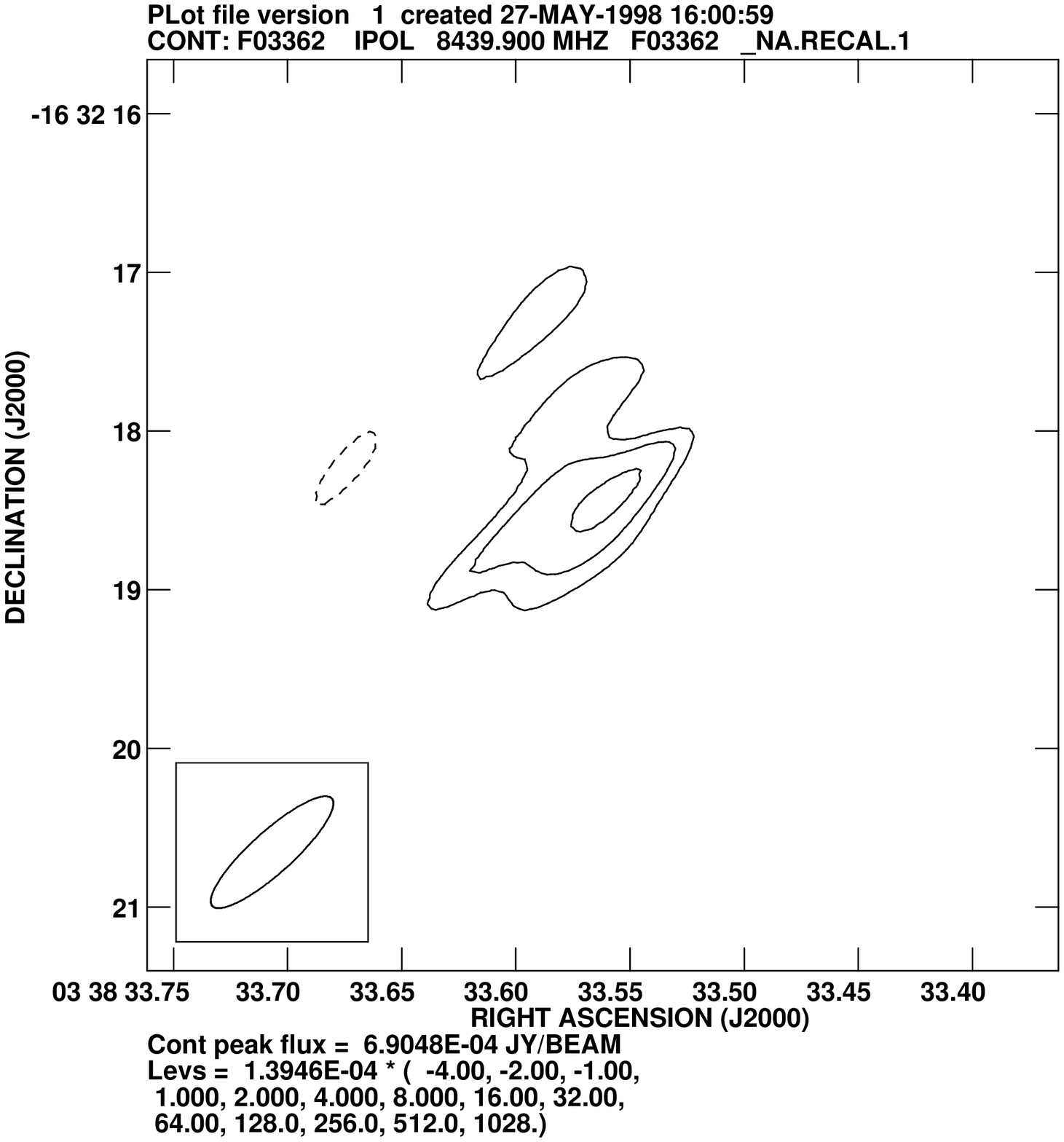}}
      \subfigure[) F03450+0055]{
        \includegraphics[width=4.8cm,clip,trim=0 62 0 35]{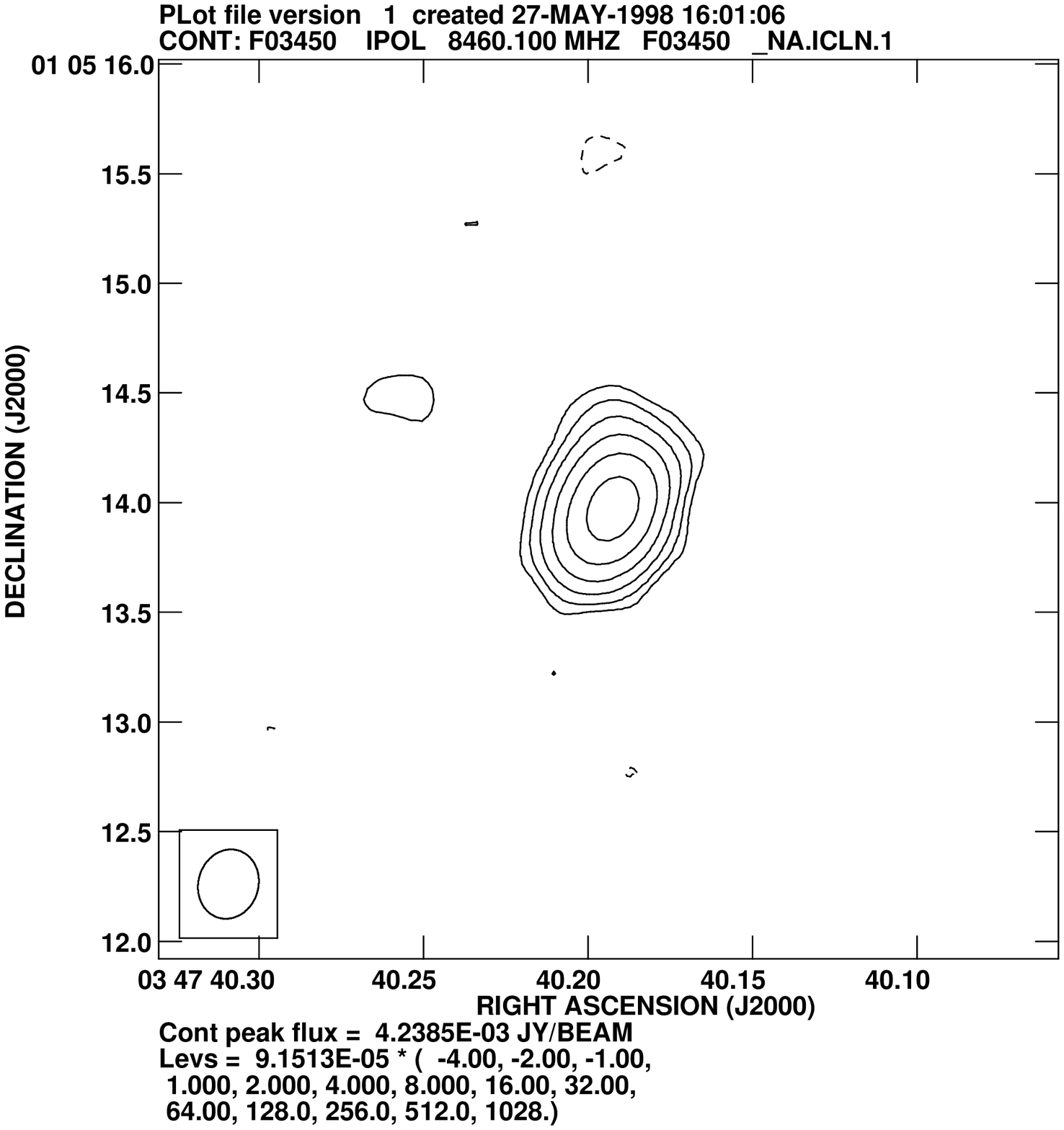}}
      \subfigure[) Markarian 618]{
        \includegraphics[width=5.0cm,clip,trim=0 47 0 35]{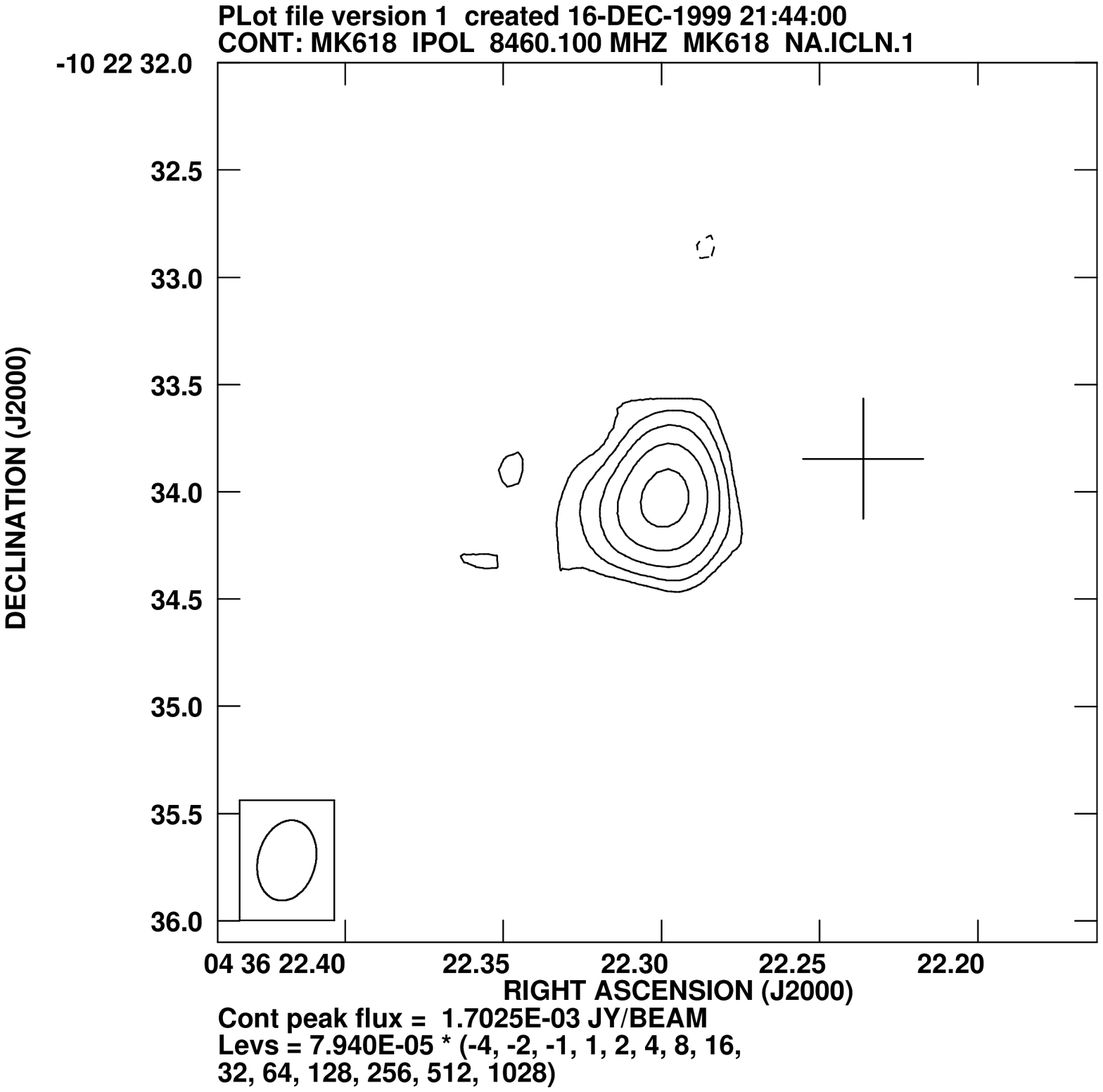}}
      \subfigure[) NGC 1667]{
        \includegraphics[width=4.9cm,clip,trim=0 47 0 35]{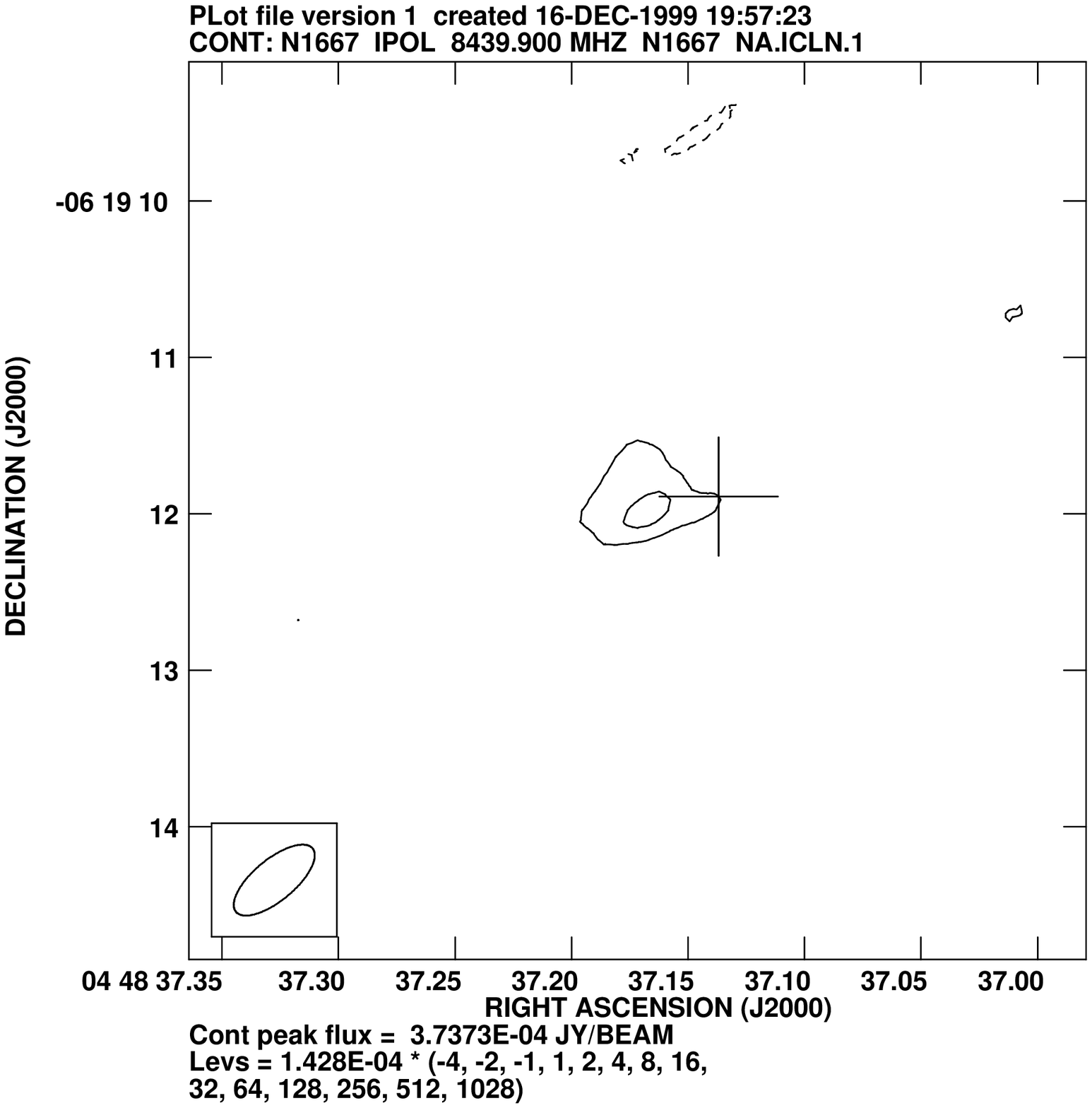}}
      \subfigure[) F04385-0828]{
        \includegraphics[width=5.0cm,clip,trim=0 62 0 35]{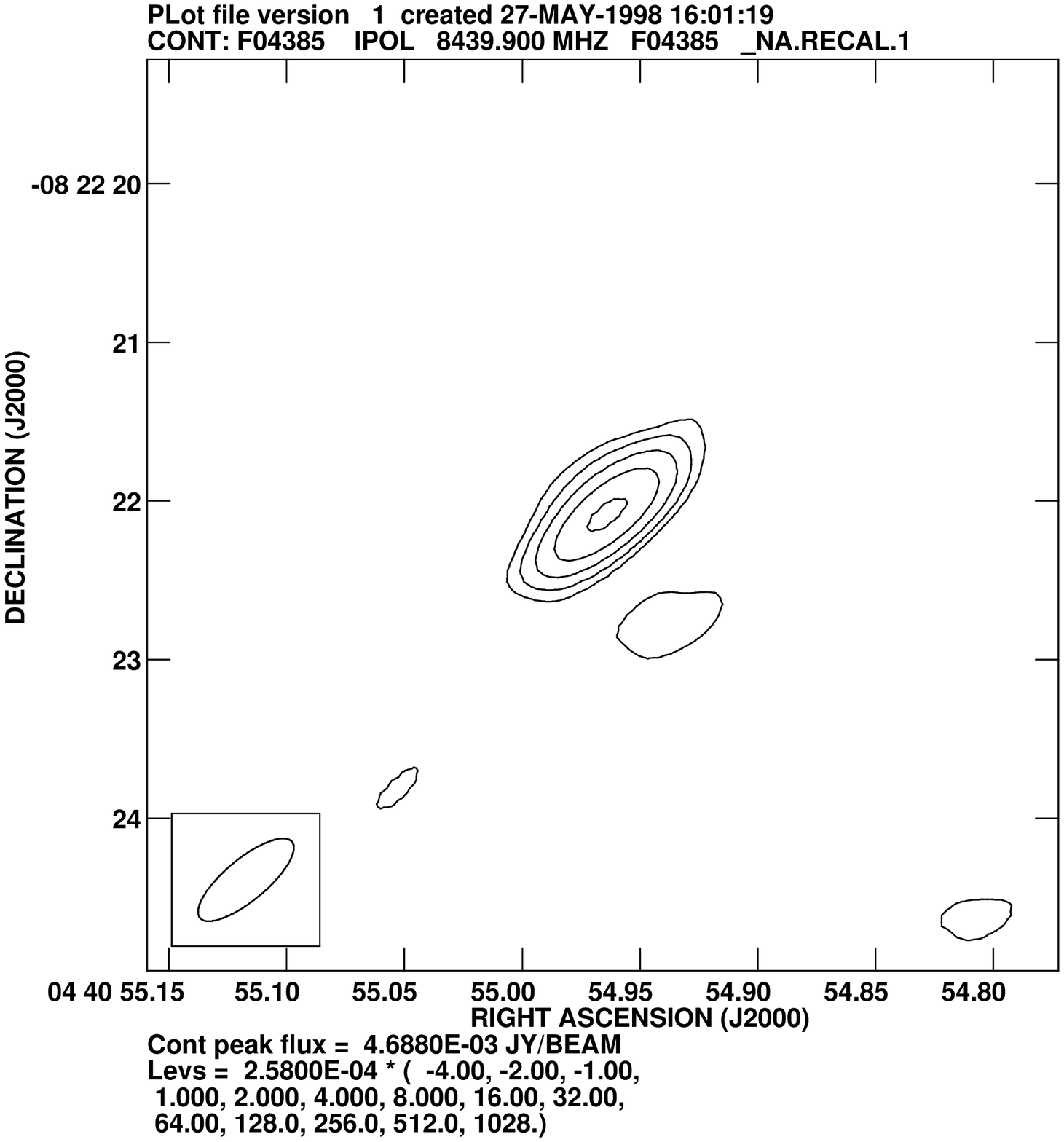}}
      \subfigure[) MCG-5-13-1]{
        \includegraphics[width=5.0cm,clip,trim=0 62 0 35]{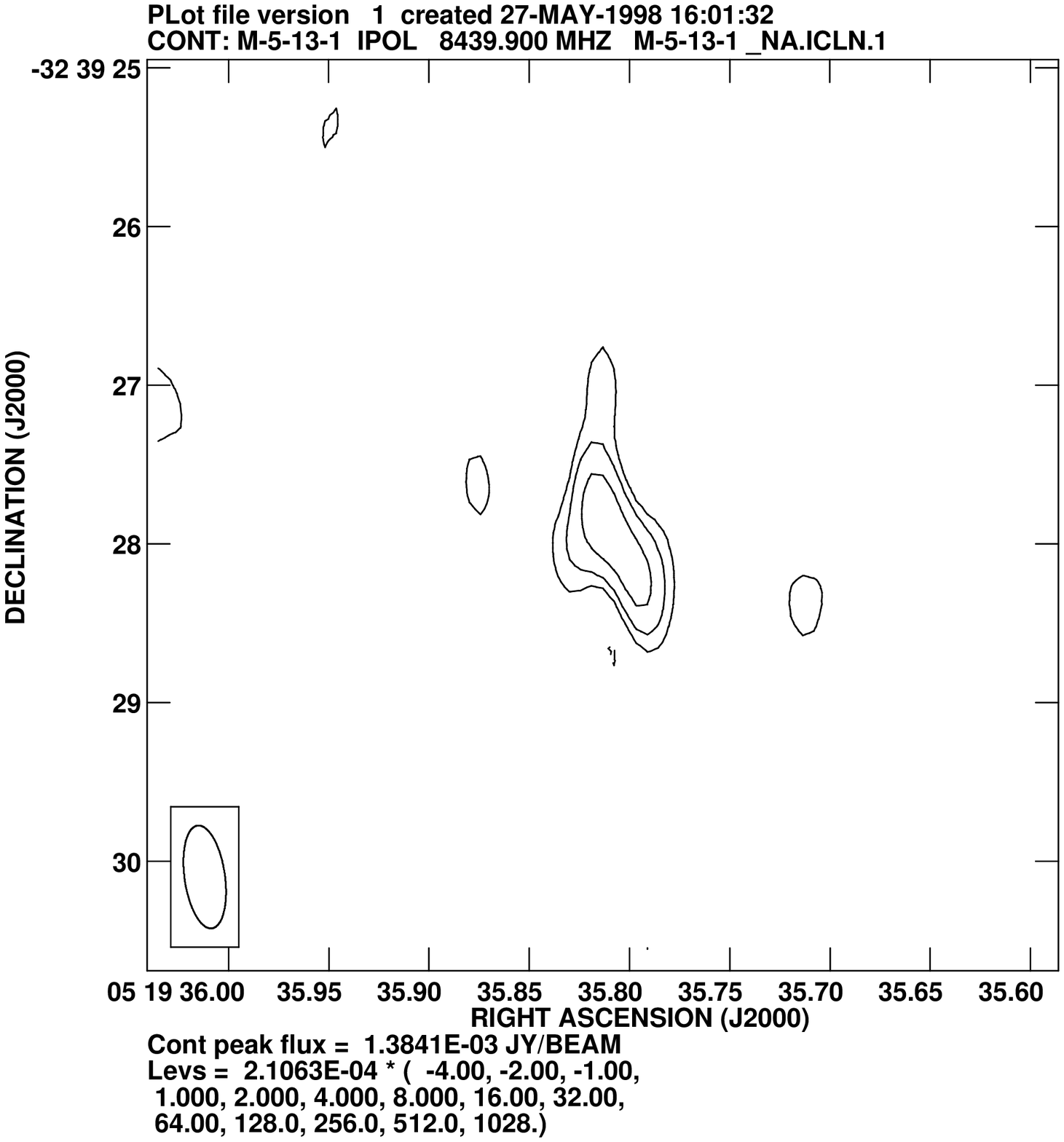}}
      \subfigure[) F05189-2524]{
        \includegraphics[width=5.0cm,clip,trim=0 62 0 35]{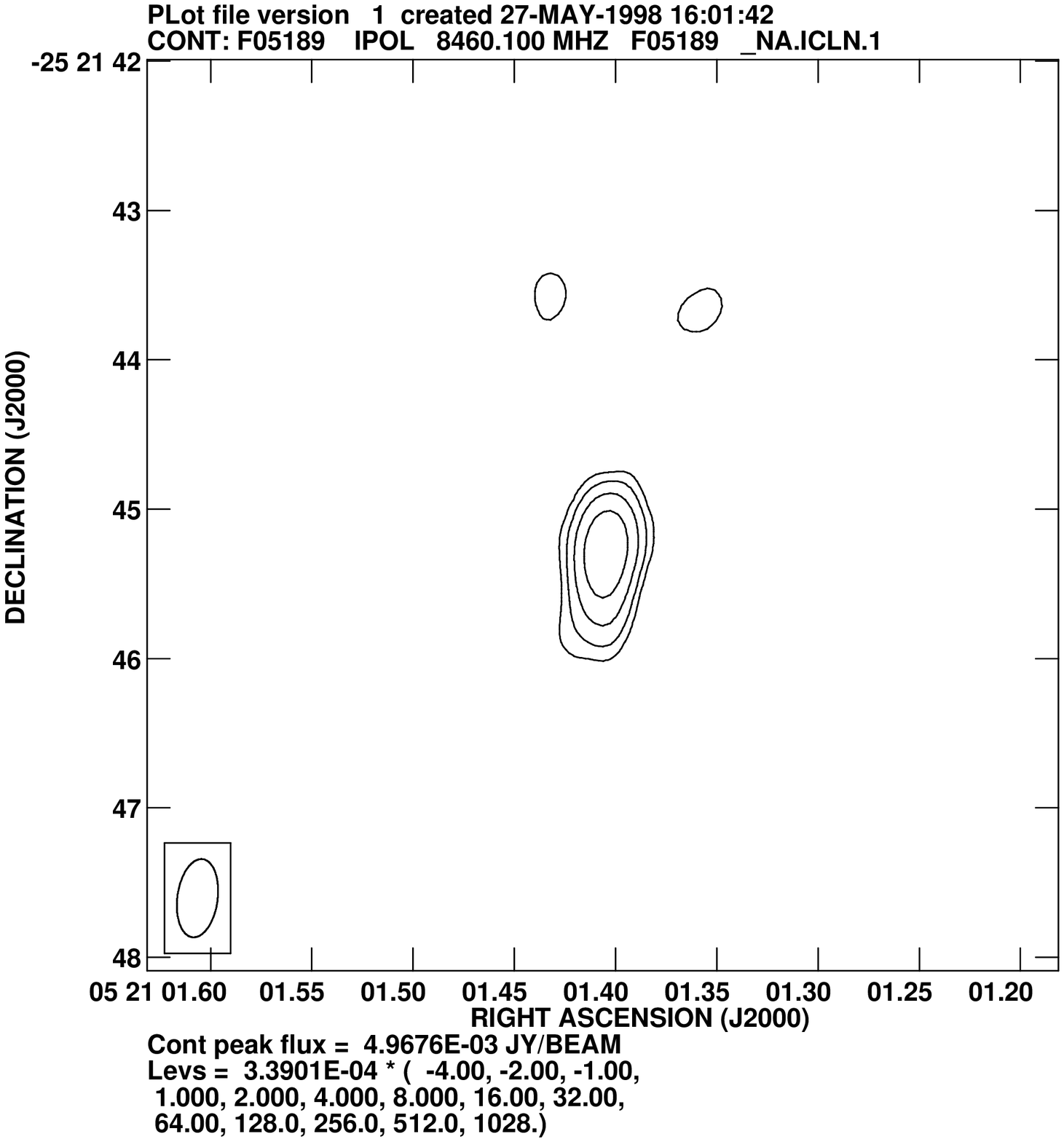}}
\caption{A-configuration 8.4 GHz images ({\it continued}).}
\end{figure*}

\setcounter{figure}{0}
\setcounter{subfigure}{24}

\begin{figure*}
\centering
       \subfigure[) F05563-3820]{
        \includegraphics[width=4.9cm,clip,trim=0 62 0 35]{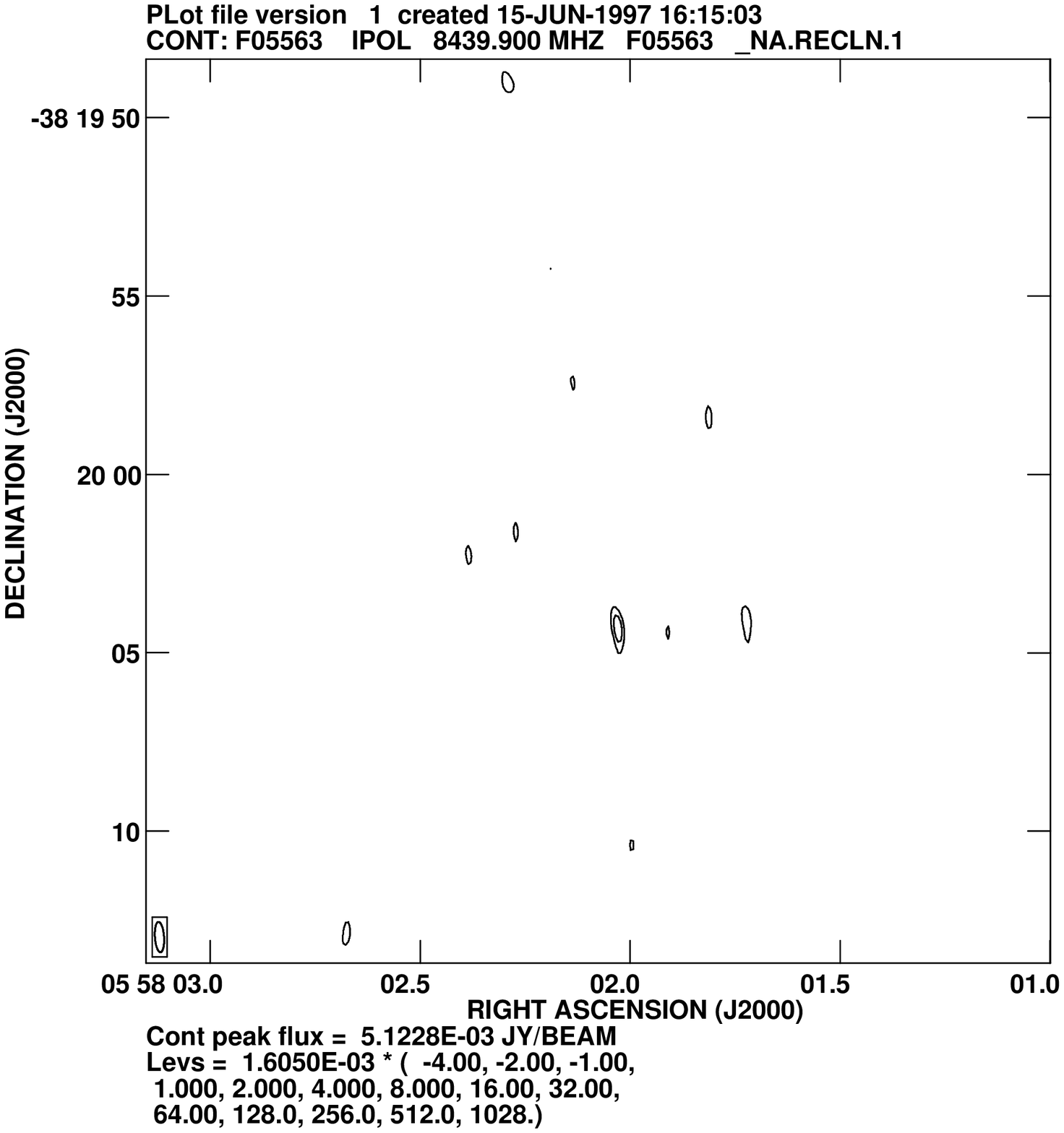}}
      \subfigure[) Markarian 6]{
        \includegraphics[width=5.0cm,clip,trim=0 47 0 35]{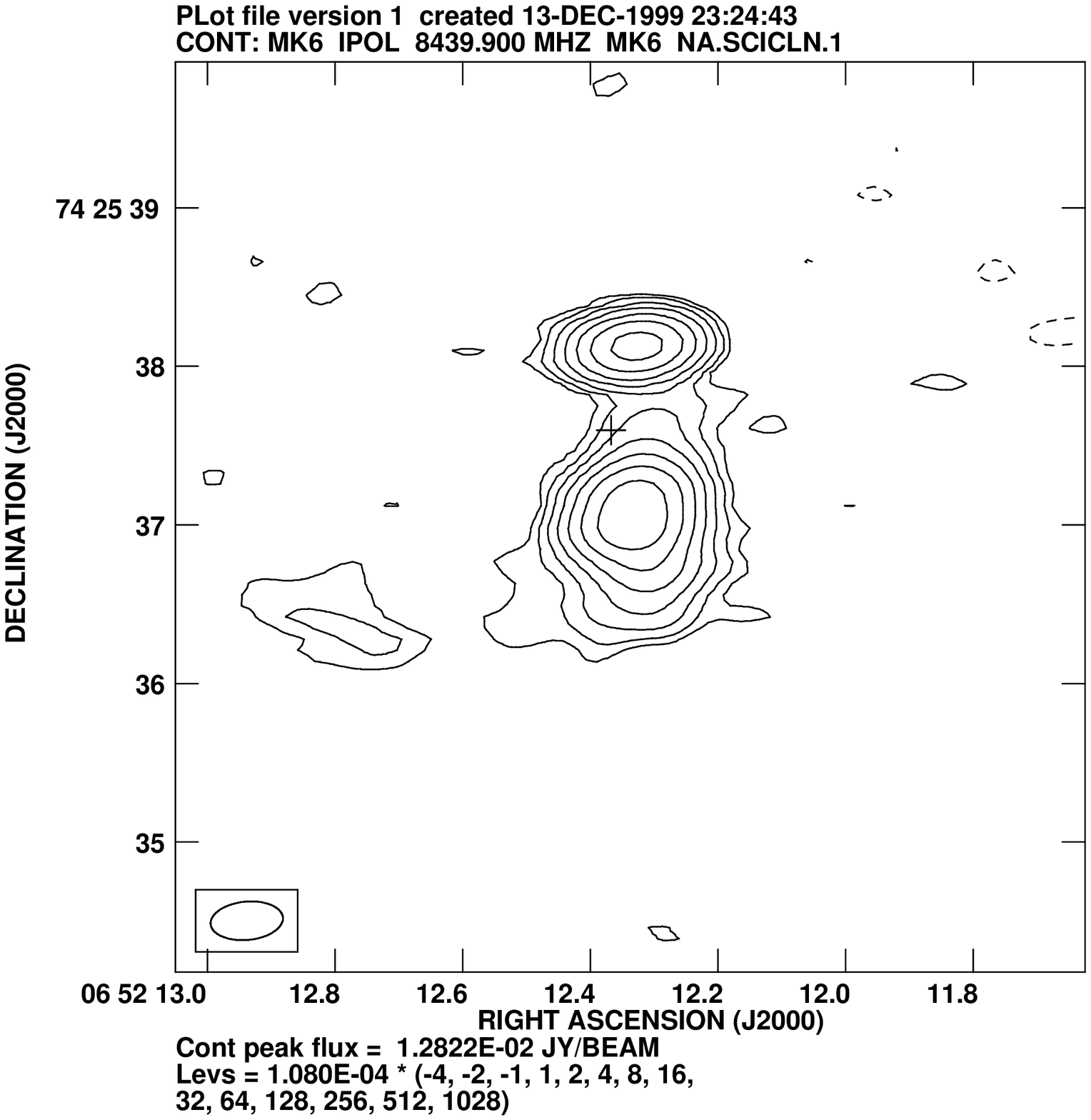}}
      \subfigure[) Markarian 9]{
        \includegraphics[width=5.0cm,clip,trim=0 47 0 35]{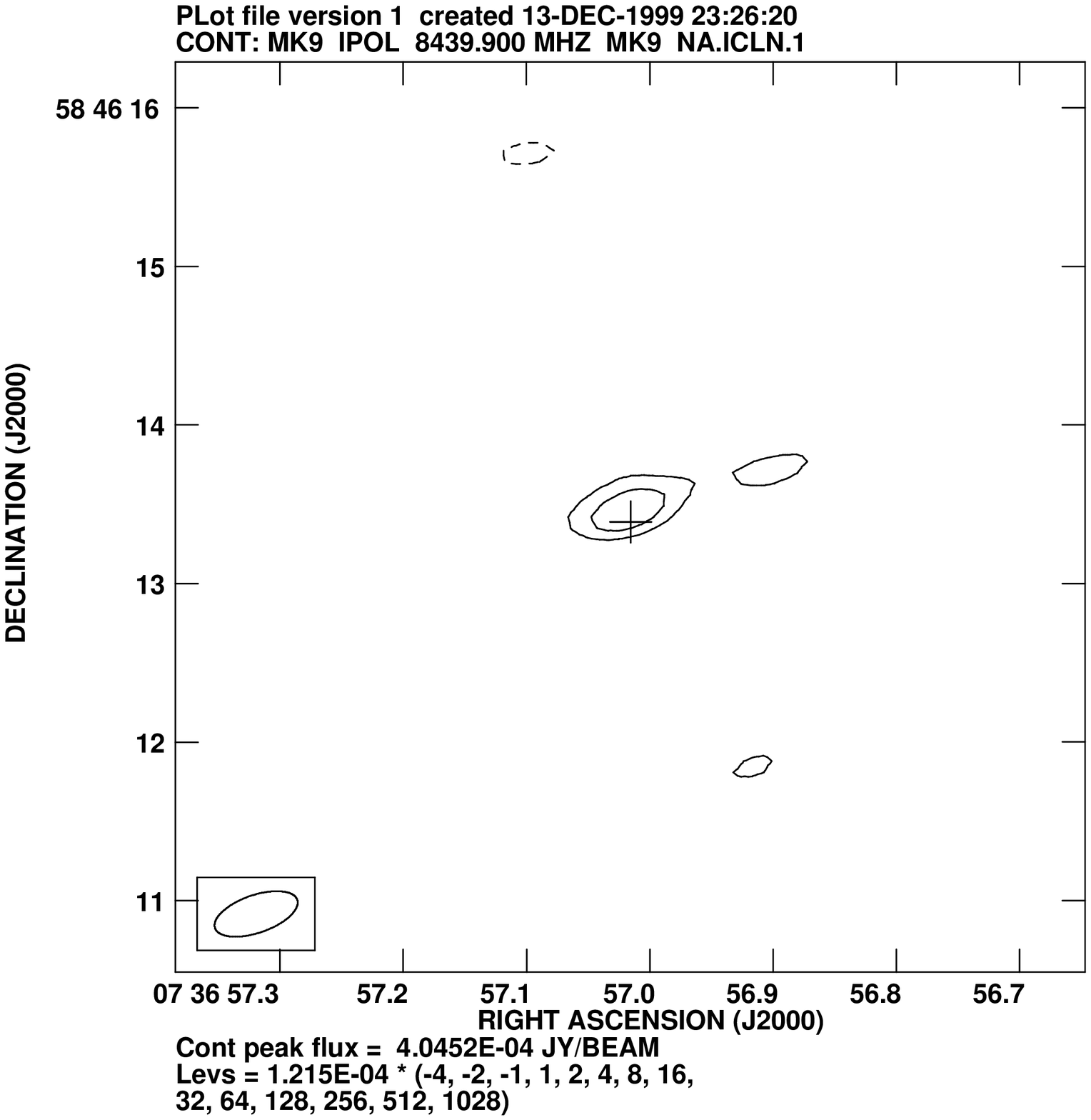}}
      \subfigure[) Markarian 79]{
        \includegraphics[width=5.0cm,clip,trim=0 47 0 35]{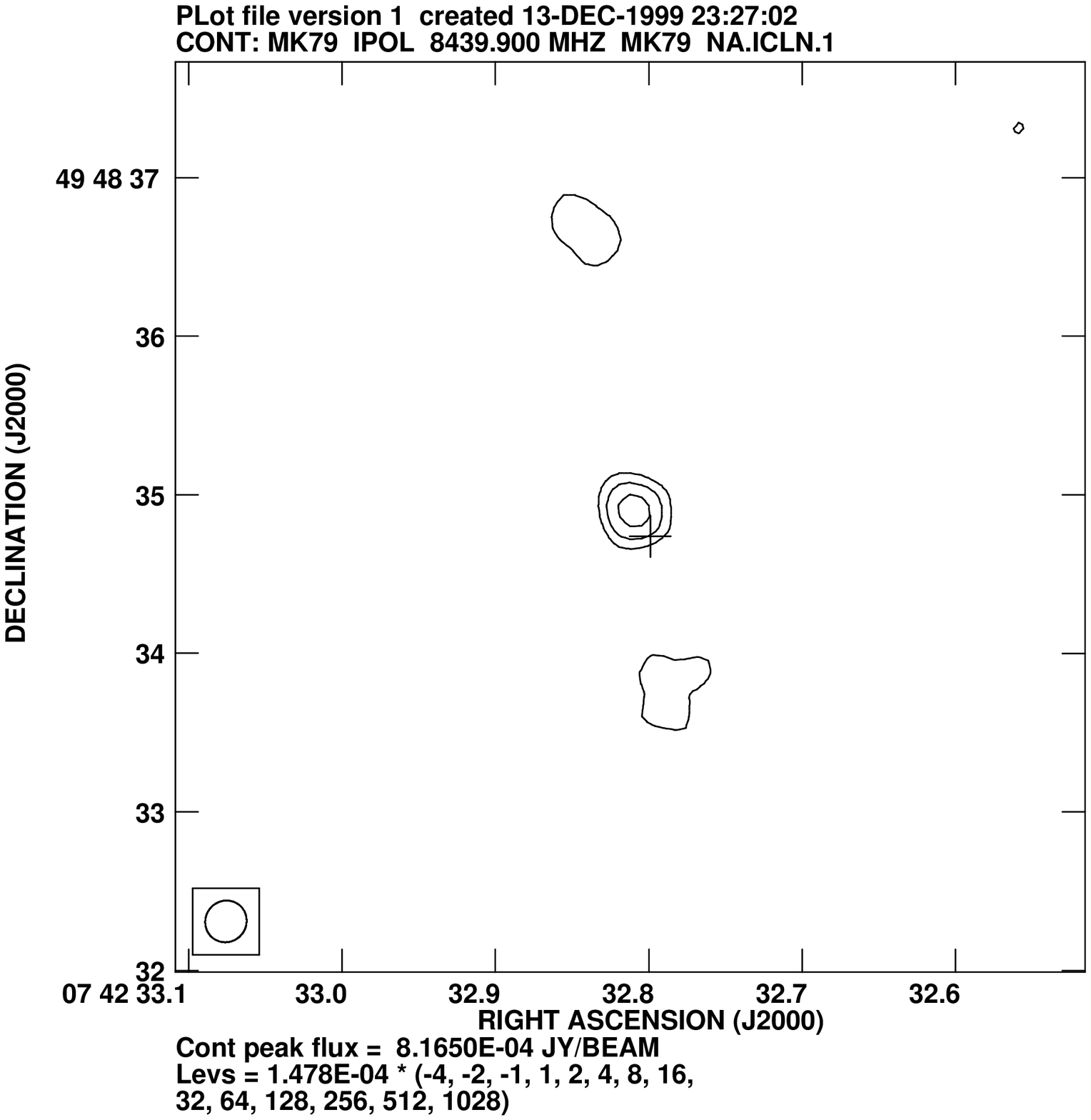}}
      \subfigure[) F07599+6508]{
        \includegraphics[width=5.0cm,clip,trim=0 62 0 35]{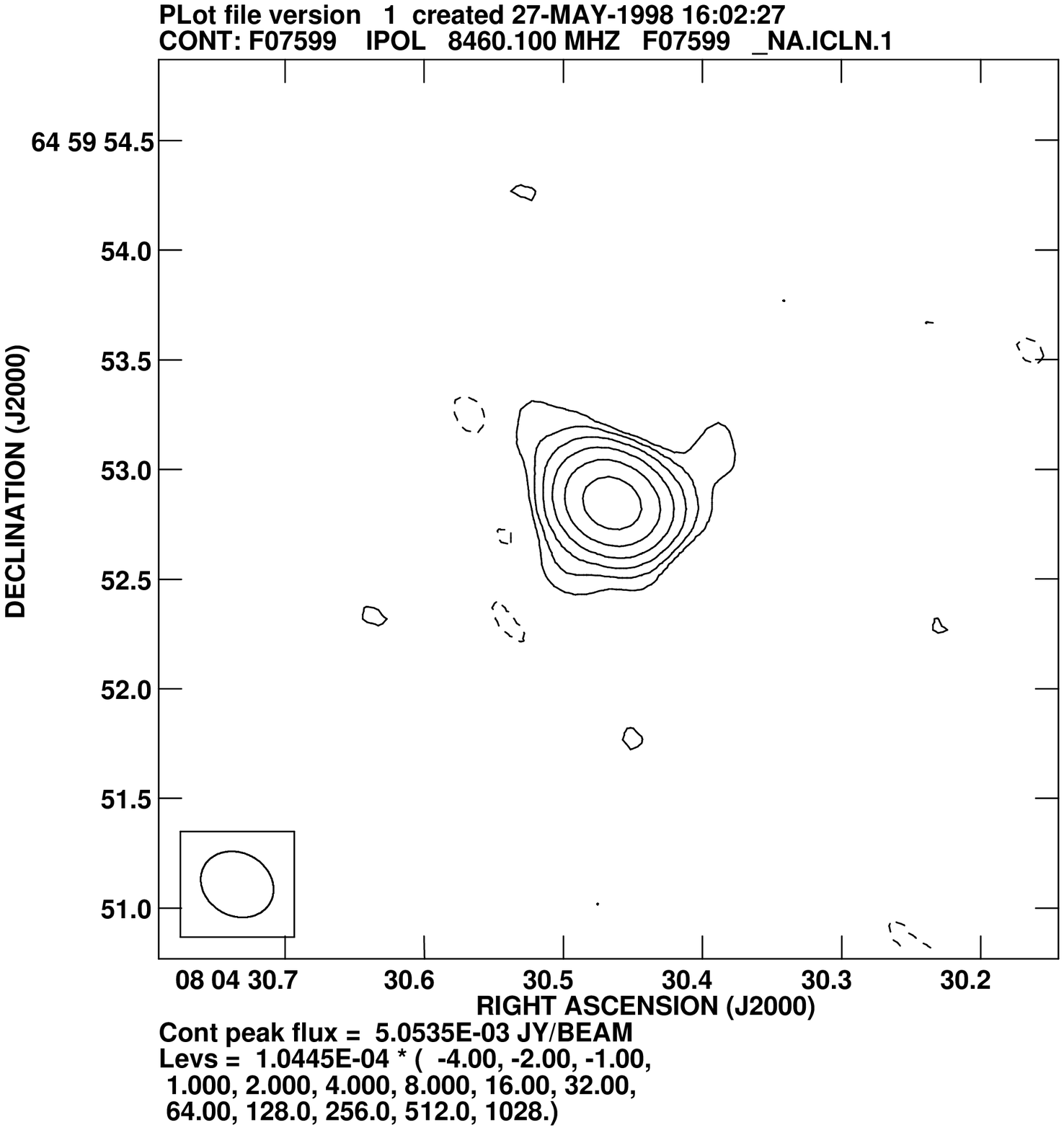}}
      \subfigure[) NGC 2639]{
        \includegraphics[width=5.0cm,clip,trim=0 47 0 35]{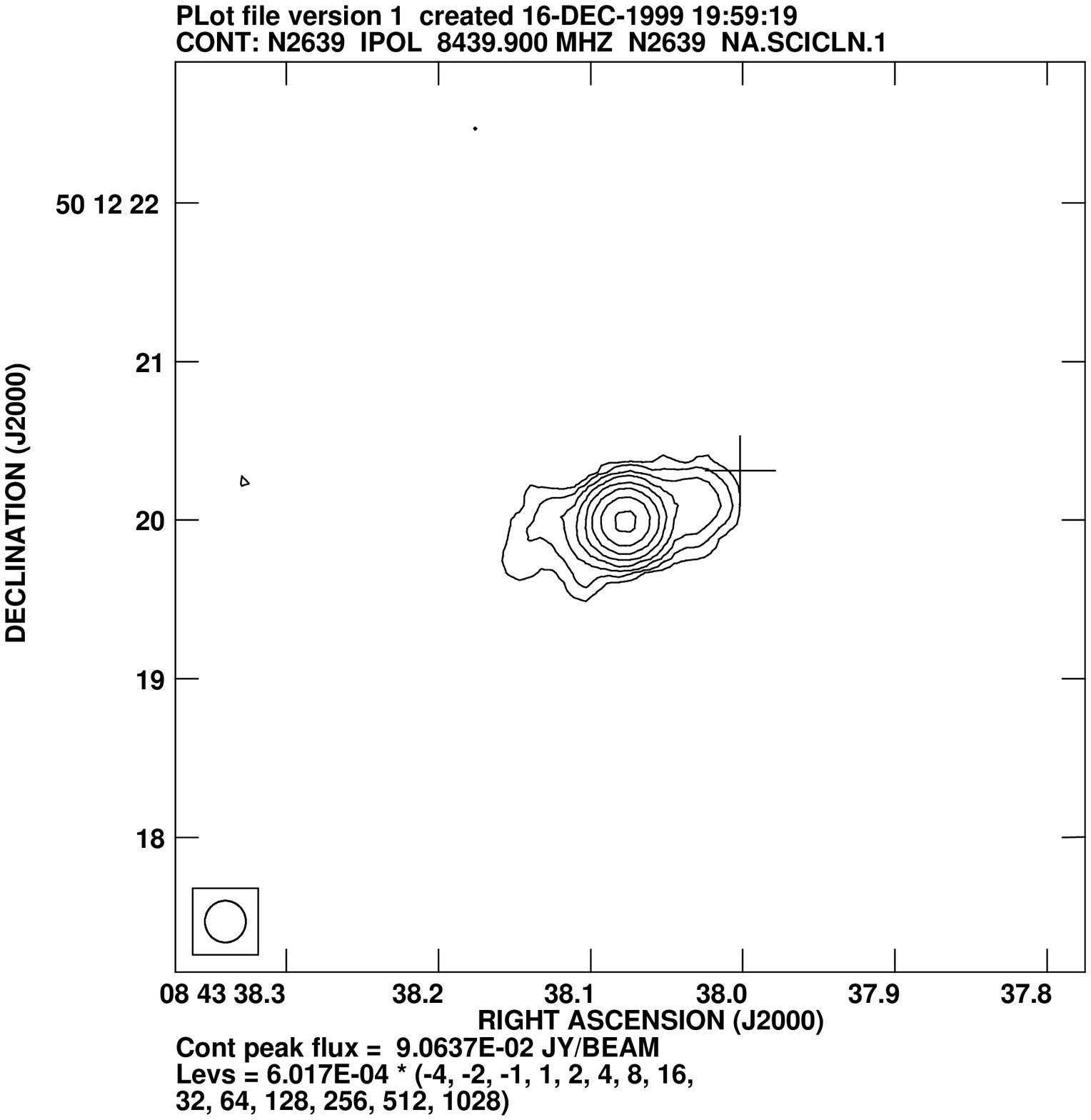}}
      \subfigure[) OJ287]{
        \includegraphics[width=4.8cm,clip,trim=0 62 0 35]{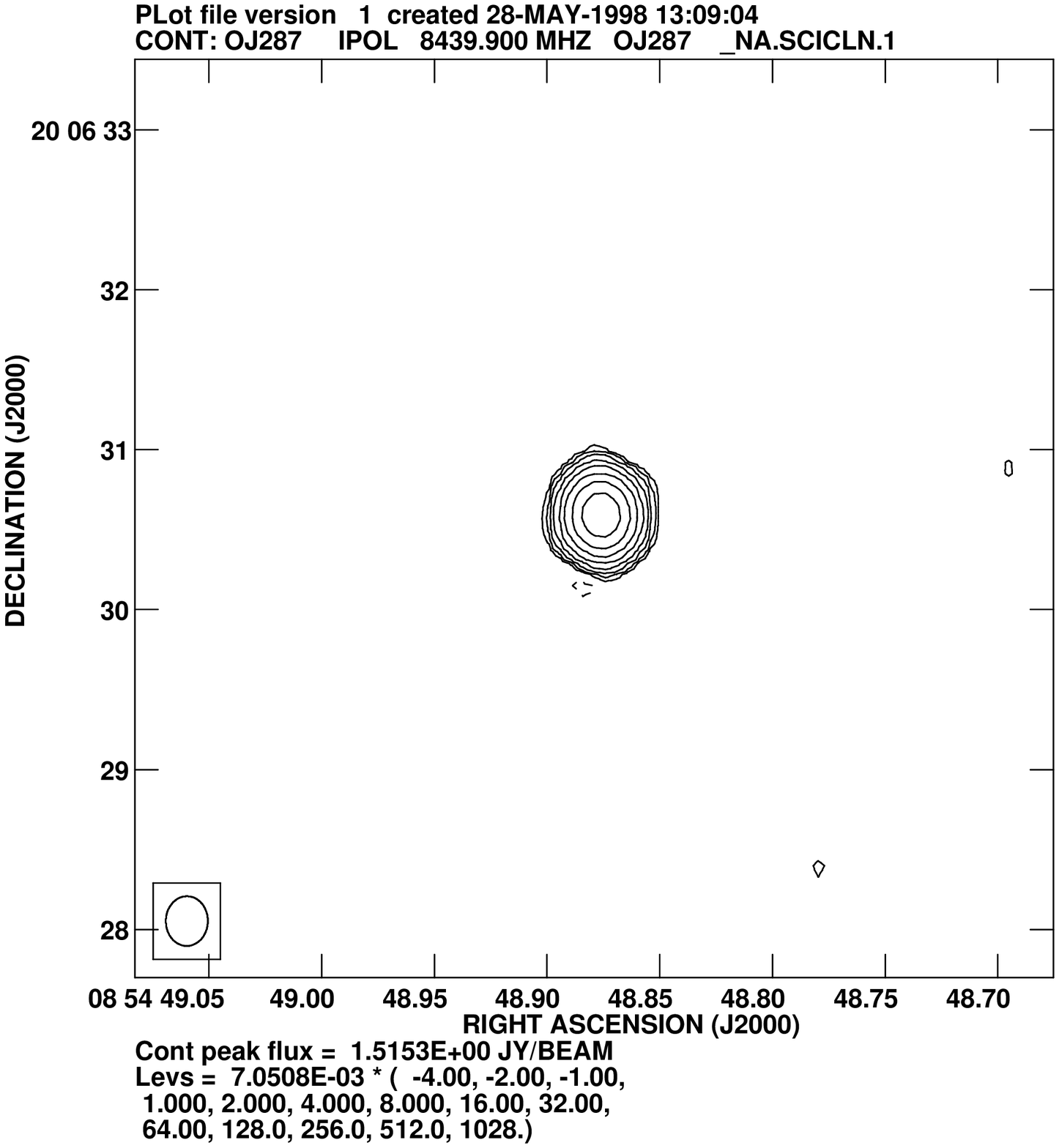}}
      \subfigure[) F08572+3915]{
        \includegraphics[width=4.8cm,clip,trim=0 62 0 35]{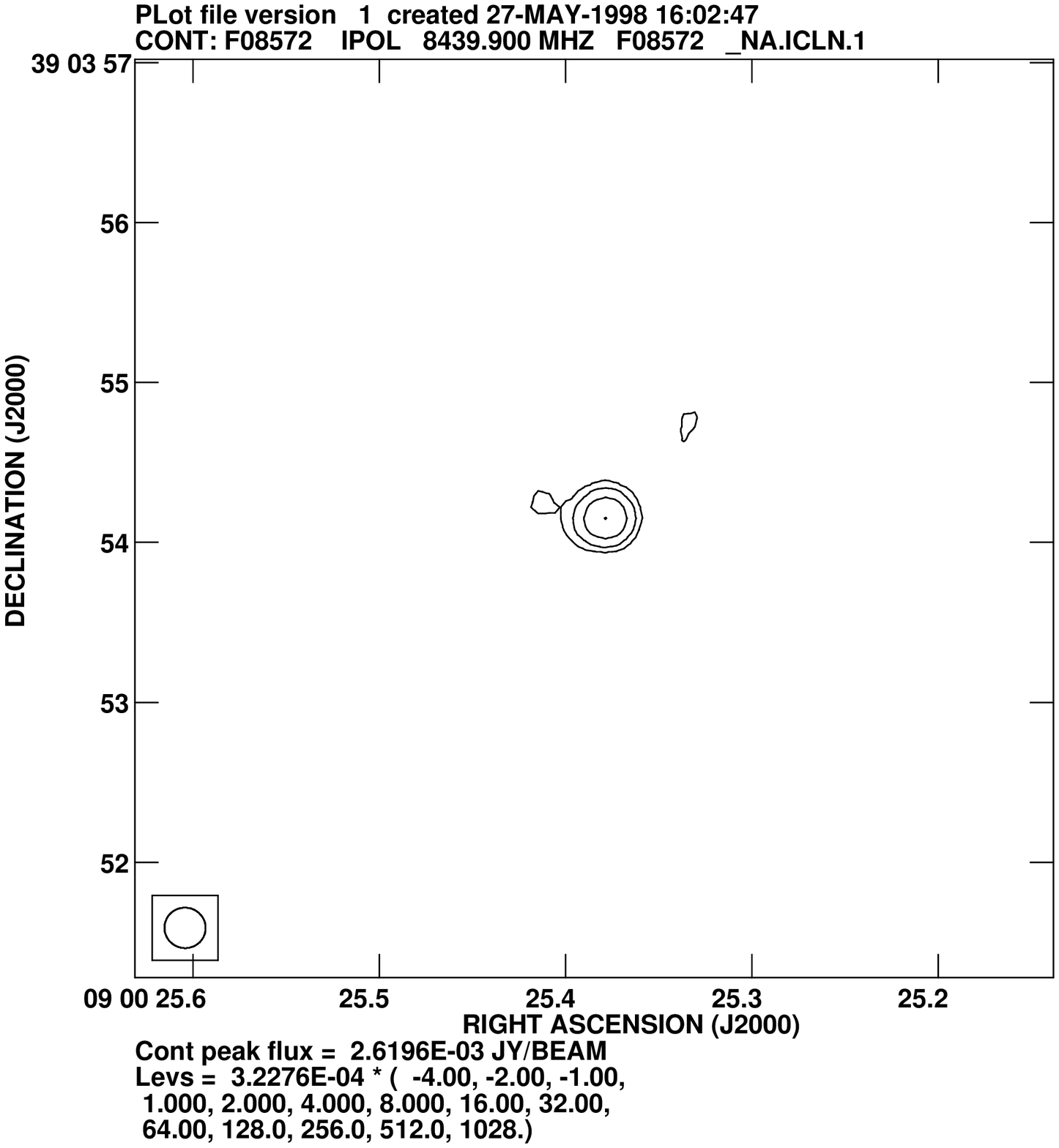}}
      \subfigure[) Markarian 704]{
        \includegraphics[width=4.8cm,clip,trim=0 62 0 35]{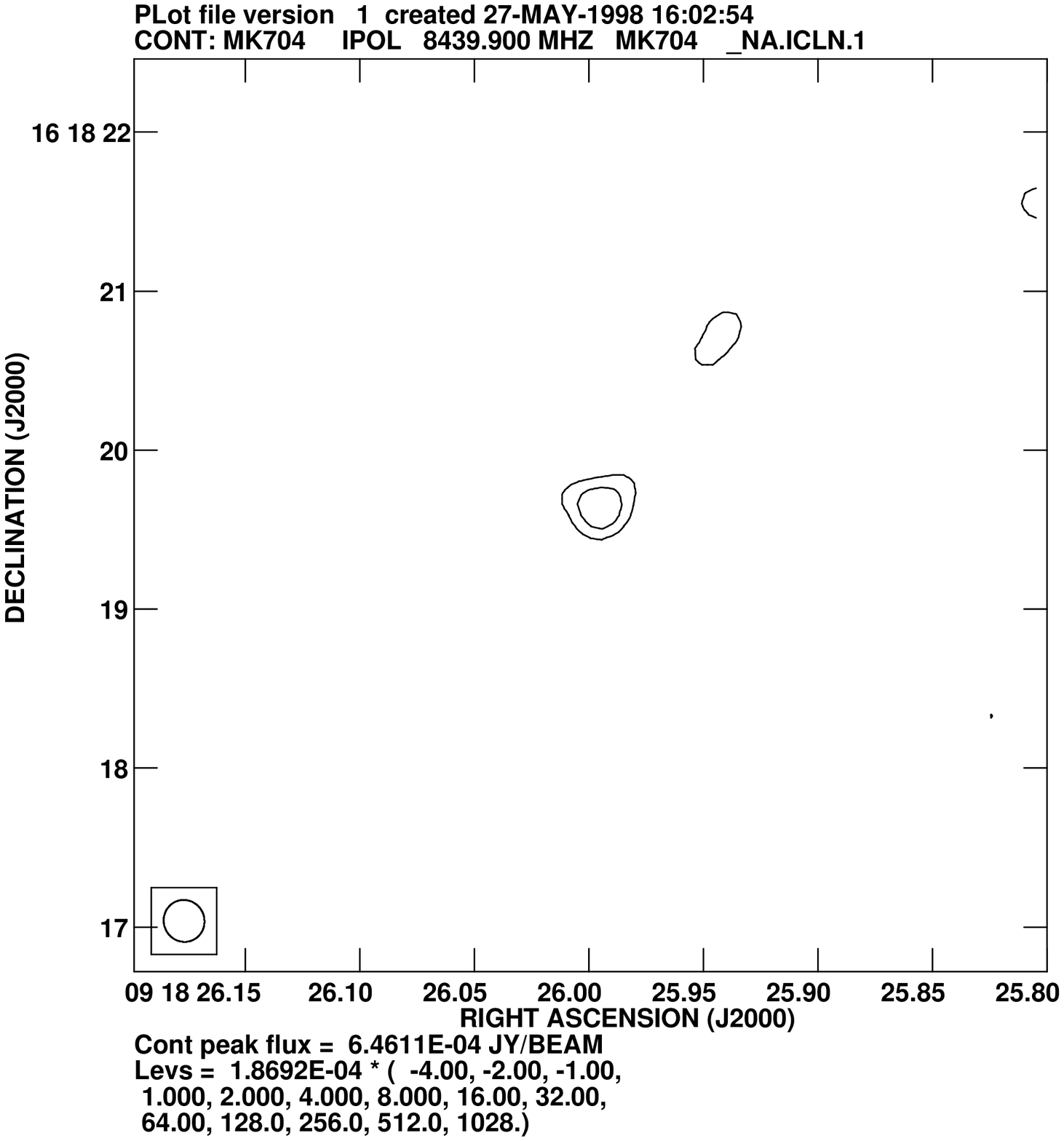}}
      \subfigure[) UGC 5101]{
        \includegraphics[width=5.1cm,clip,trim=0 62 0 35]{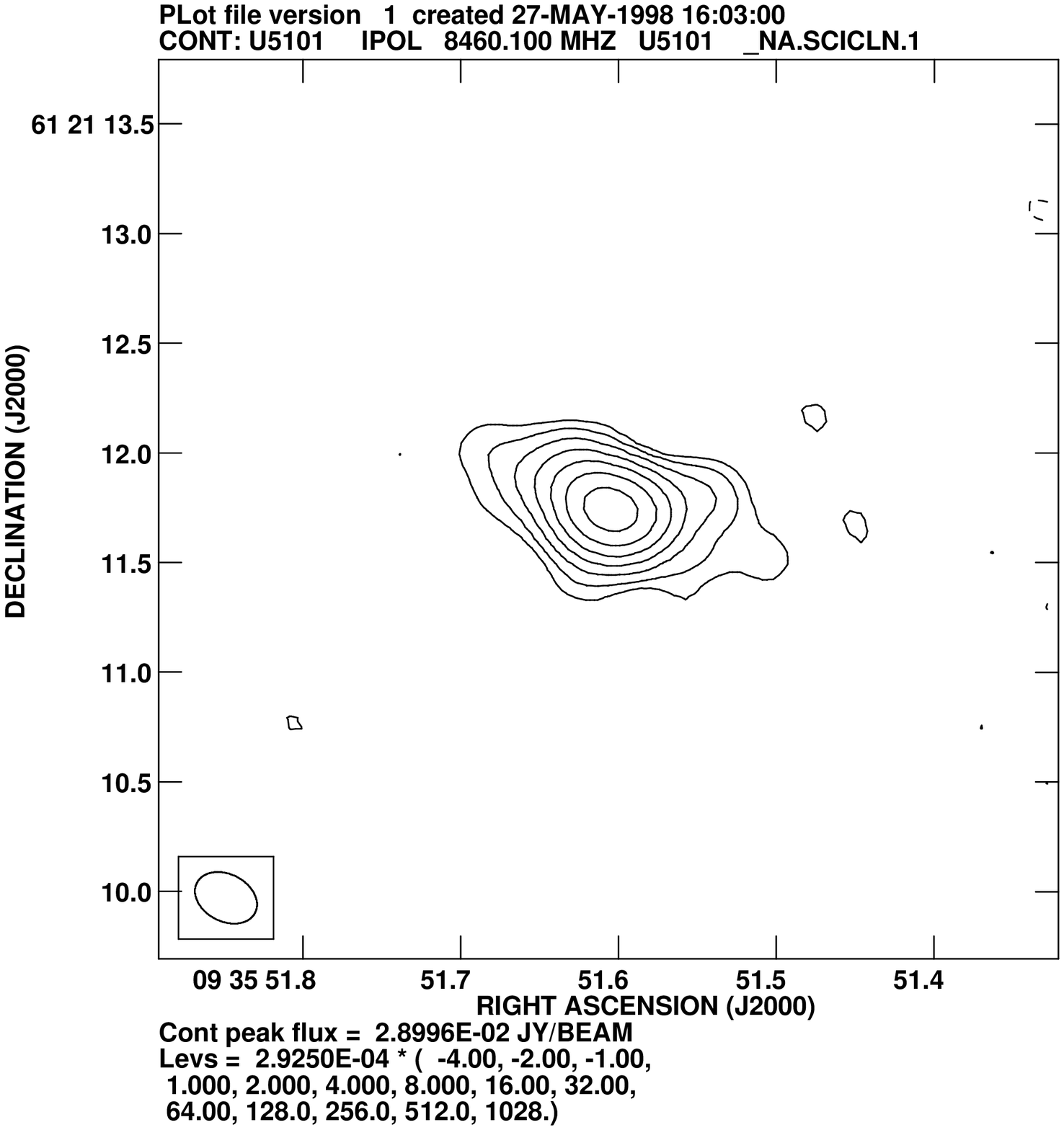}}
      \subfigure[) NGC 2992]{
        \includegraphics[width=5.2cm,clip,trim=0 47 0 35]{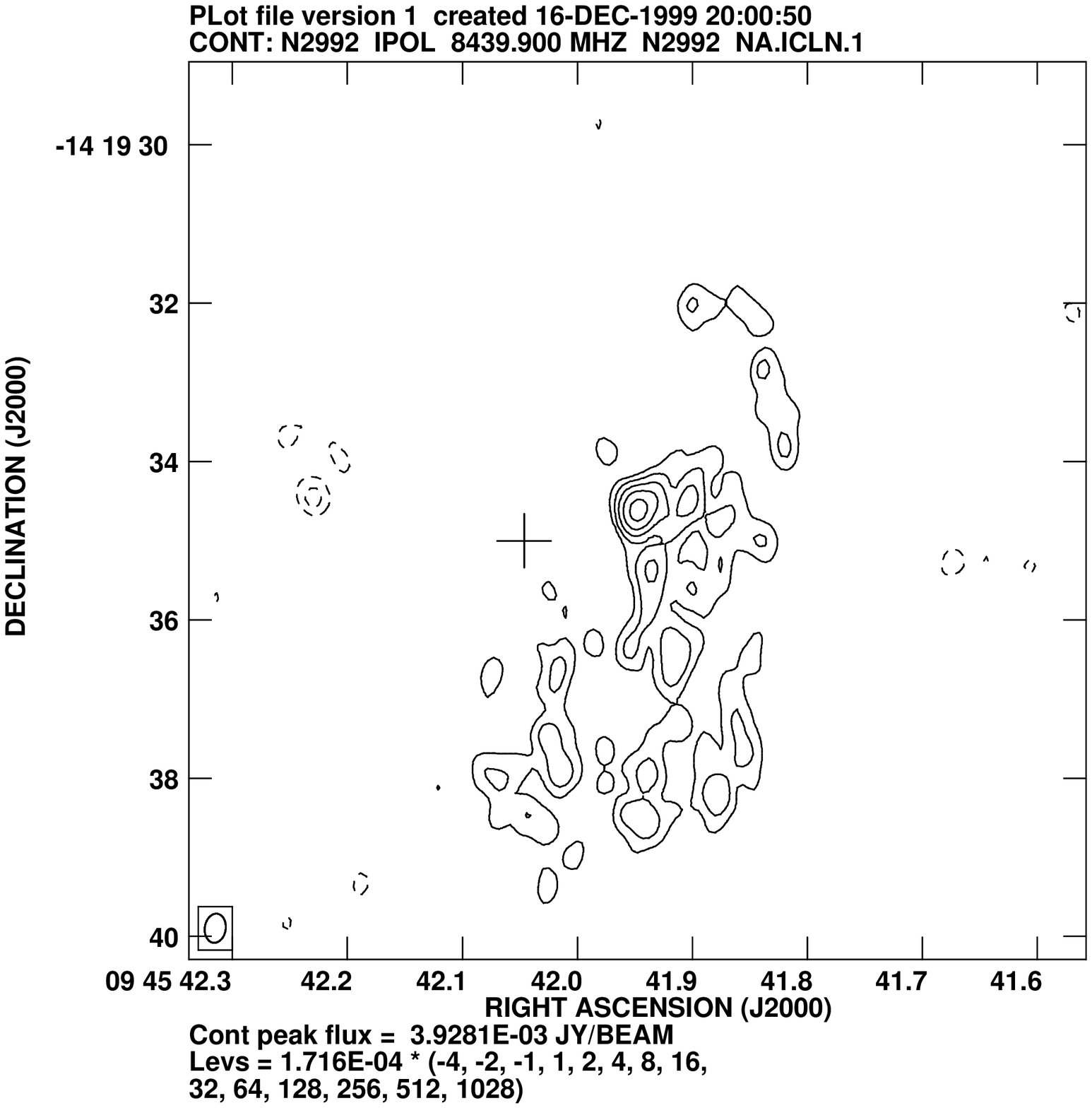}}
      \subfigure[) Markarian 1239]{
        \includegraphics[width=5.2cm,clip,trim=0 47 0 35]{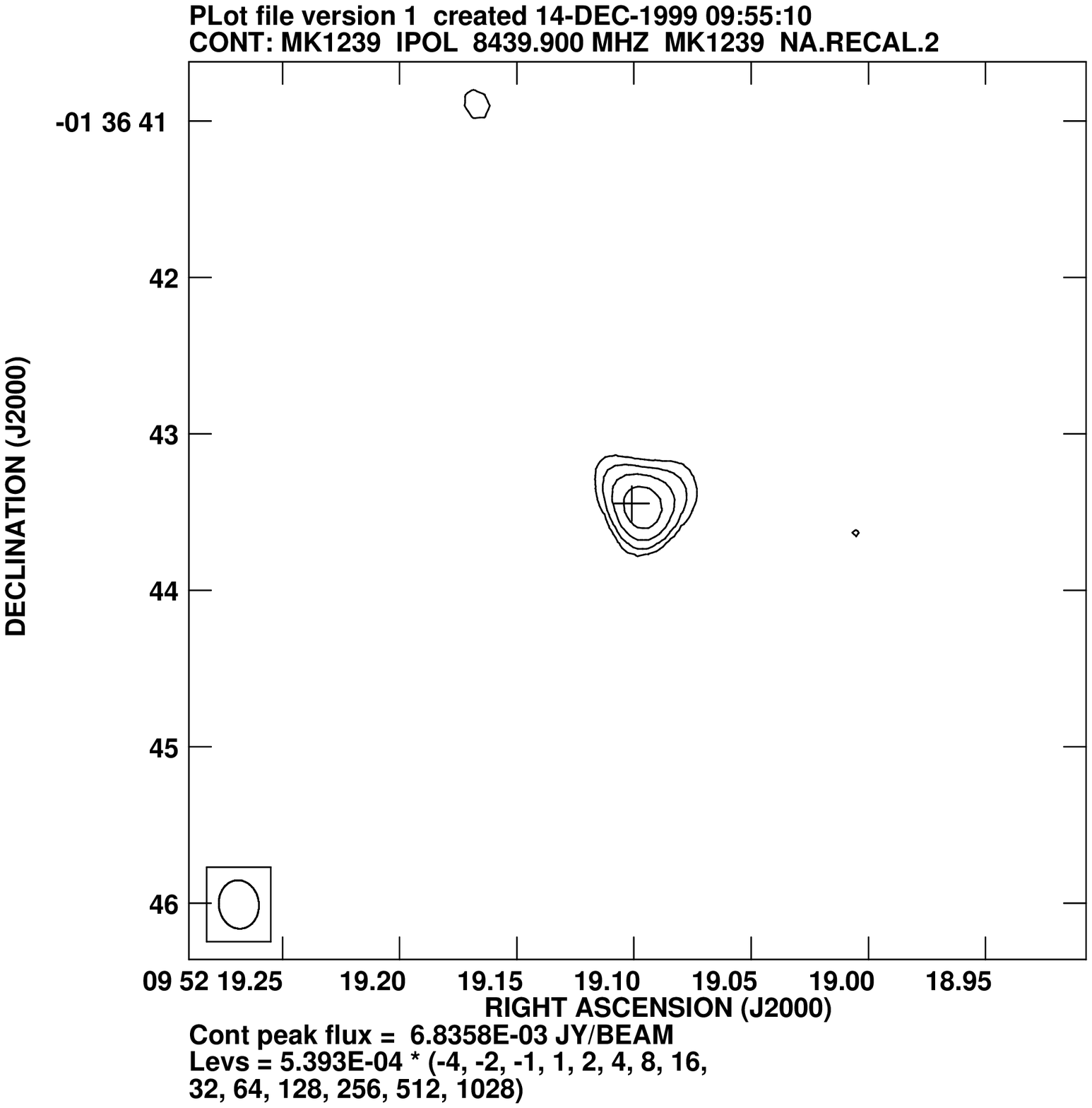}}
\caption{A-configuration 8.4 GHz images ({\it continued}).}
\end{figure*}

\setcounter{figure}{0}
\setcounter{subfigure}{36}
\begin{figure*}
\centering
      \subfigure[) NGC 3031]{
        \includegraphics[width=5.1cm,clip,trim=0 47 0 35]{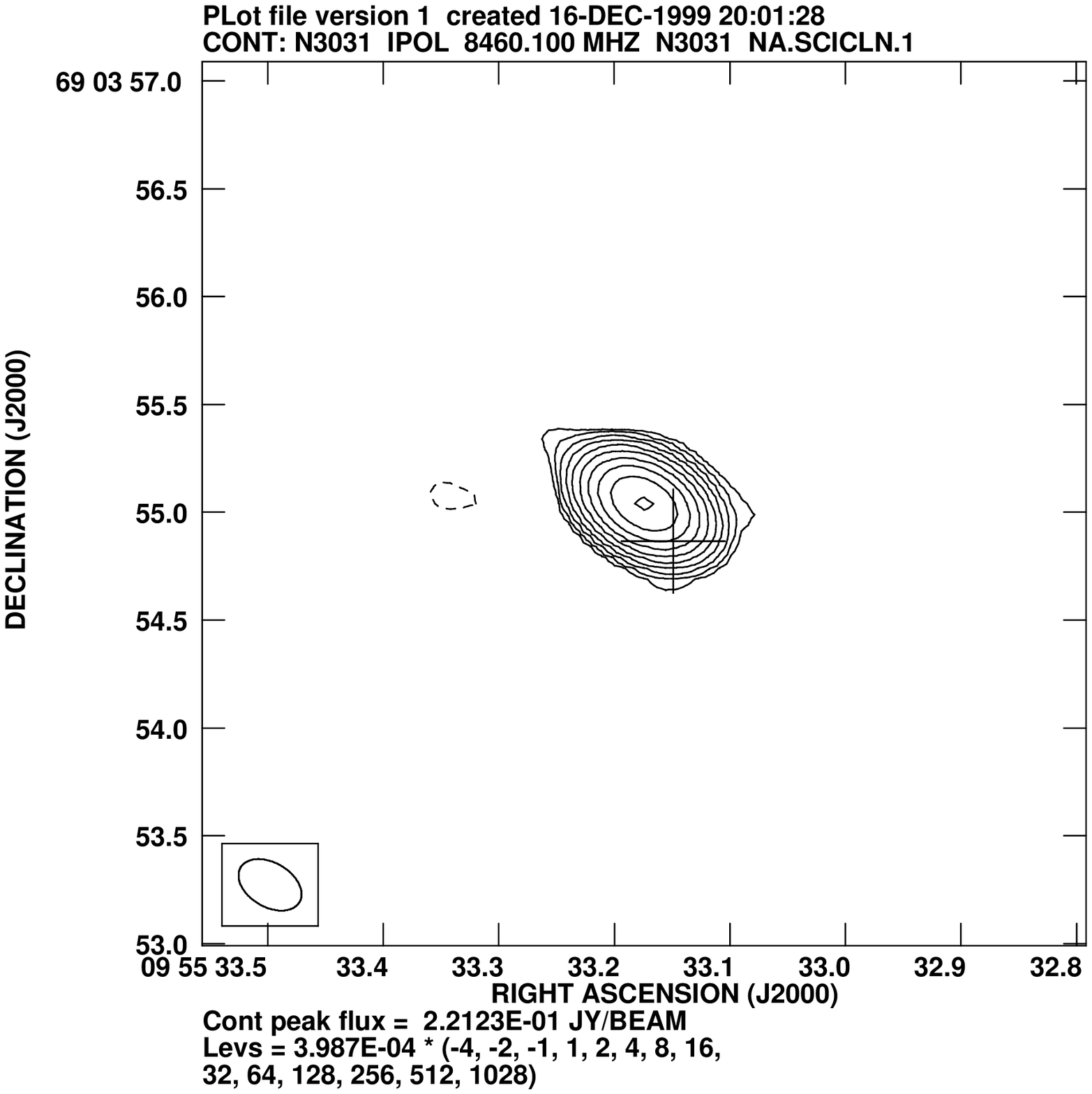}}
      \subfigure[) 3C 234]{
        \includegraphics[width=4.8cm,clip,trim=0 62 0 35]{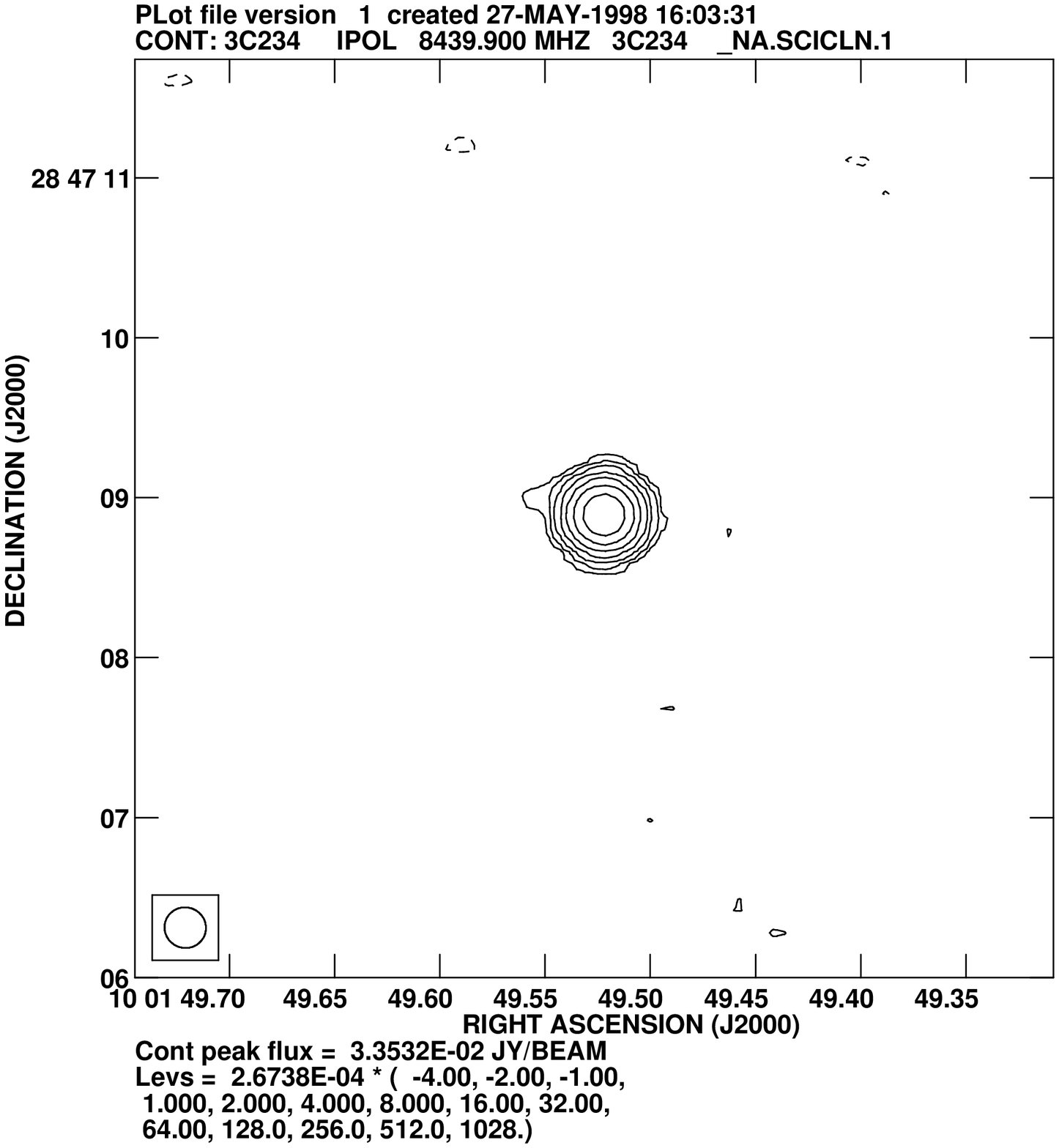}}
      \subfigure[) NGC 4579]{
        \includegraphics[width=5.0cm,clip,trim=0 47 0 35]{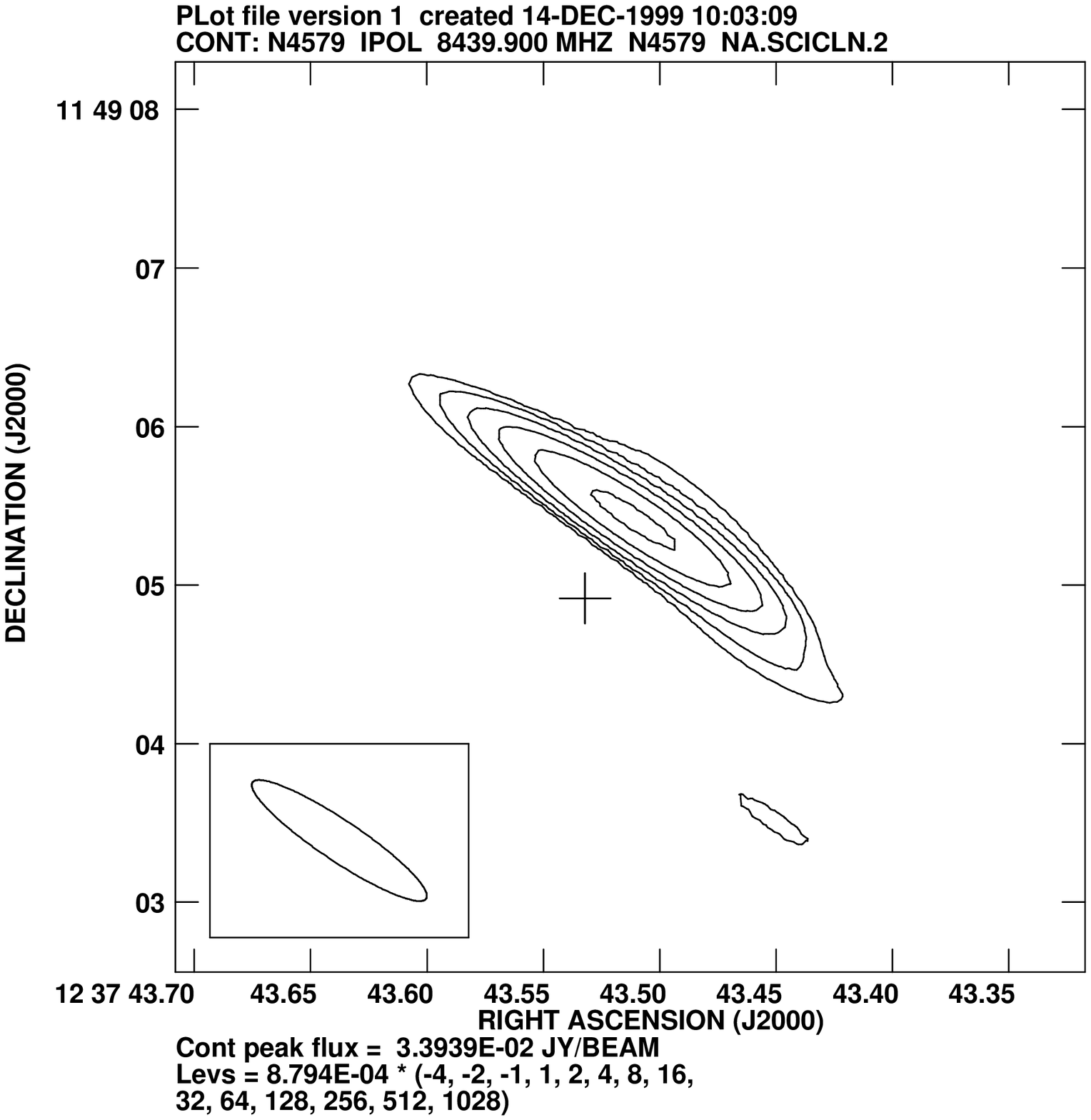}}
      \subfigure[) NGC 4593]{
        \includegraphics[width=5.0cm,clip,trim=0 47 0 35]{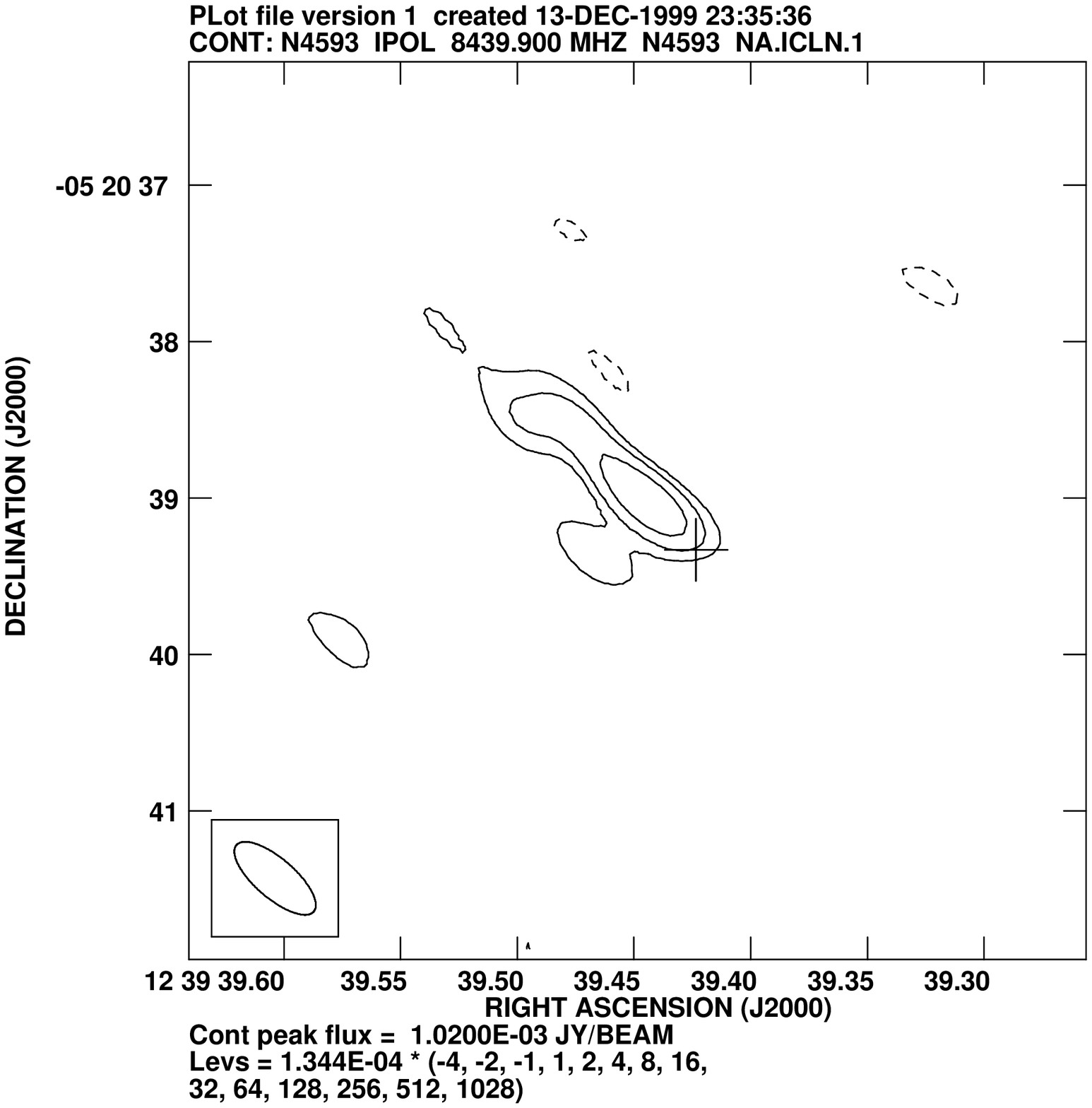}}
      \subfigure[) NGC 4594]{
        \includegraphics[width=4.8cm,clip,trim=0 62 0 35]{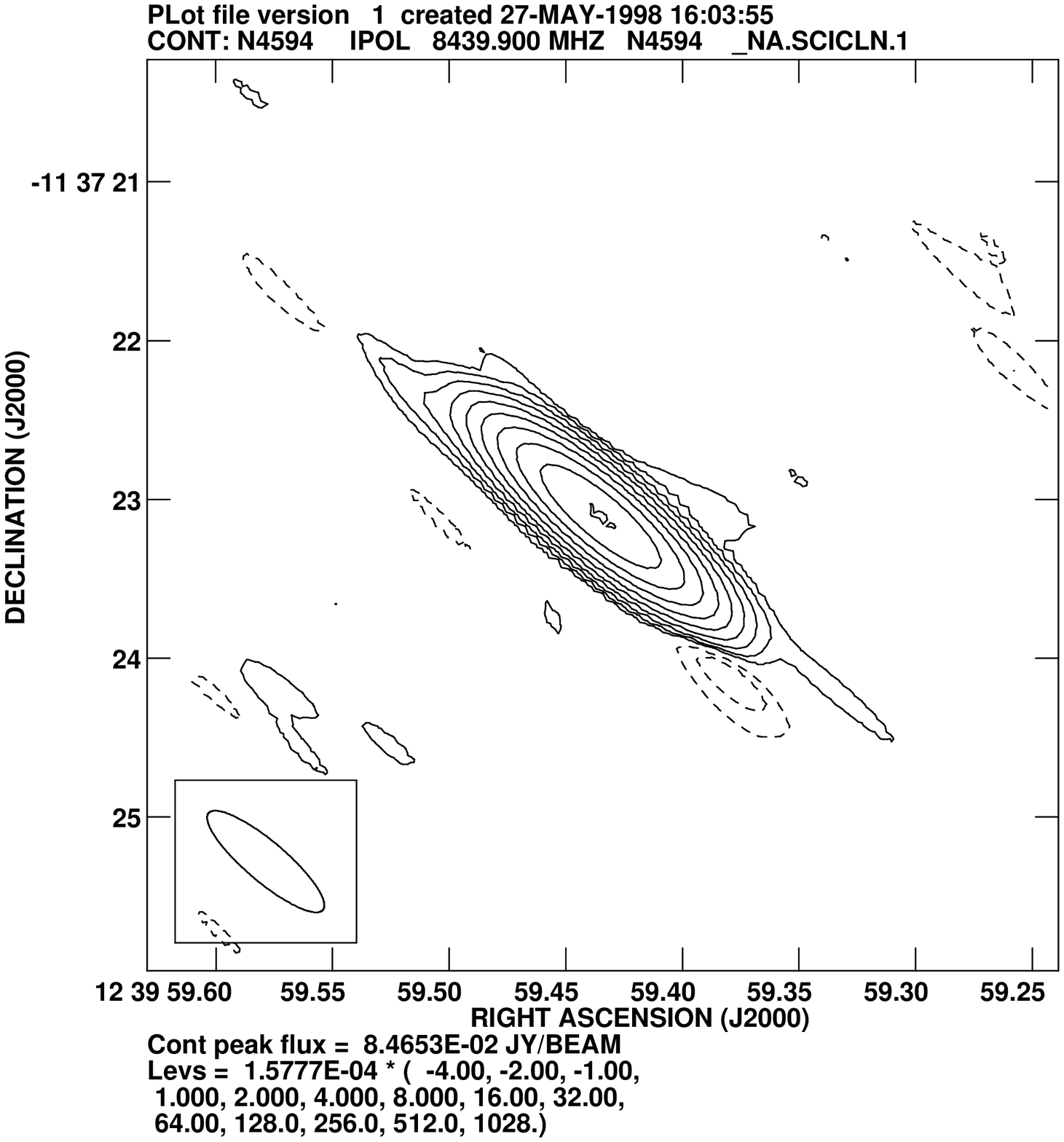}}
      \subfigure[) TOL1238]{
        \includegraphics[width=4.8cm,clip,trim=0 62 0 35]{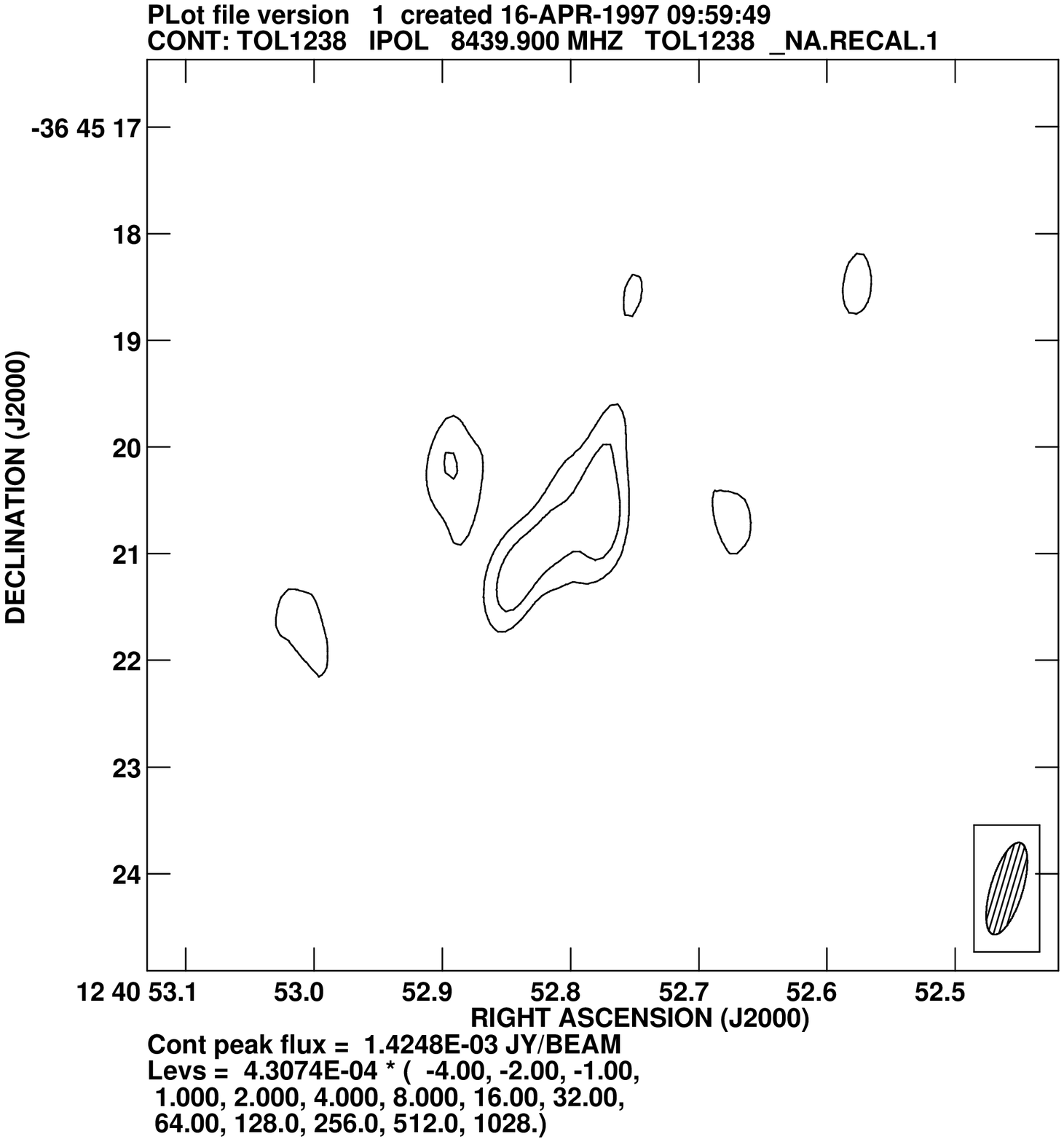}}
      \subfigure[) MCG-2-33-3]{
        \includegraphics[width=4.8cm,clip,trim=0 62 0 35]{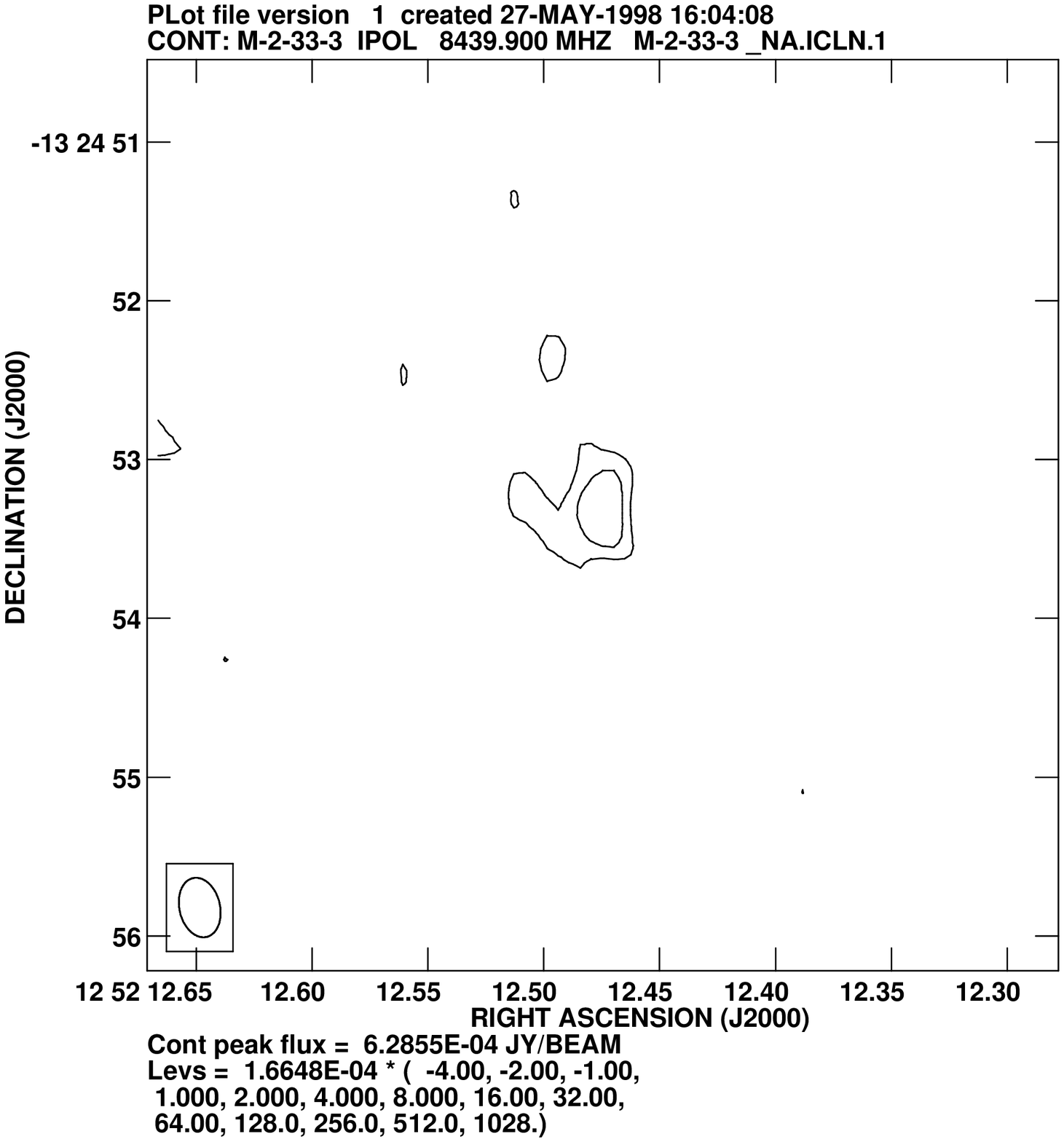}}
      \subfigure[) PGC 044896]{
        \includegraphics[width=5.1cm,clip,trim=0 48 0 35]{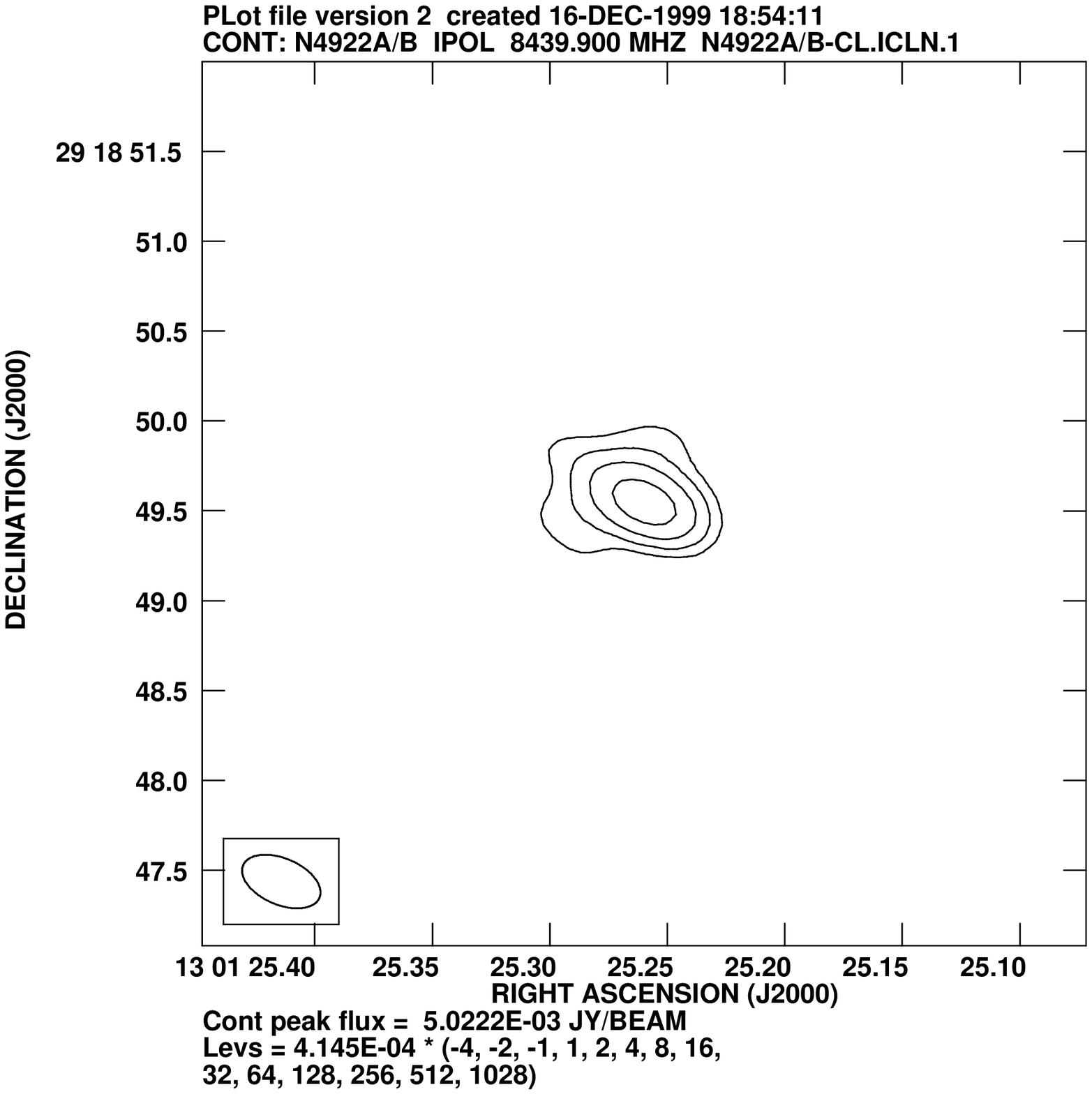}}
      \subfigure[) NGC 4941]{
        \includegraphics[width=4.8cm,clip,trim=0 62 0 35]{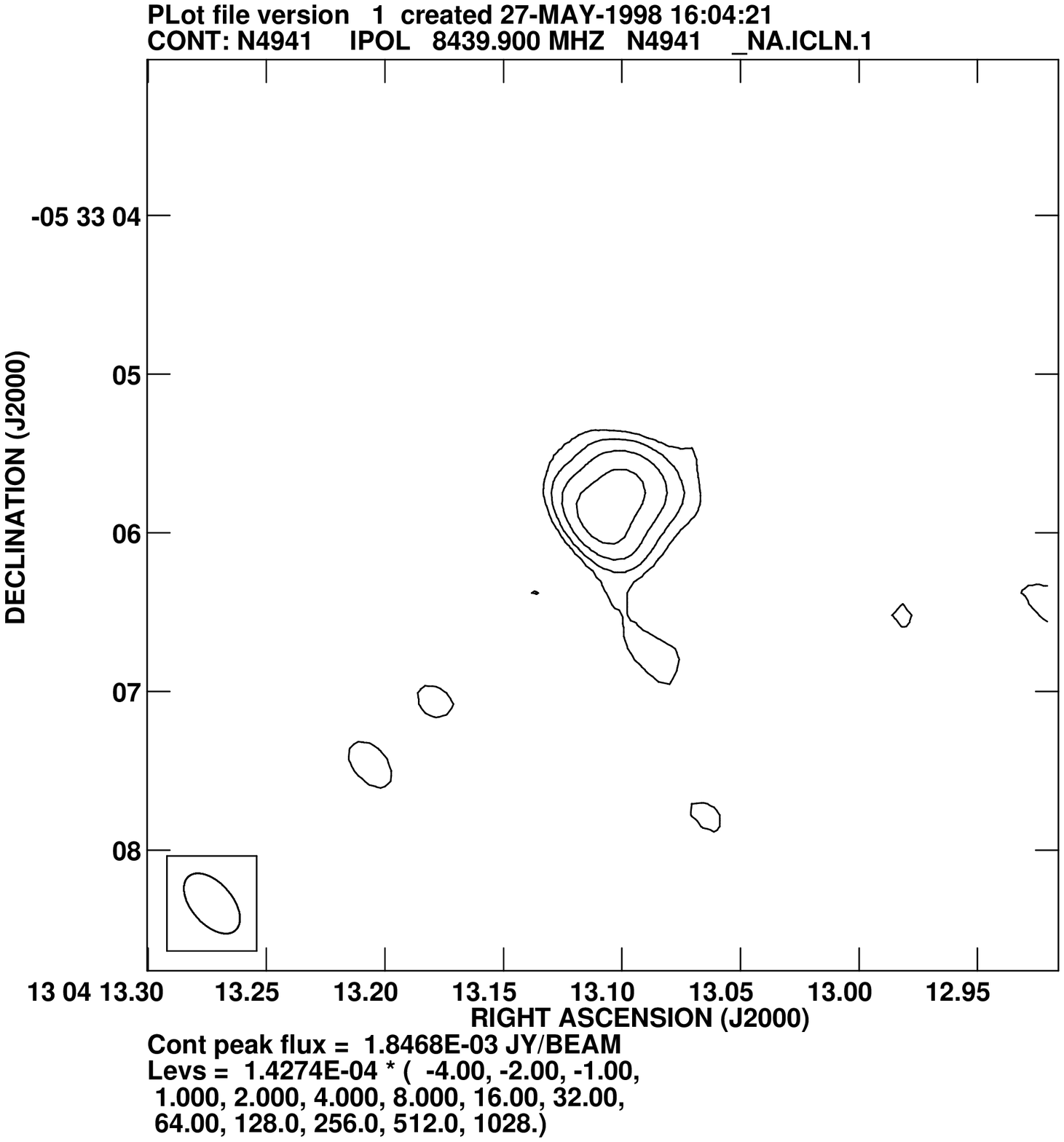}}
      \subfigure[) NGC 4968]{
        \includegraphics[width=4.8cm,clip,trim=0 62 0 35]{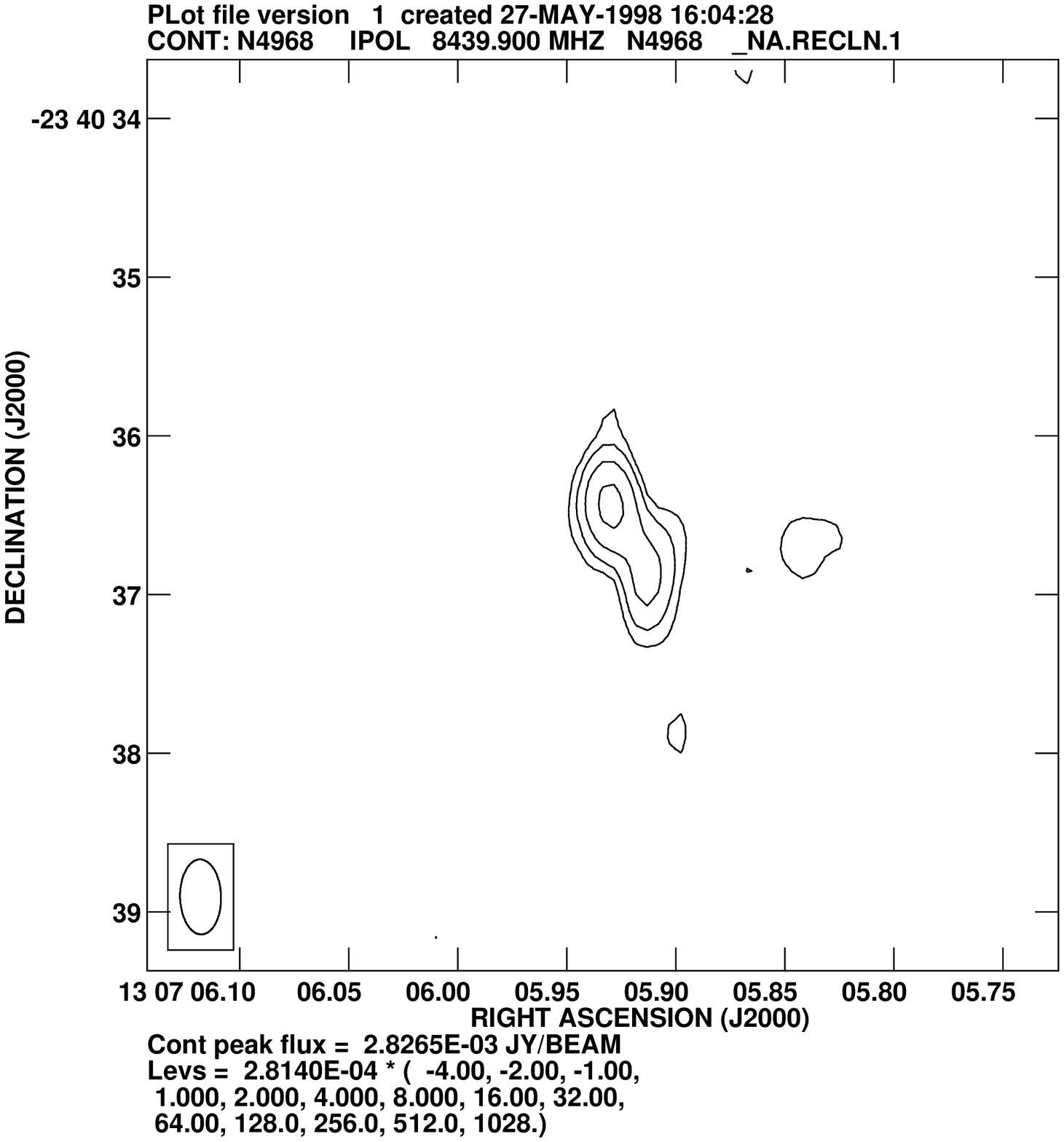}}
      \subfigure[) NGC 5005]{
        \includegraphics[width=4.7cm,clip,trim=0 62 0 35]{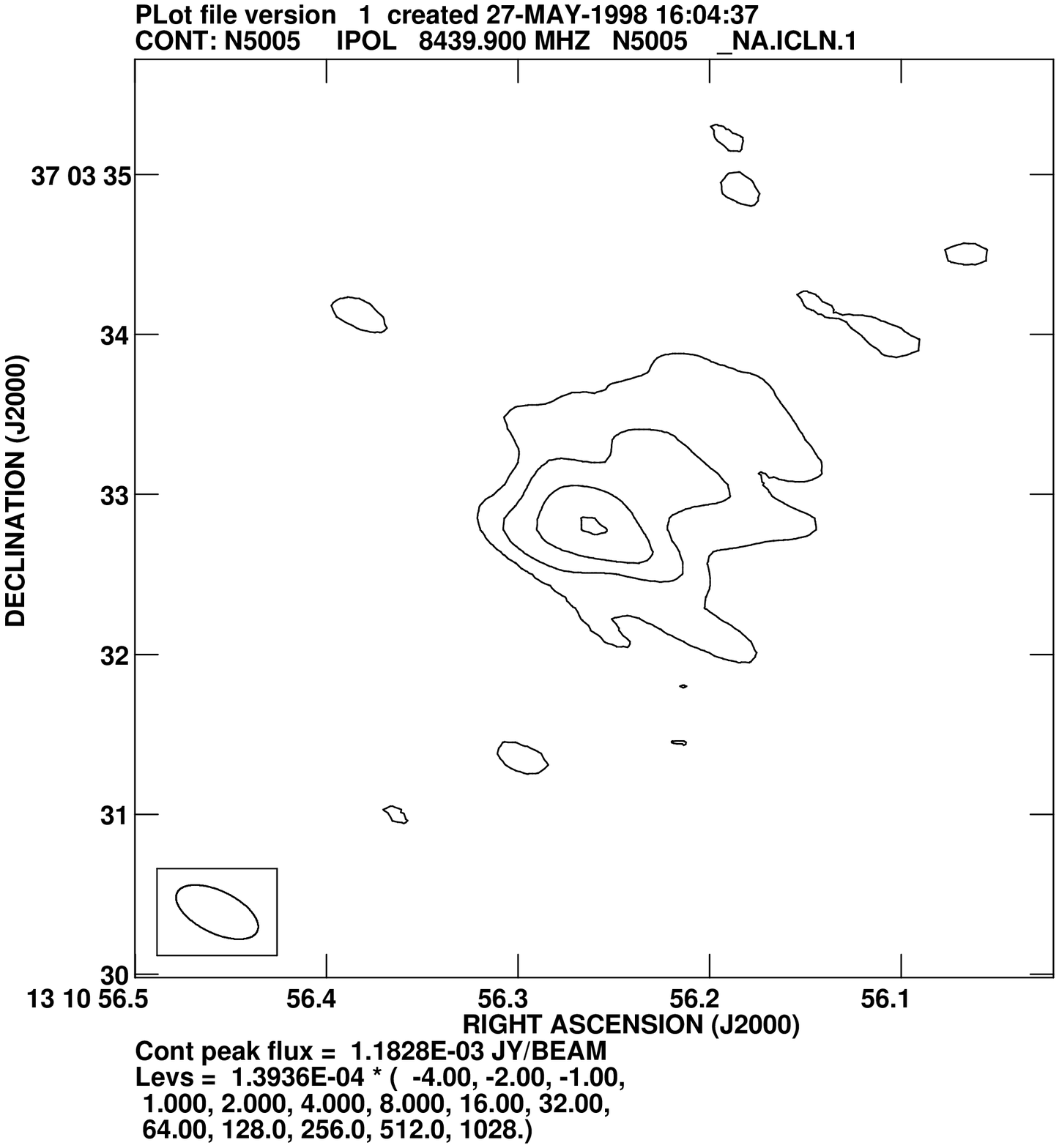}}
      \subfigure[) MCG-3-34-6]{
        \includegraphics[width=4.8cm,clip,trim=0 62 0 35]{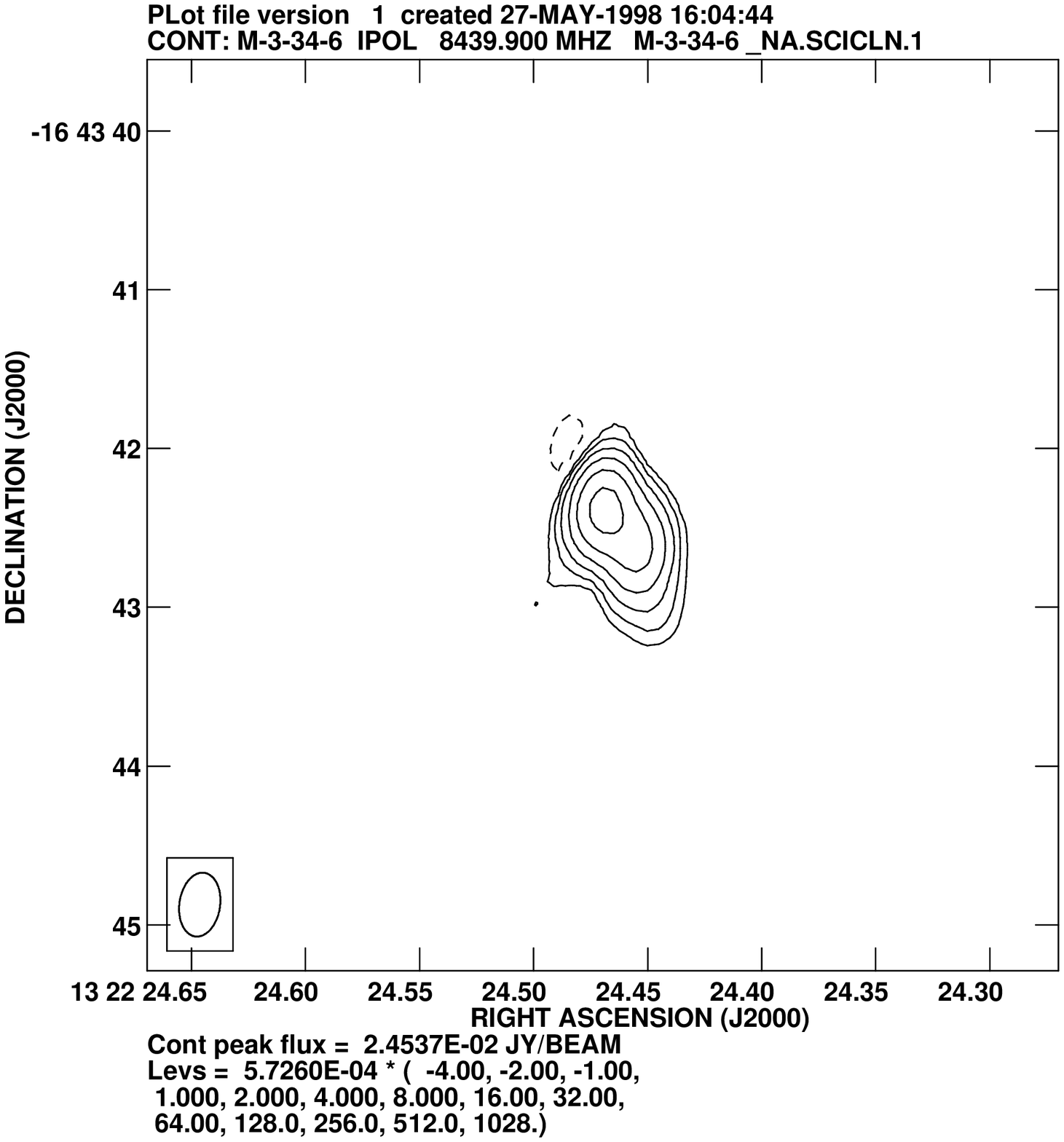}}
\caption{A-configuration 8.4 GHz images ({\it continued}).}
\end{figure*}

\setcounter{figure}{0}
\setcounter{subfigure}{48}
\begin{figure*}
\centering
      \subfigure[) NGC 5194]{
        \includegraphics[width=4.8cm,clip,trim=0 62 0 35]{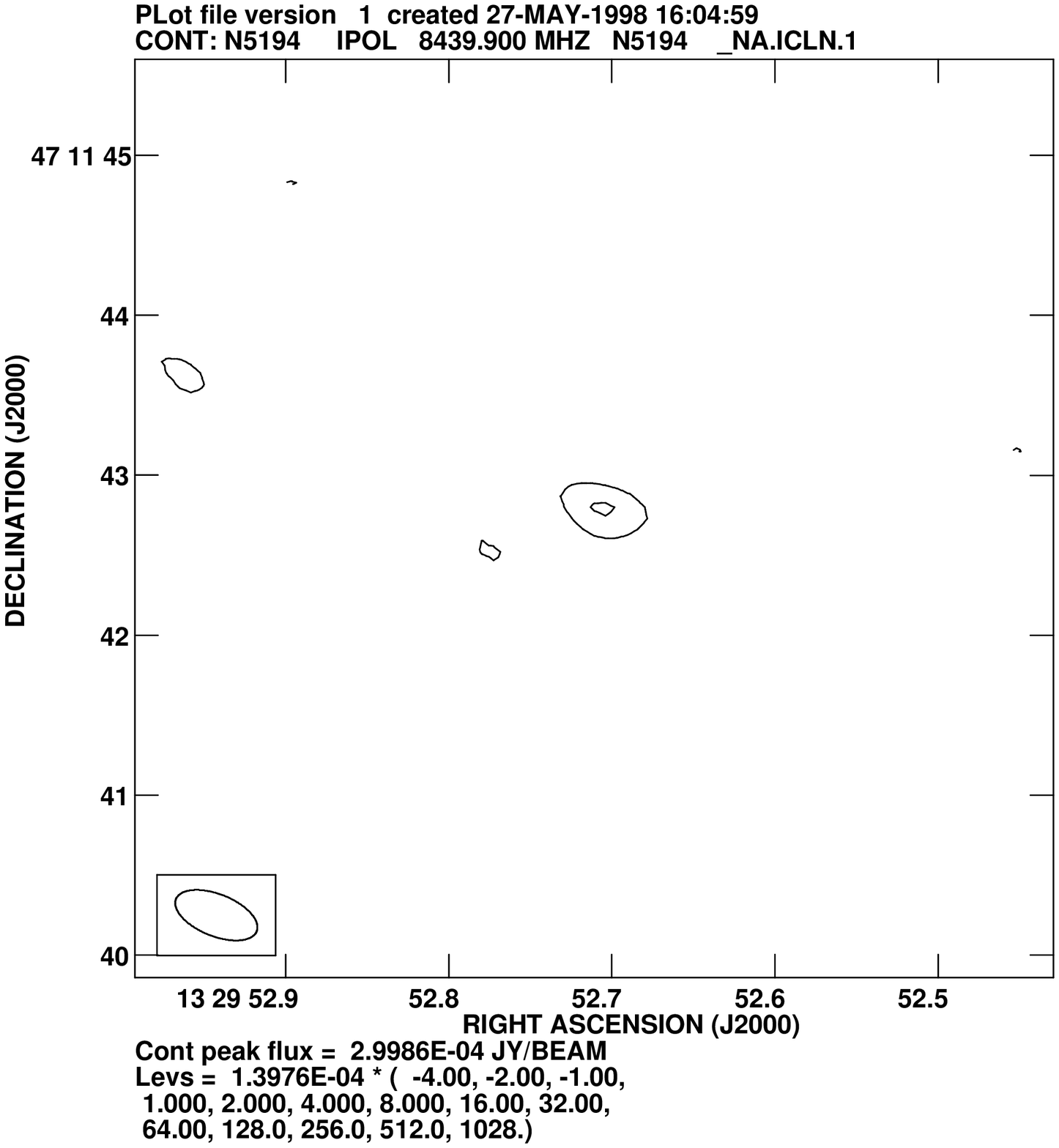}}
      \subfigure[) MCG-6-30-1]{
        \includegraphics[width=4.8cm,clip,trim=0 62 0 35]{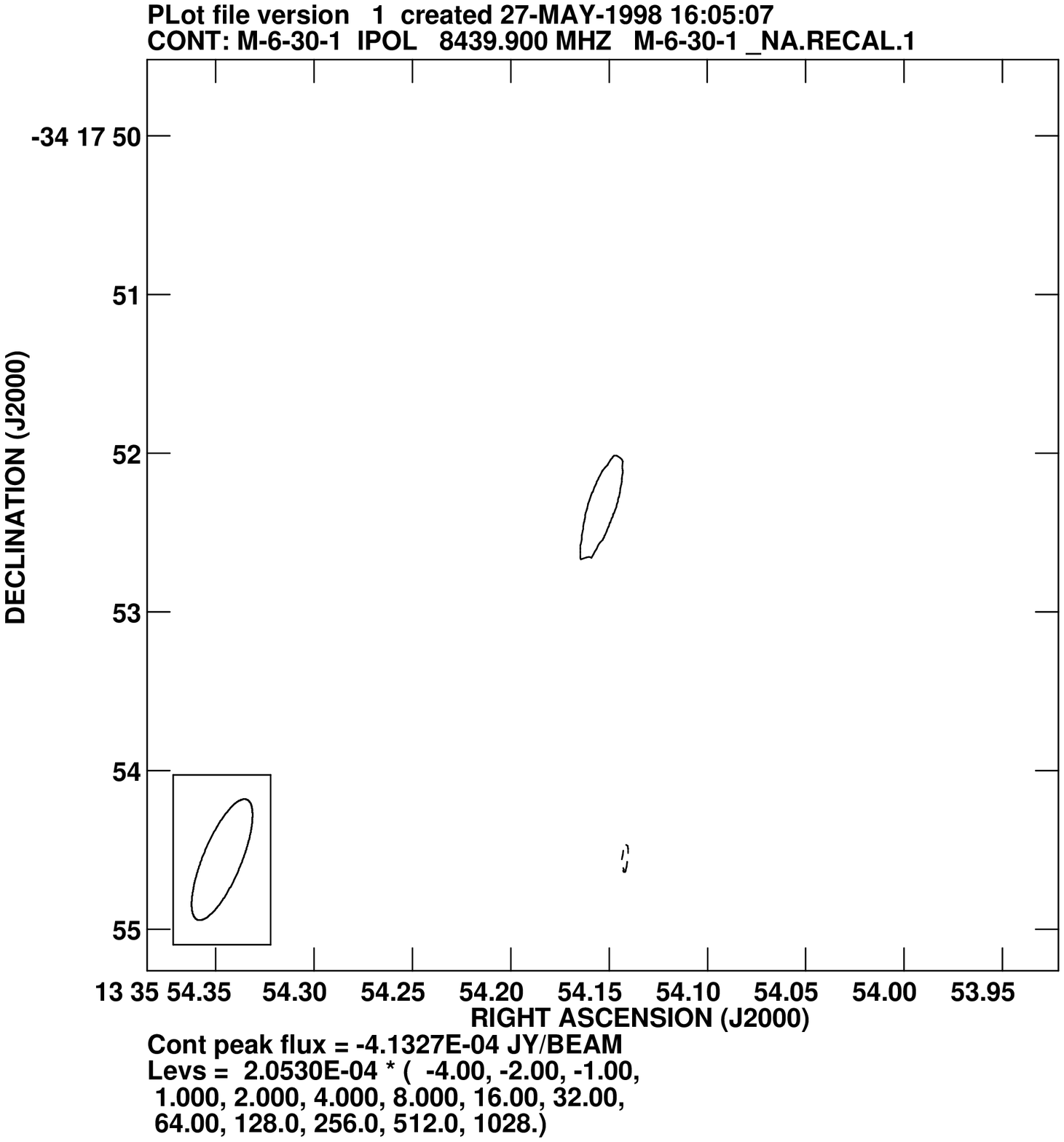}}
      \subfigure[) F13349+2438]{
        \includegraphics[width=4.9cm,clip,trim=0 34 0 35]{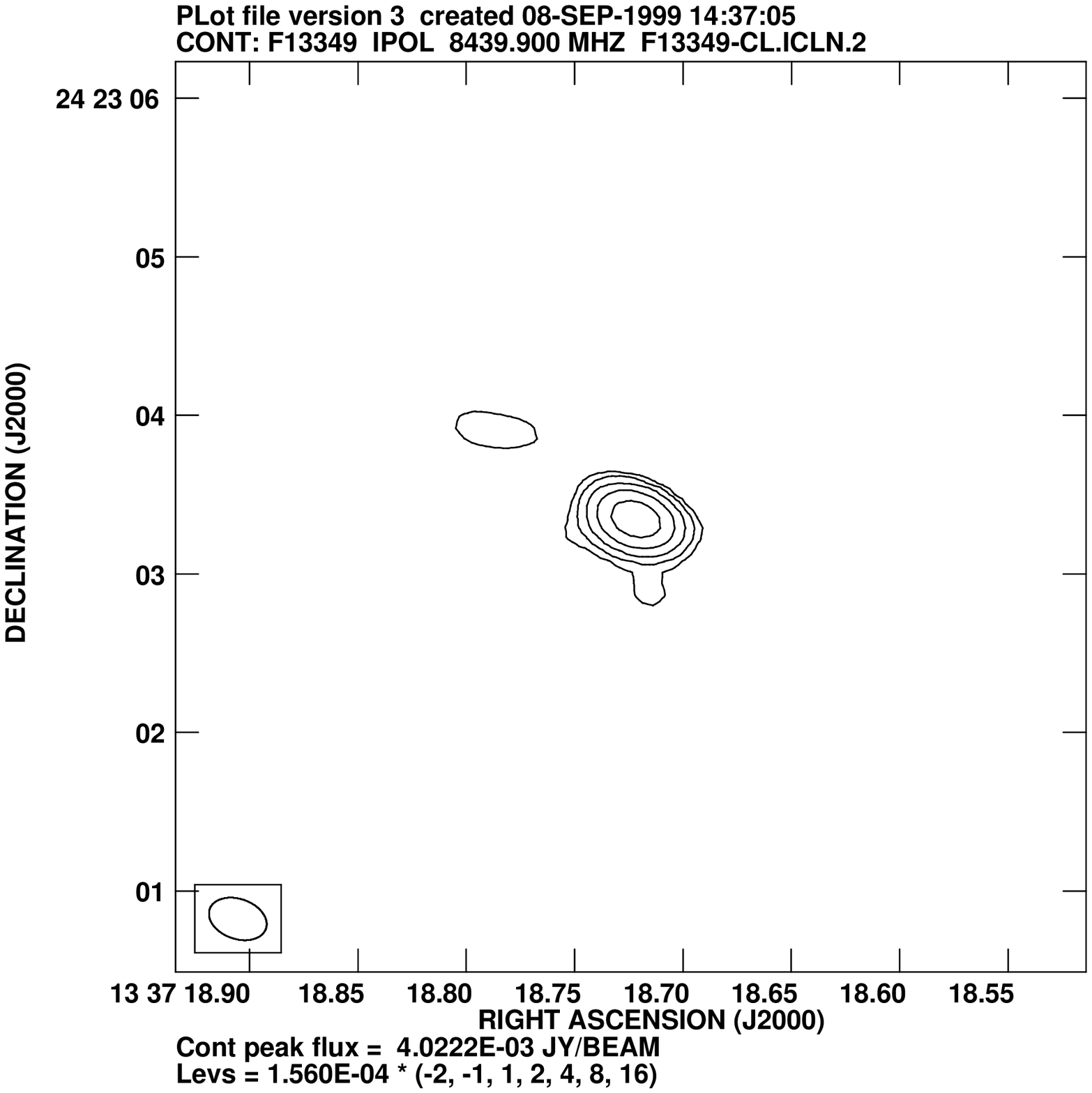}}
      \subfigure[) NGC 5256]{
        \includegraphics[width=5.0cm,clip,trim=0 47 0 35]{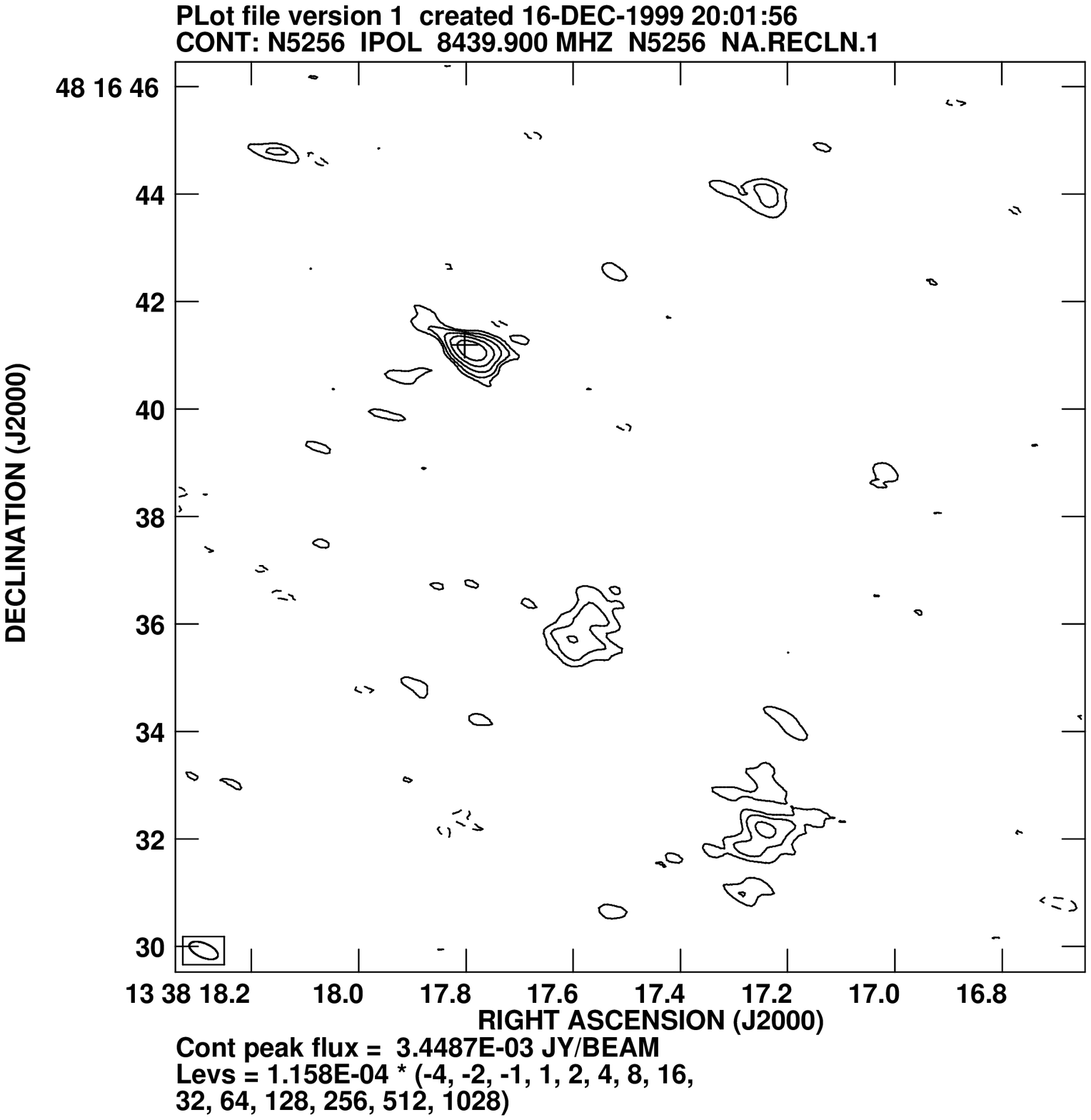}}
      \subfigure[) Markarian 273]{
        \includegraphics[width=5.0cm,clip,trim=0 47 0 35]{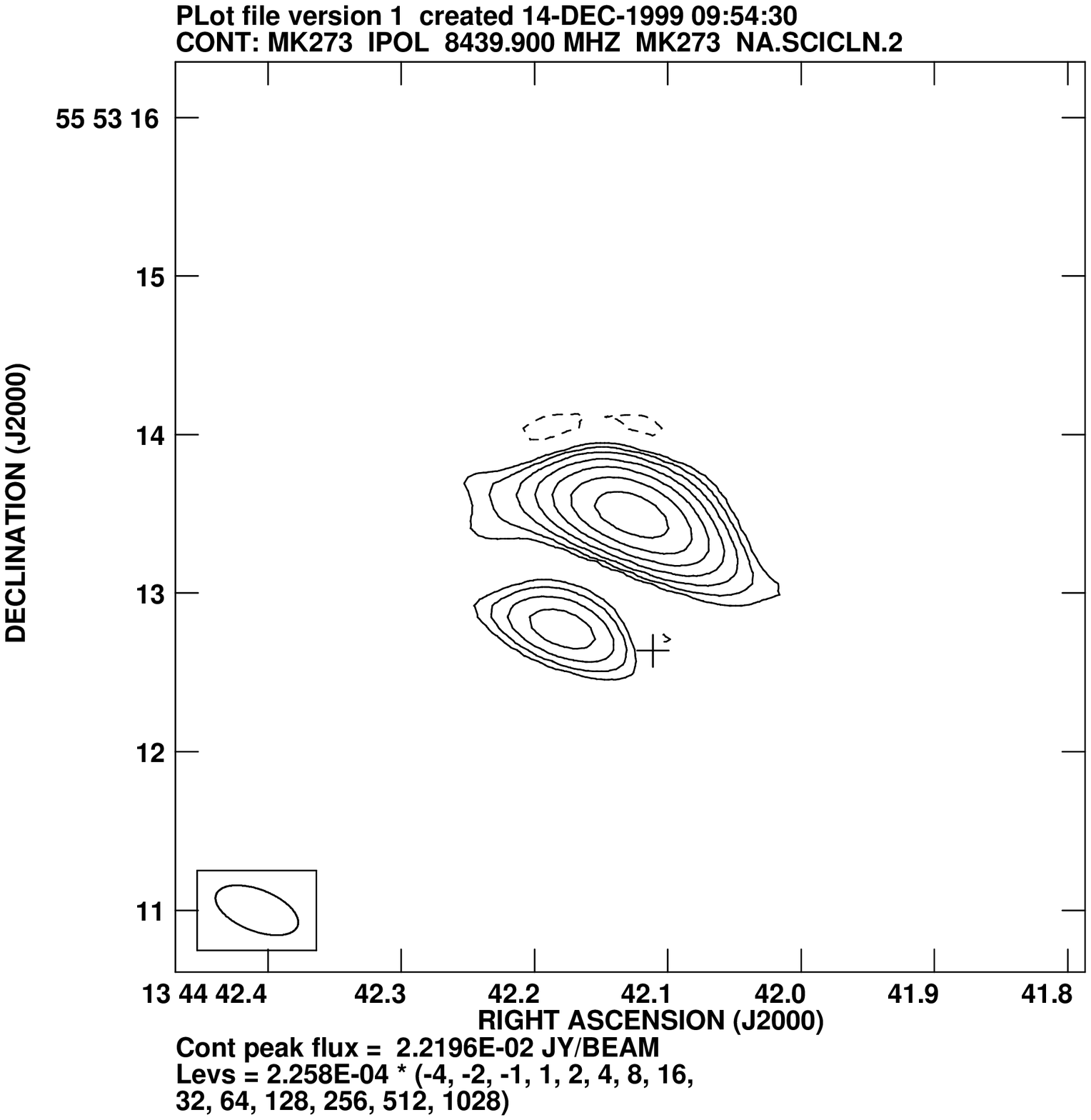}}
      \subfigure[) IC 4329A]{
        \includegraphics[width=4.9cm,clip,trim=0 62 0 35]{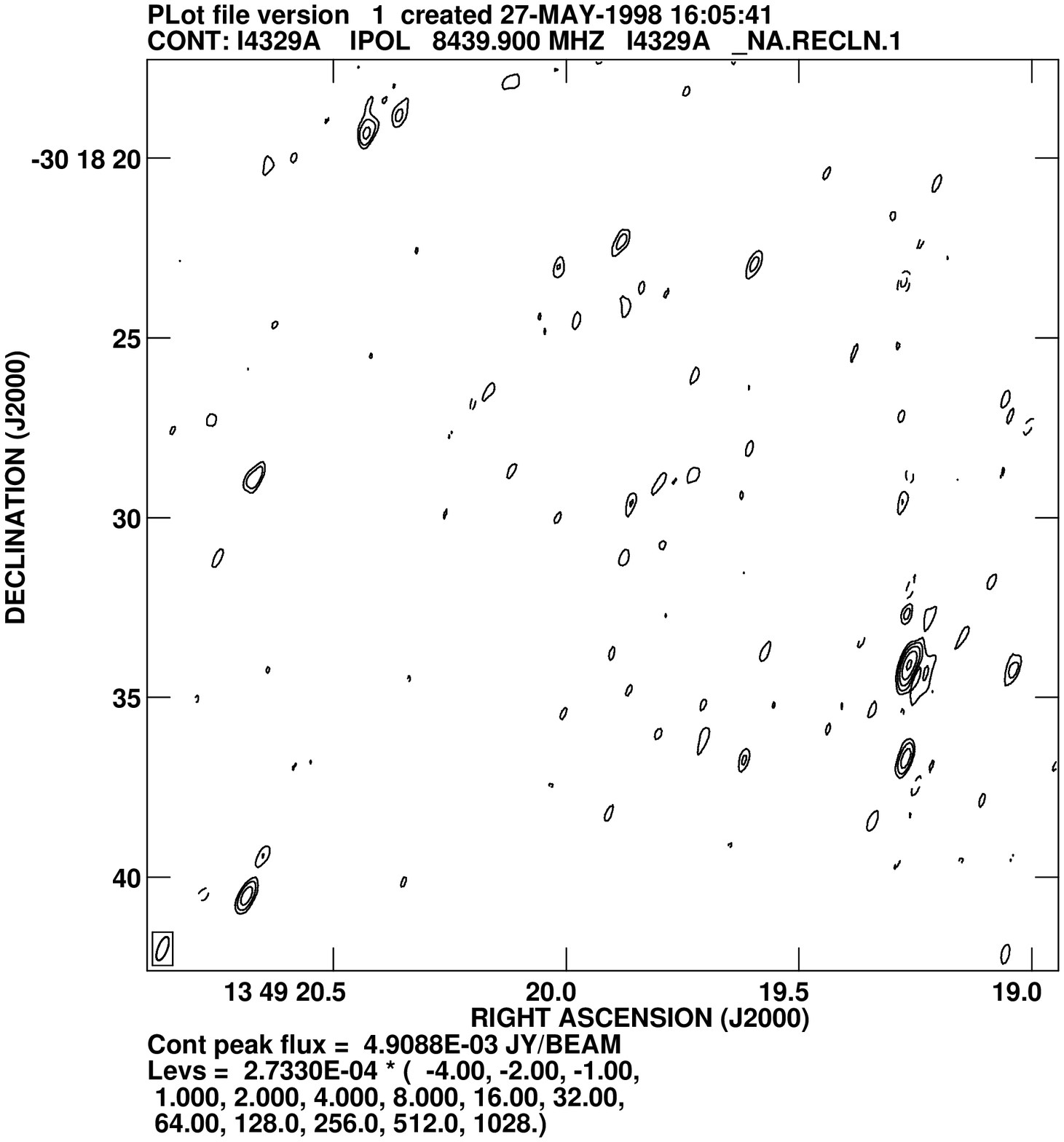}}
      \subfigure[) NGC 5347]{
        \includegraphics[width=5.0cm,clip,trim=0 47 0 35]{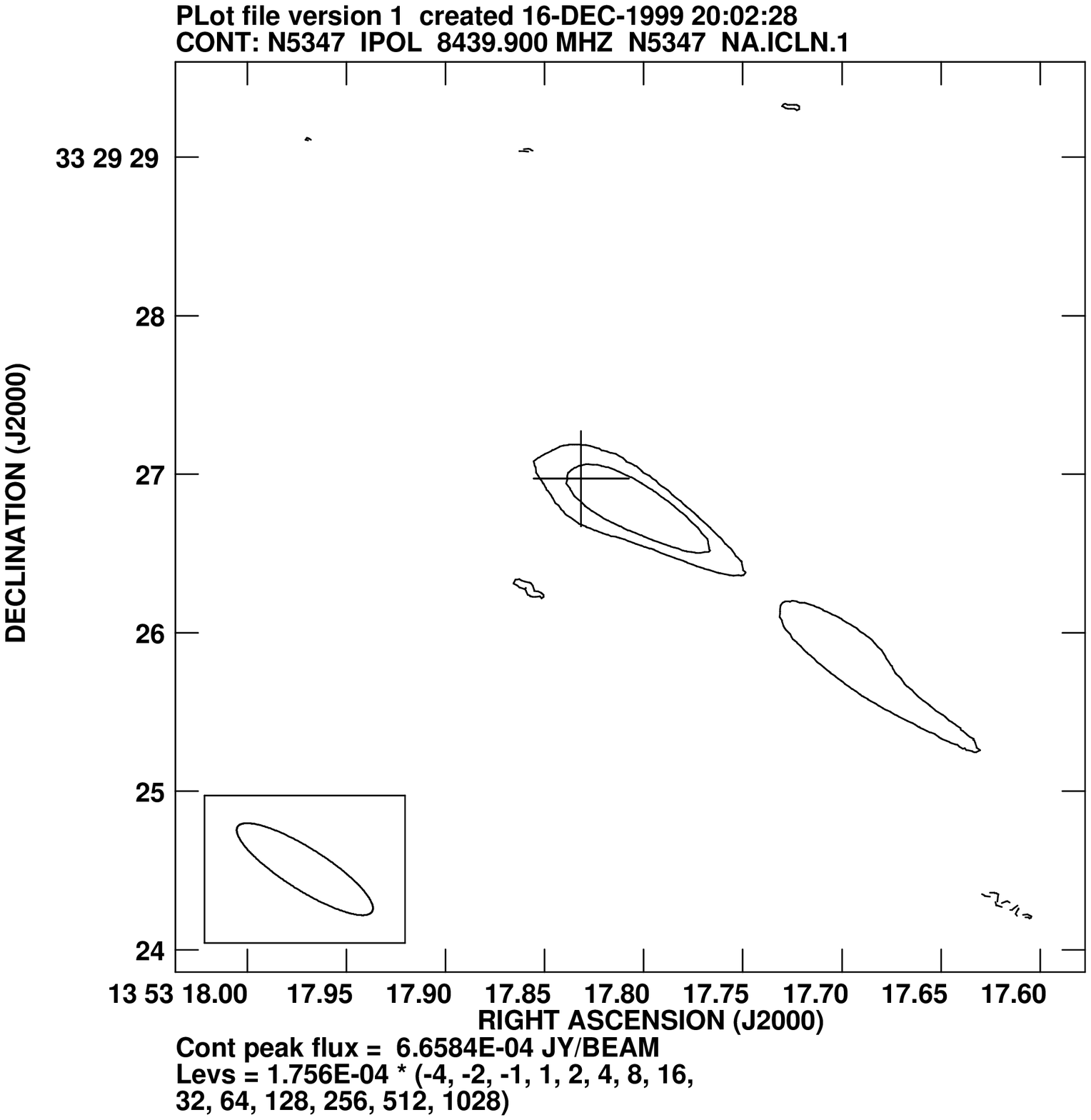}}
      \subfigure[) Markarian 463]{
        \includegraphics[width=5.0cm,clip,trim=0 47 0 35]{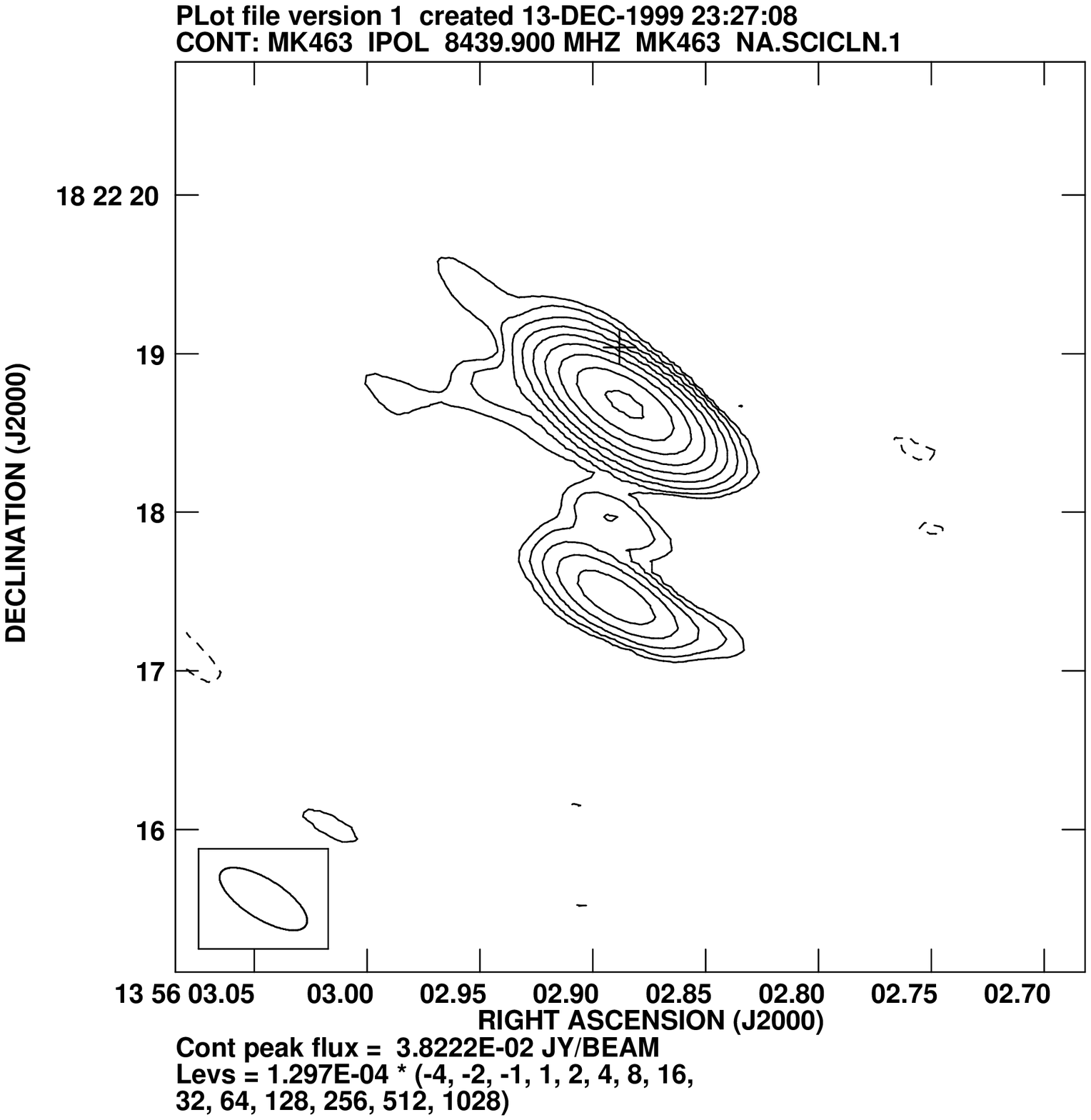}}
      \subfigure[) NGC 5506]{
        \includegraphics[width=5.0cm,clip,trim=0 62 0 35]{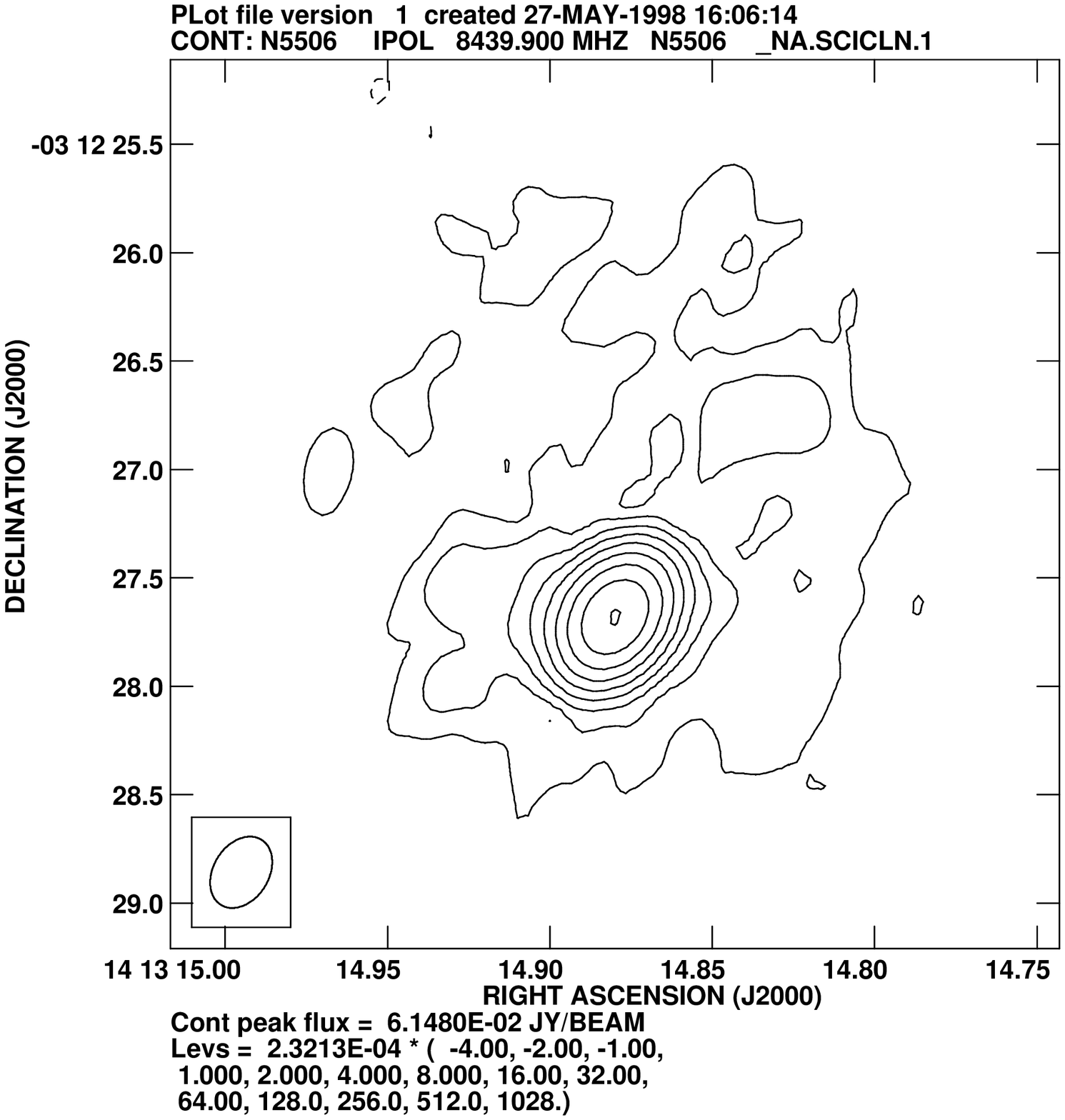}}
      \subfigure[) Markarian 817]{
        \includegraphics[width=5.0cm,clip,trim=0 47 0 35]{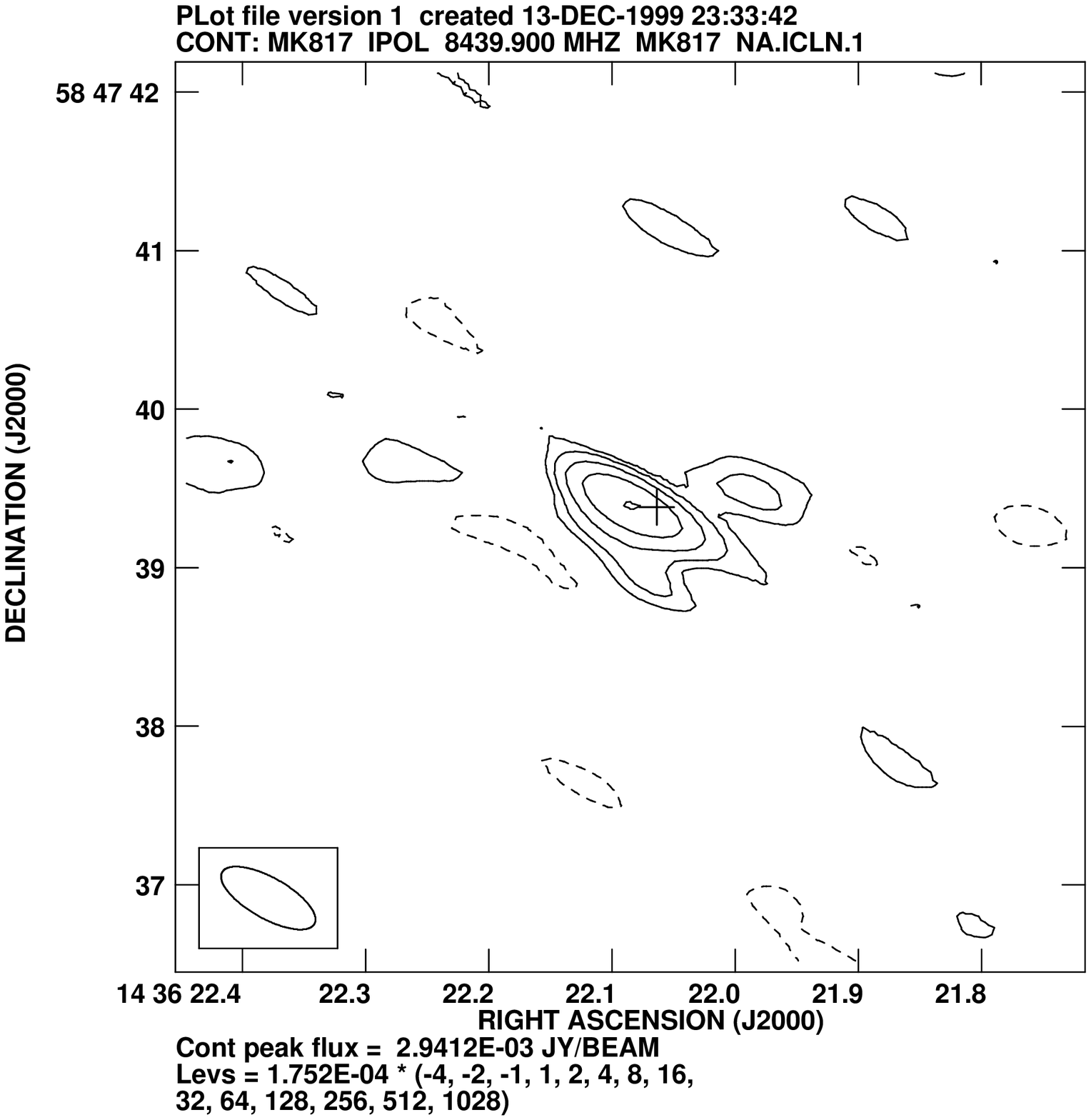}}
      \subfigure[) F15091-2107]{
        \includegraphics[width=5cm,clip,trim=0 47 0 35]{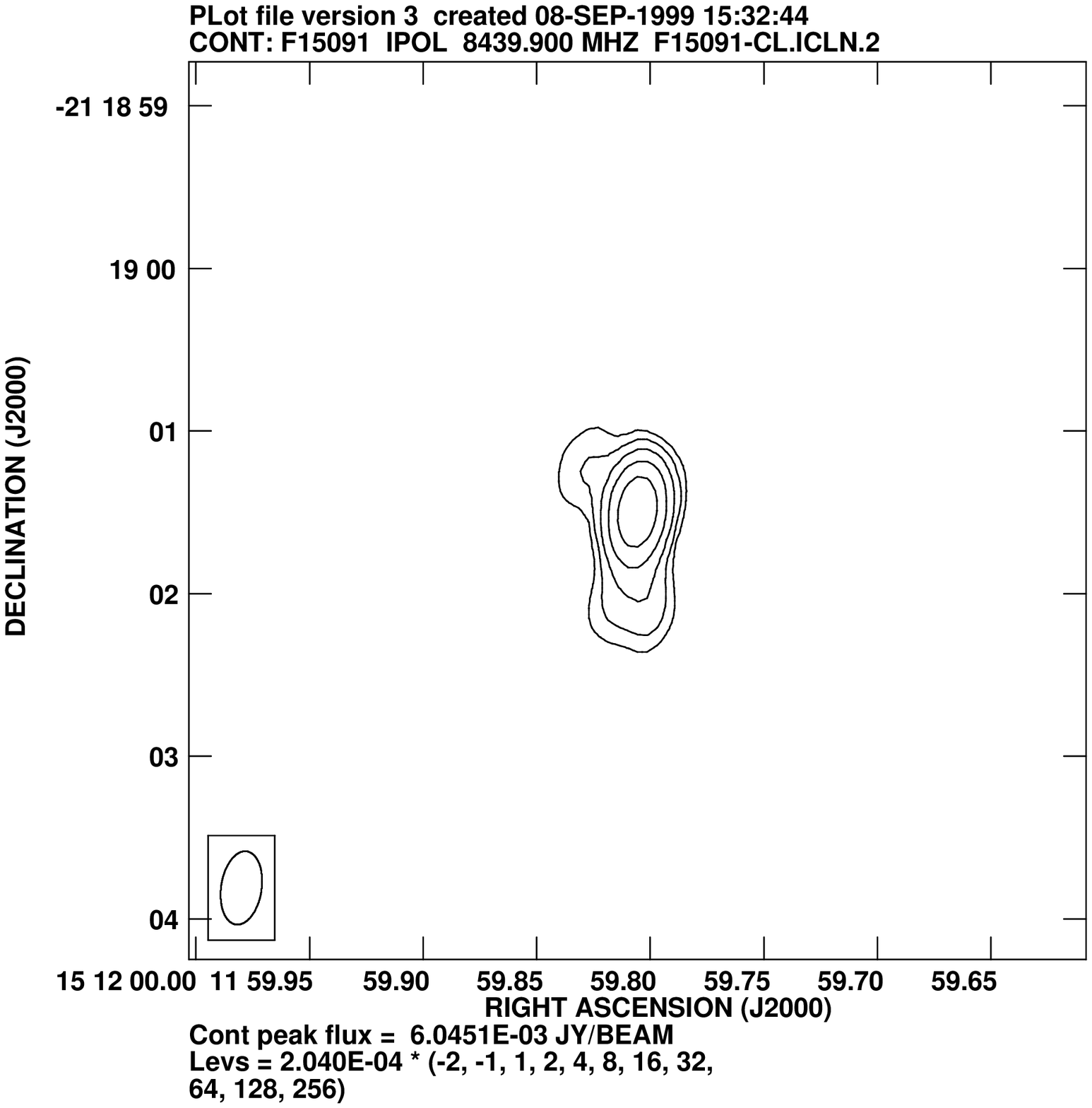}}
      \subfigure[) NGC 5953]{
        \includegraphics[width=4.8cm,clip,trim=0 62 0 35]{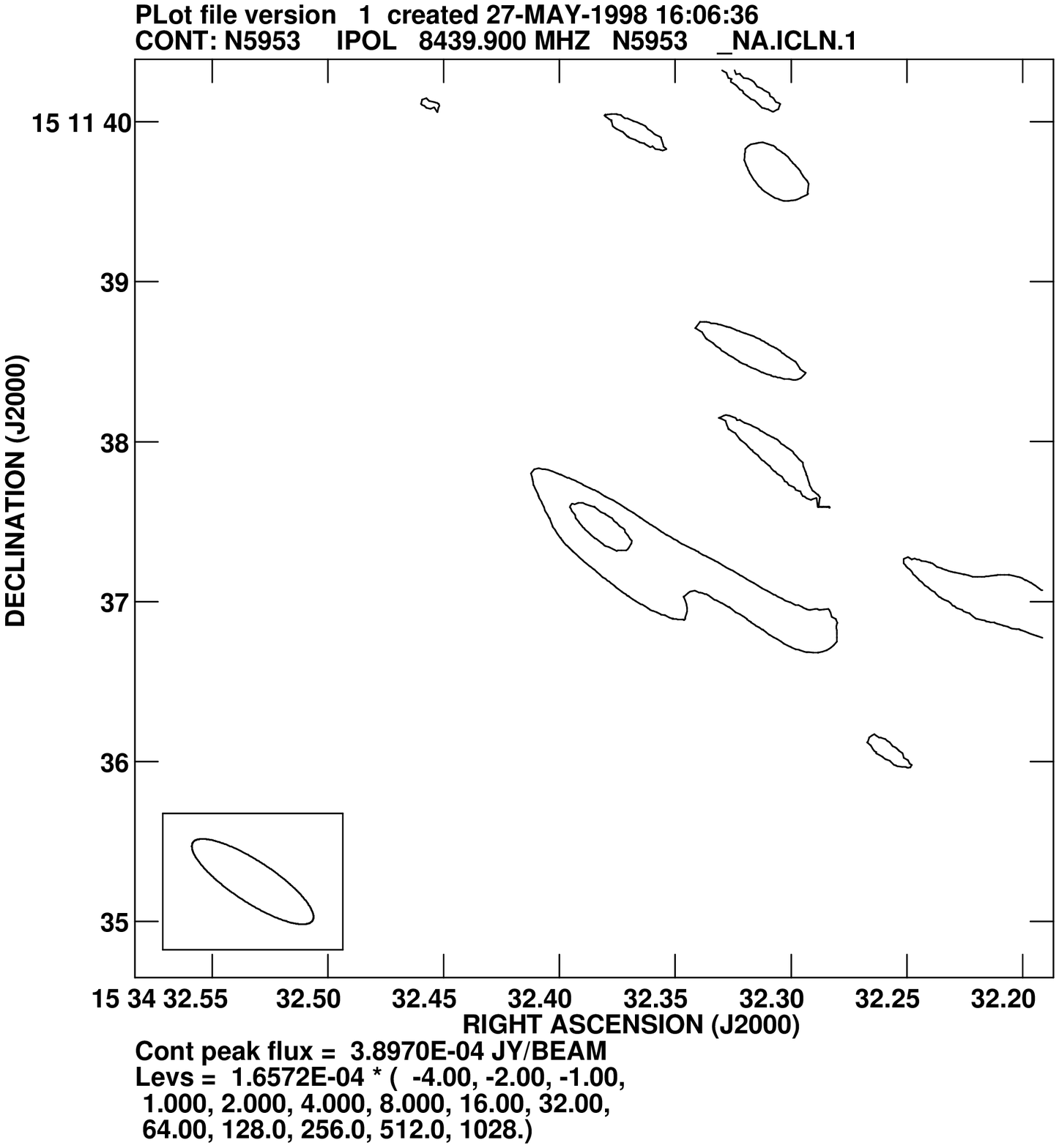}}
\caption{A-configuration 8.4 GHz images ({\it continued}).}
\end{figure*}

\setcounter{figure}{0}
\setcounter{subfigure}{60}

\begin{figure*}
\centering
       \subfigure[) UGC 9913 = Arp 220]{
        \includegraphics[width=4.8cm,clip,trim=0 62 0 35]{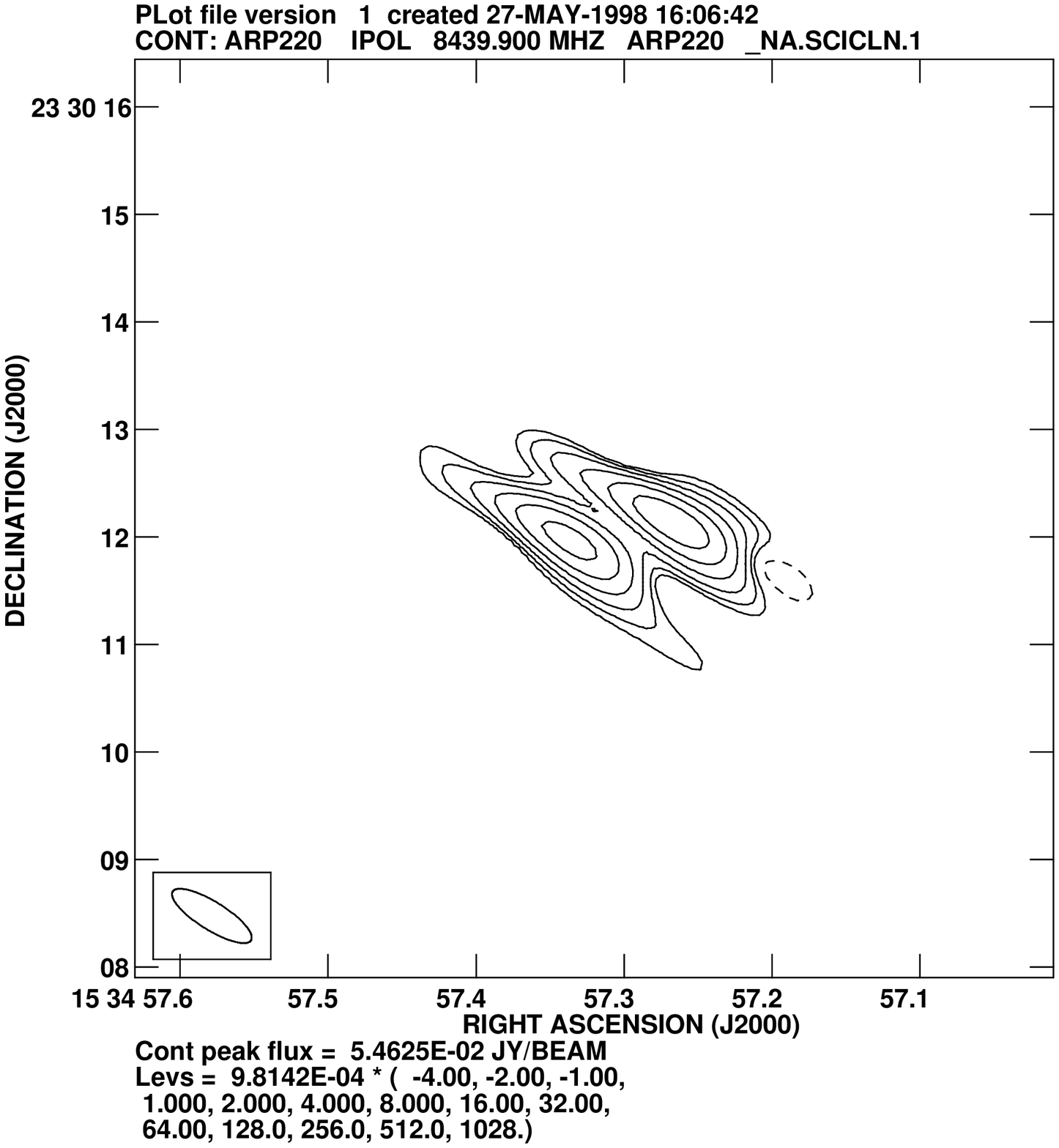}}
      \subfigure[) MCG-2-40-4]{
        \includegraphics[width=4.8cm,clip,trim=0 62 0 35]{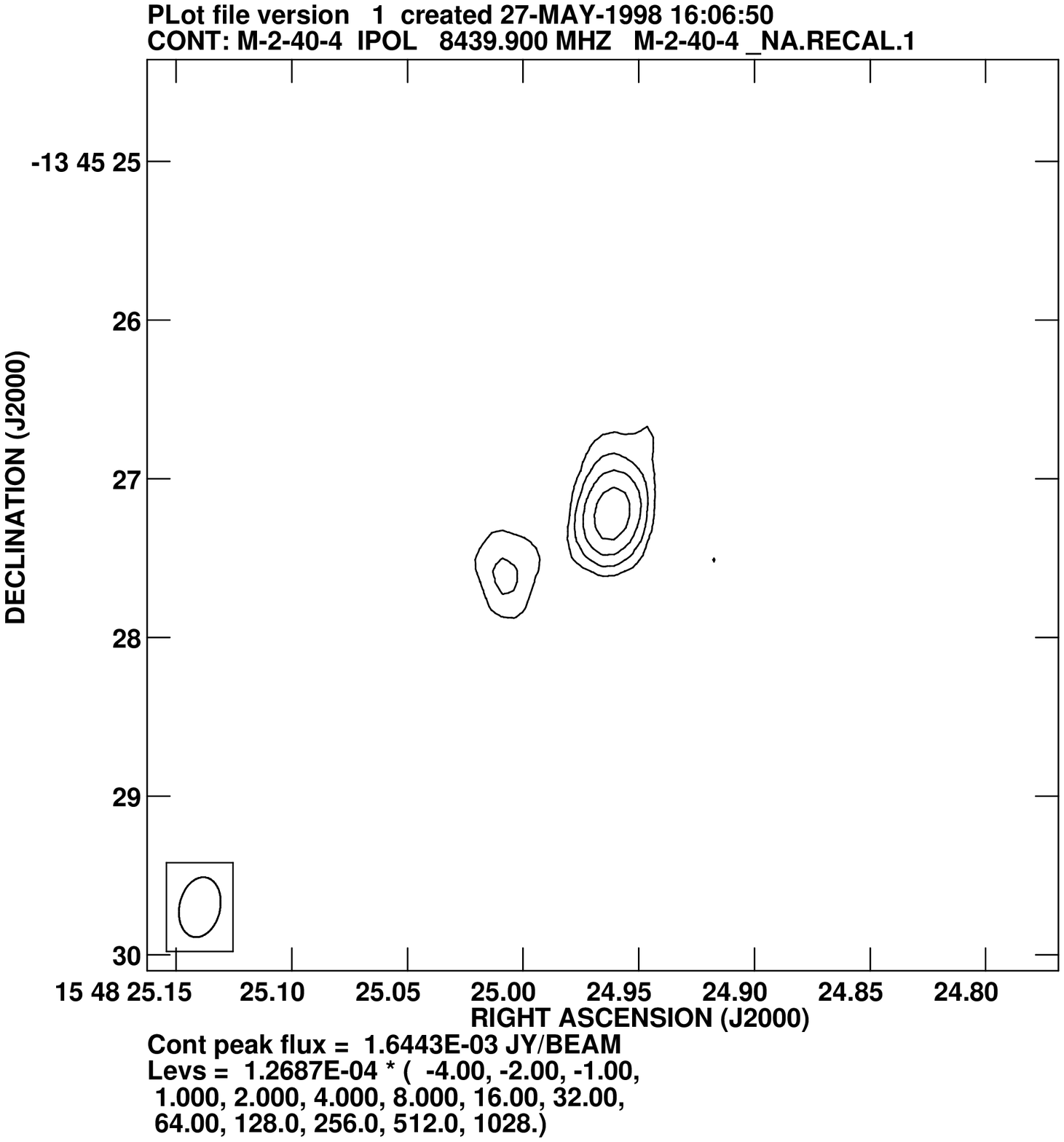}}
      \subfigure[) F15480-0344]{
        \includegraphics[width=4.8cm,clip,trim=0 62 0 35]{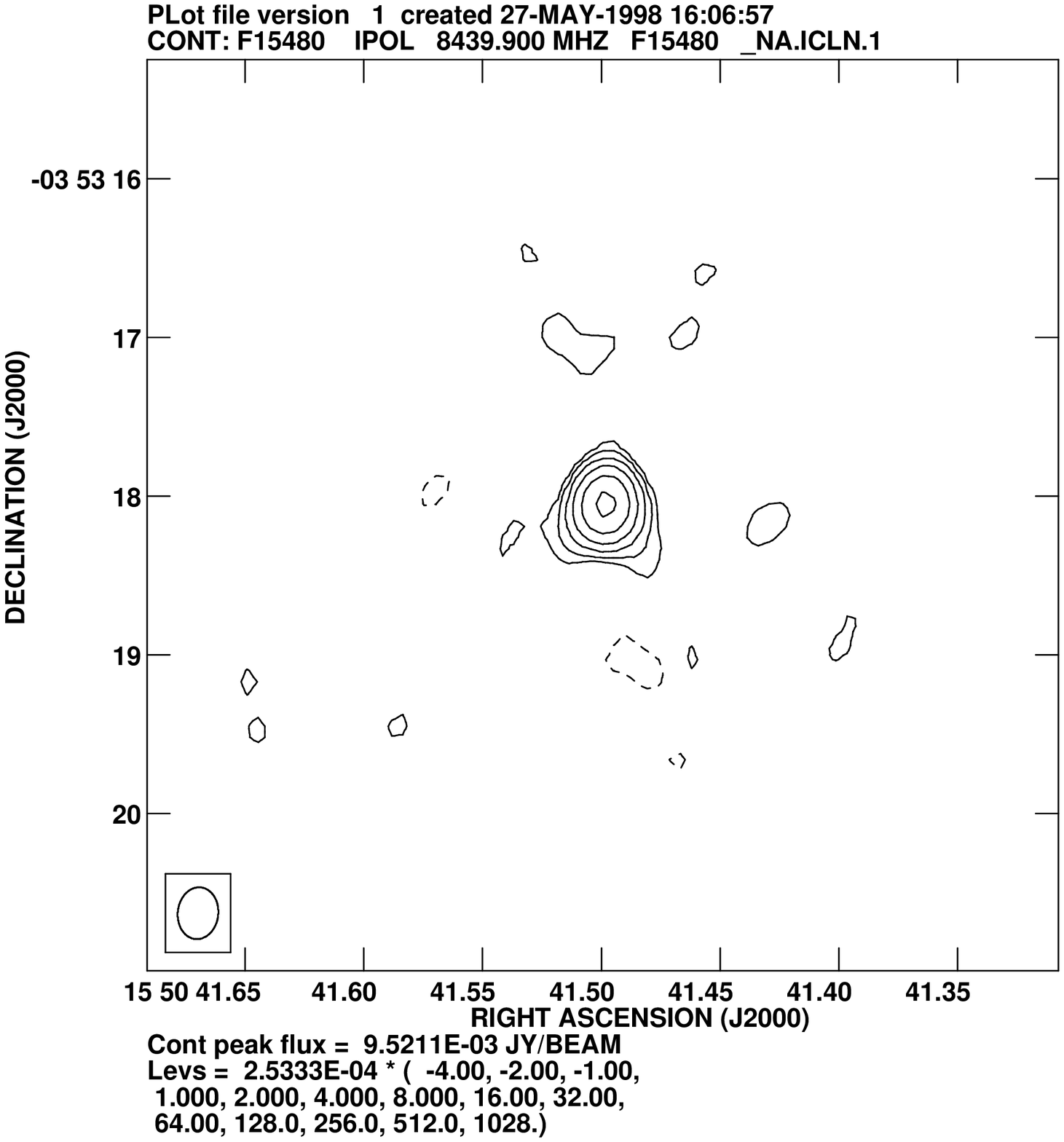}}
      \subfigure[) NGC 6890]{
        \includegraphics[width=4.75cm,clip,trim=0 62 0 35]{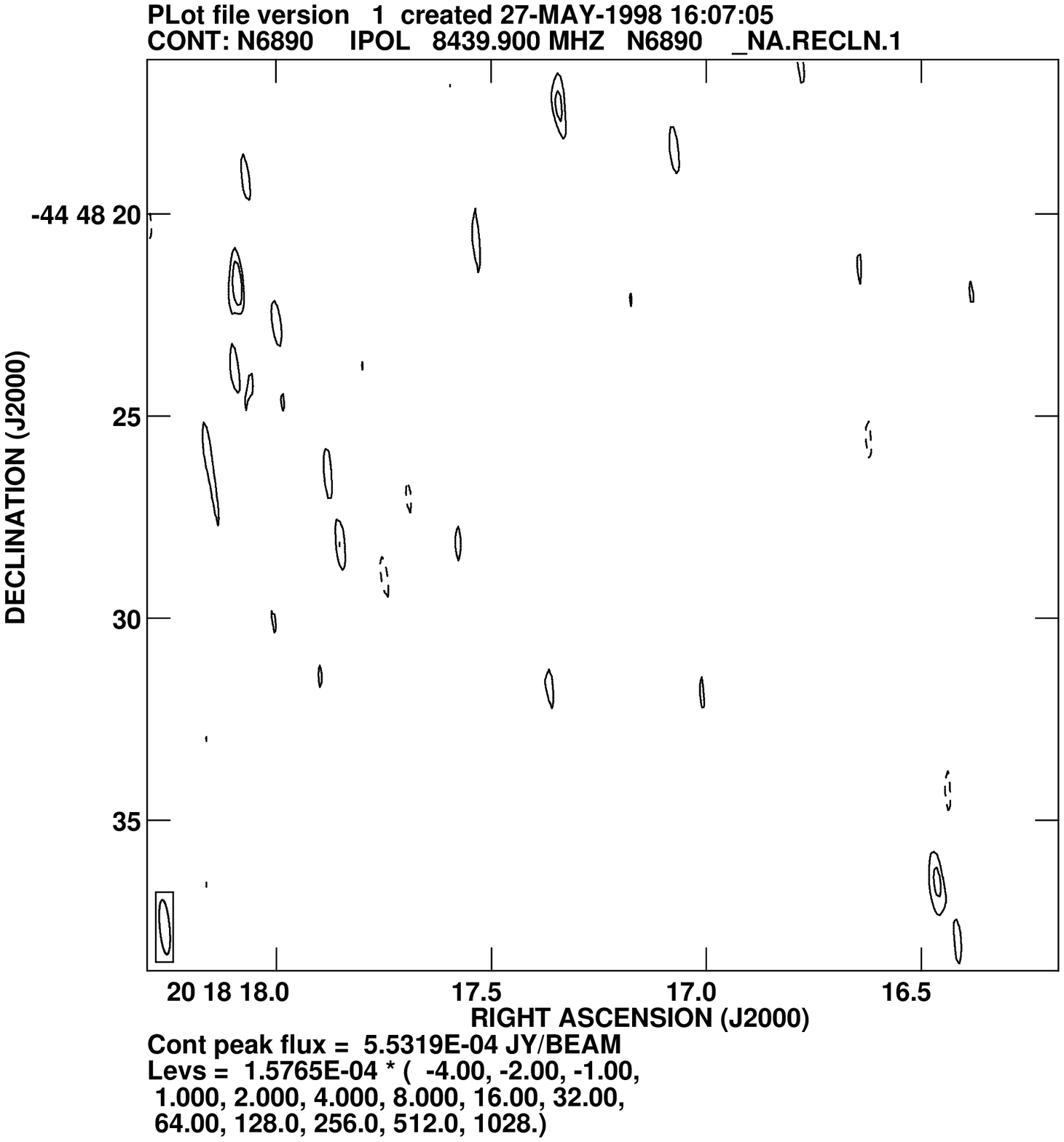}}
      \subfigure[) Markarian 509]{
        \includegraphics[width=5.0cm,clip,trim=0 47 0 35]{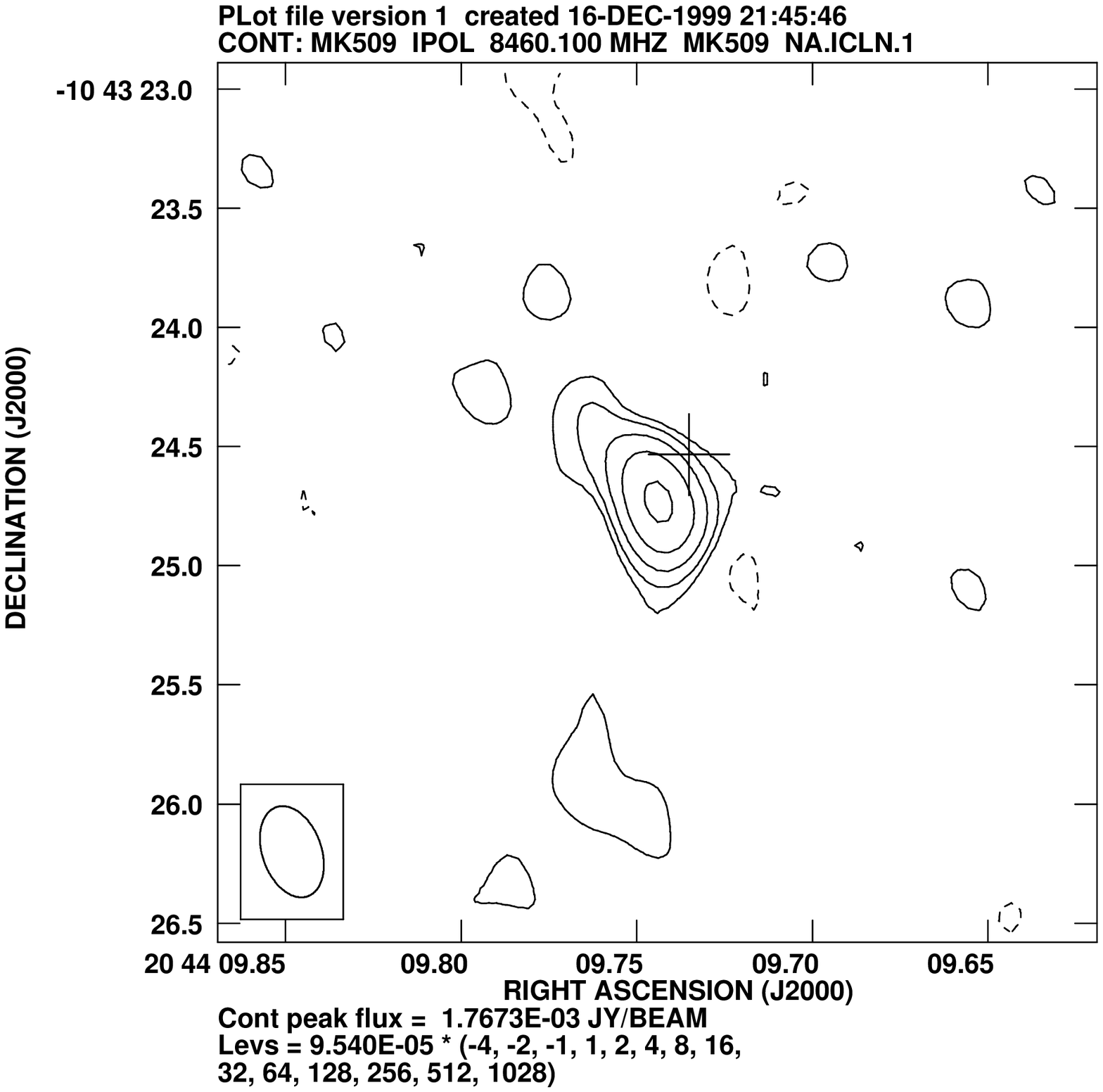}}
      \subfigure[) Markarian 897]{
        \includegraphics[width=4.8cm,clip,trim=0 62 0 35]{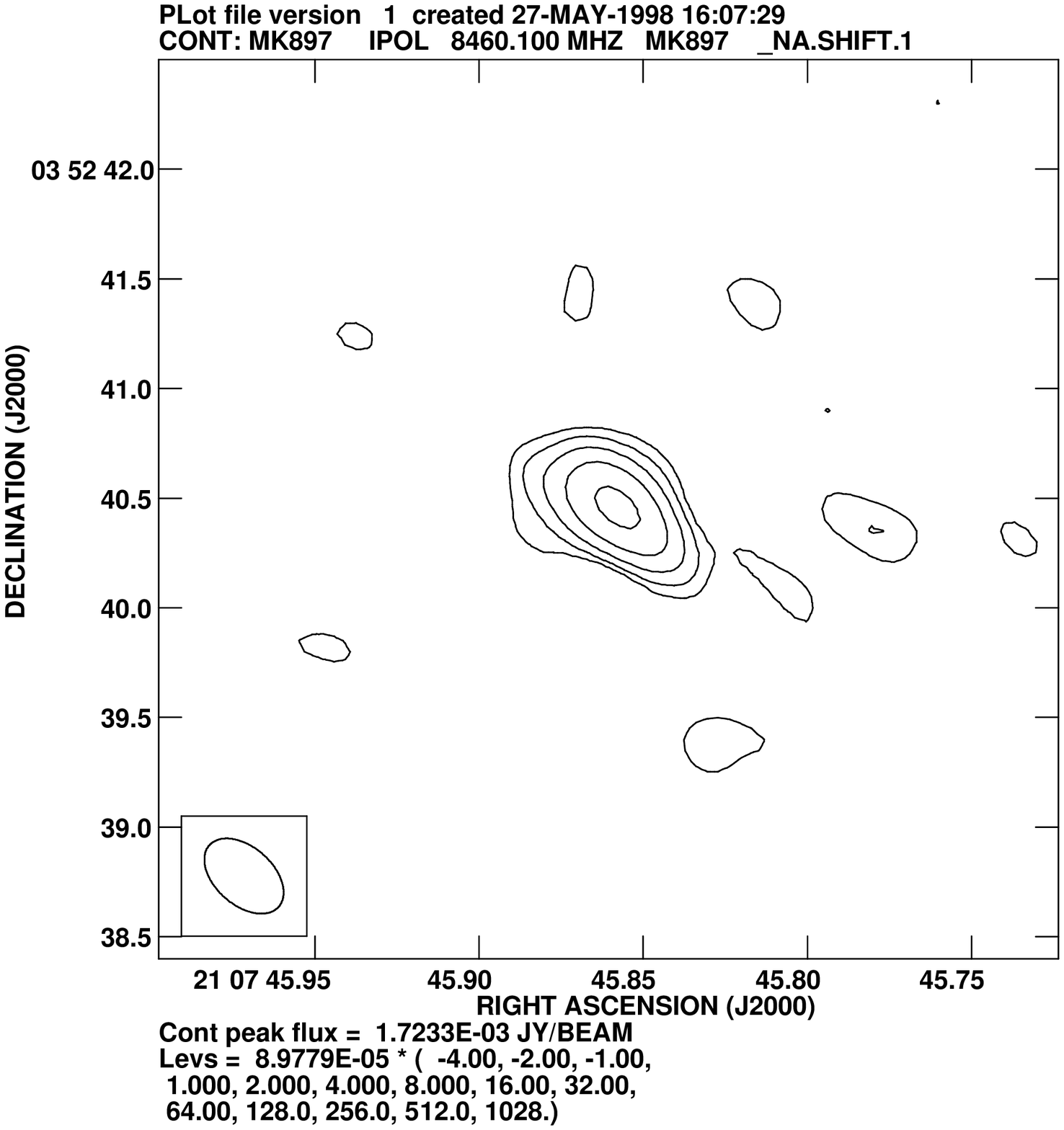}}
      \subfigure[) NGC 7130]{
        \includegraphics[width=4.8cm,clip,trim=0 62 0 35]{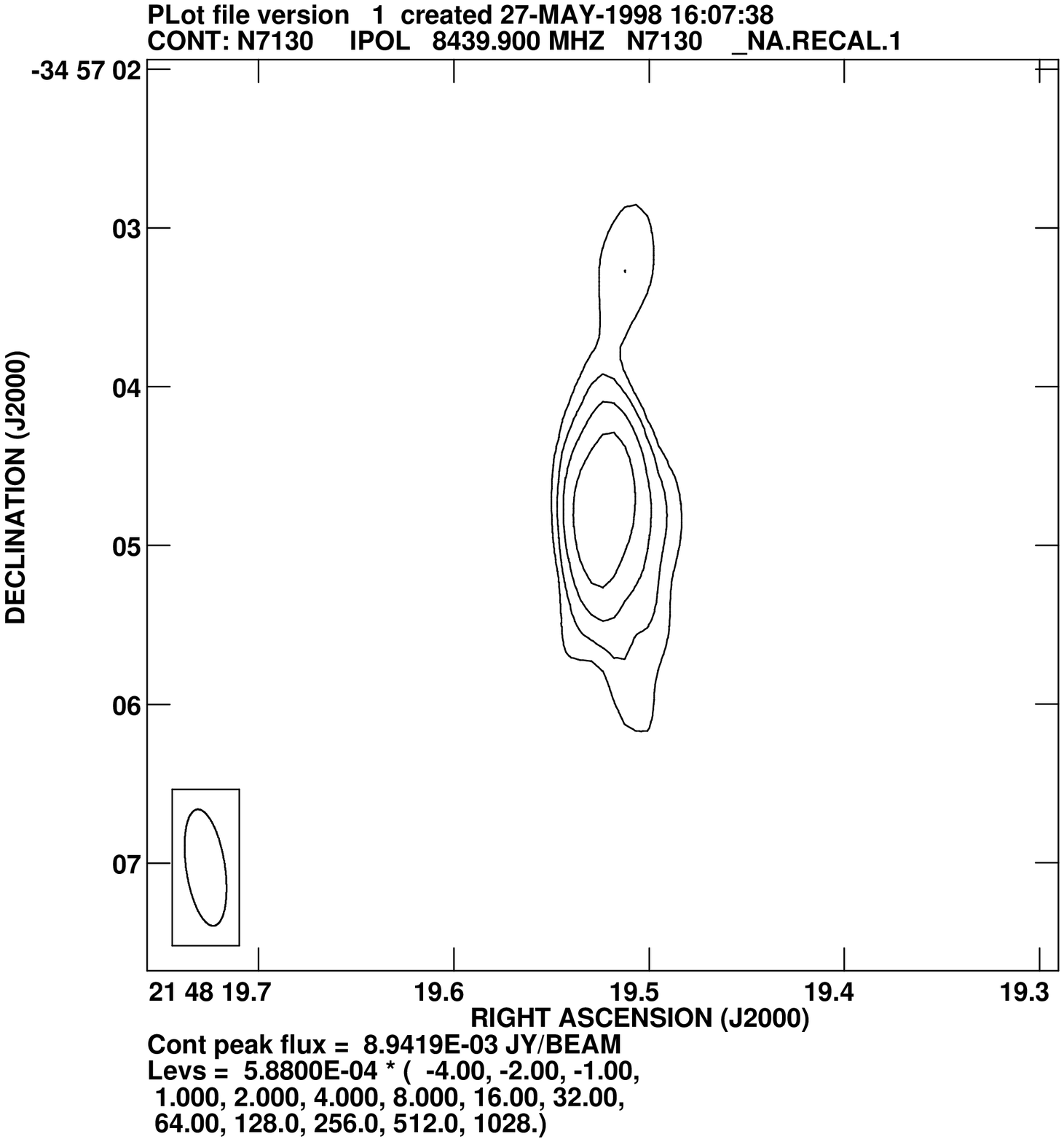}}
      \subfigure[) NGC 7172]{
        \includegraphics[width=4.8cm,clip,trim=0 62 0 35]{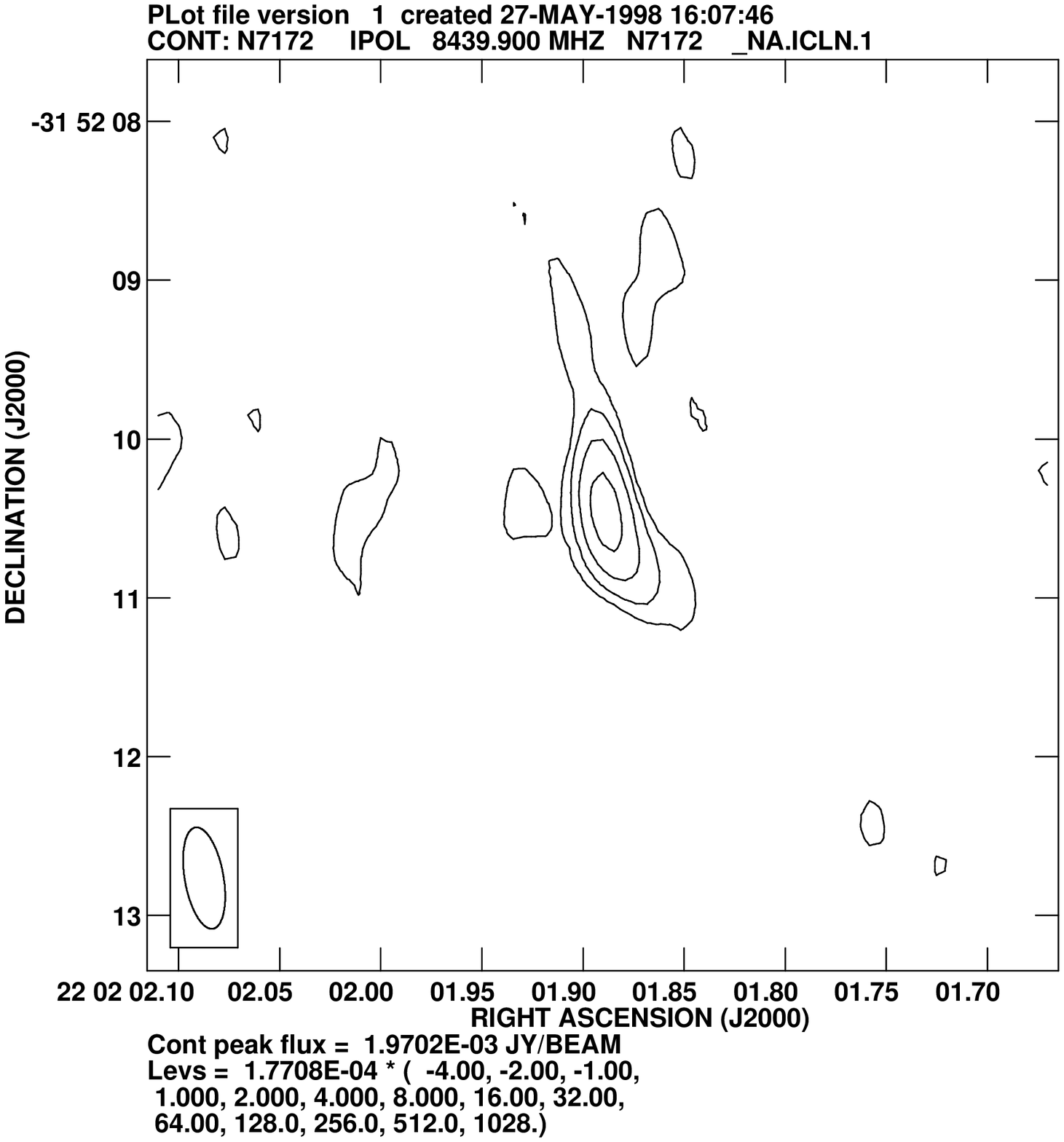}}
      \subfigure[) F22017+0319]{
        \includegraphics[width=4.9cm,clip,trim=0 62 0 35]{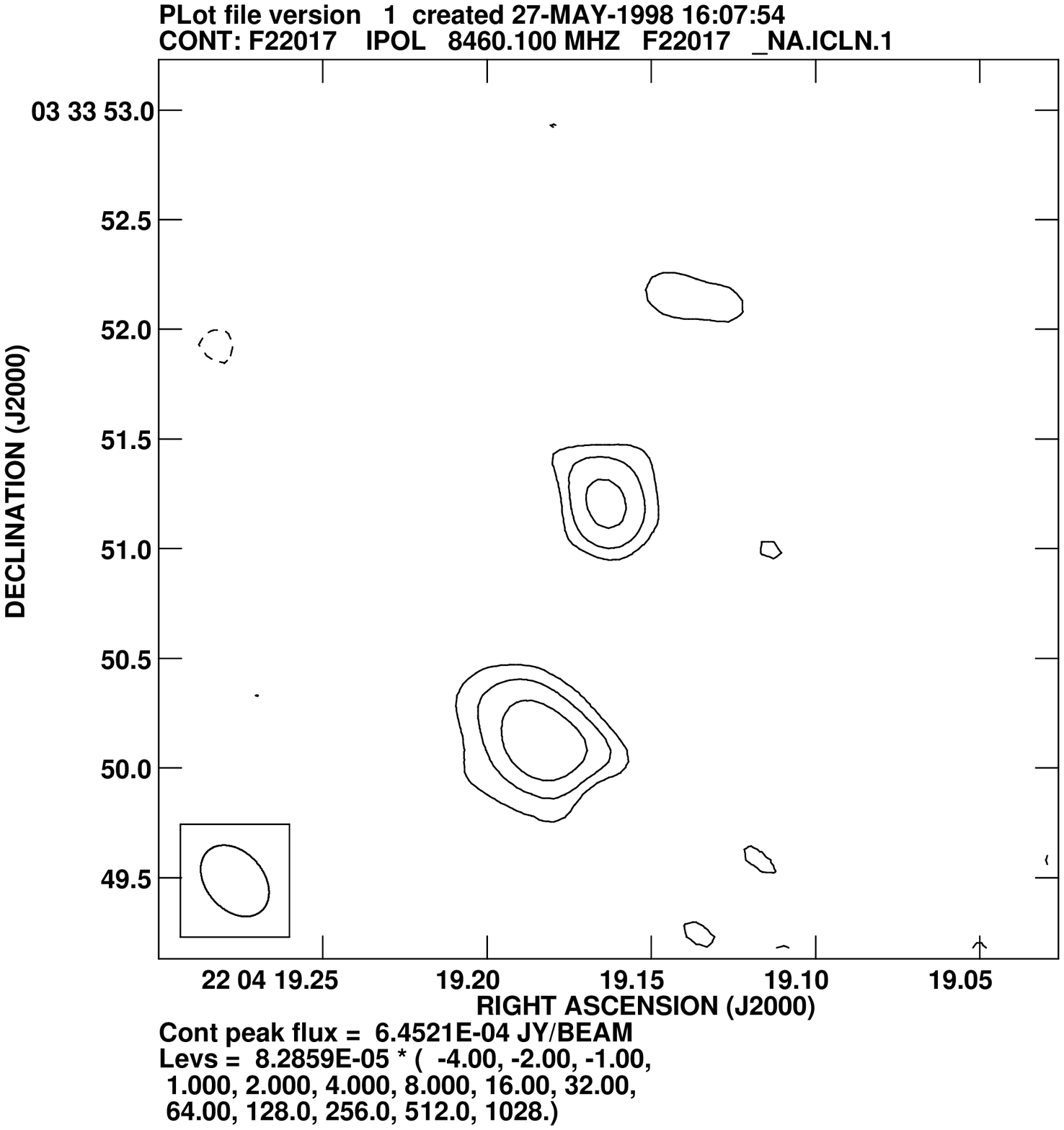}}
      \subfigure[) NGC 7213]{
        \includegraphics[width=4.8cm,clip,trim=0 62 0 35]{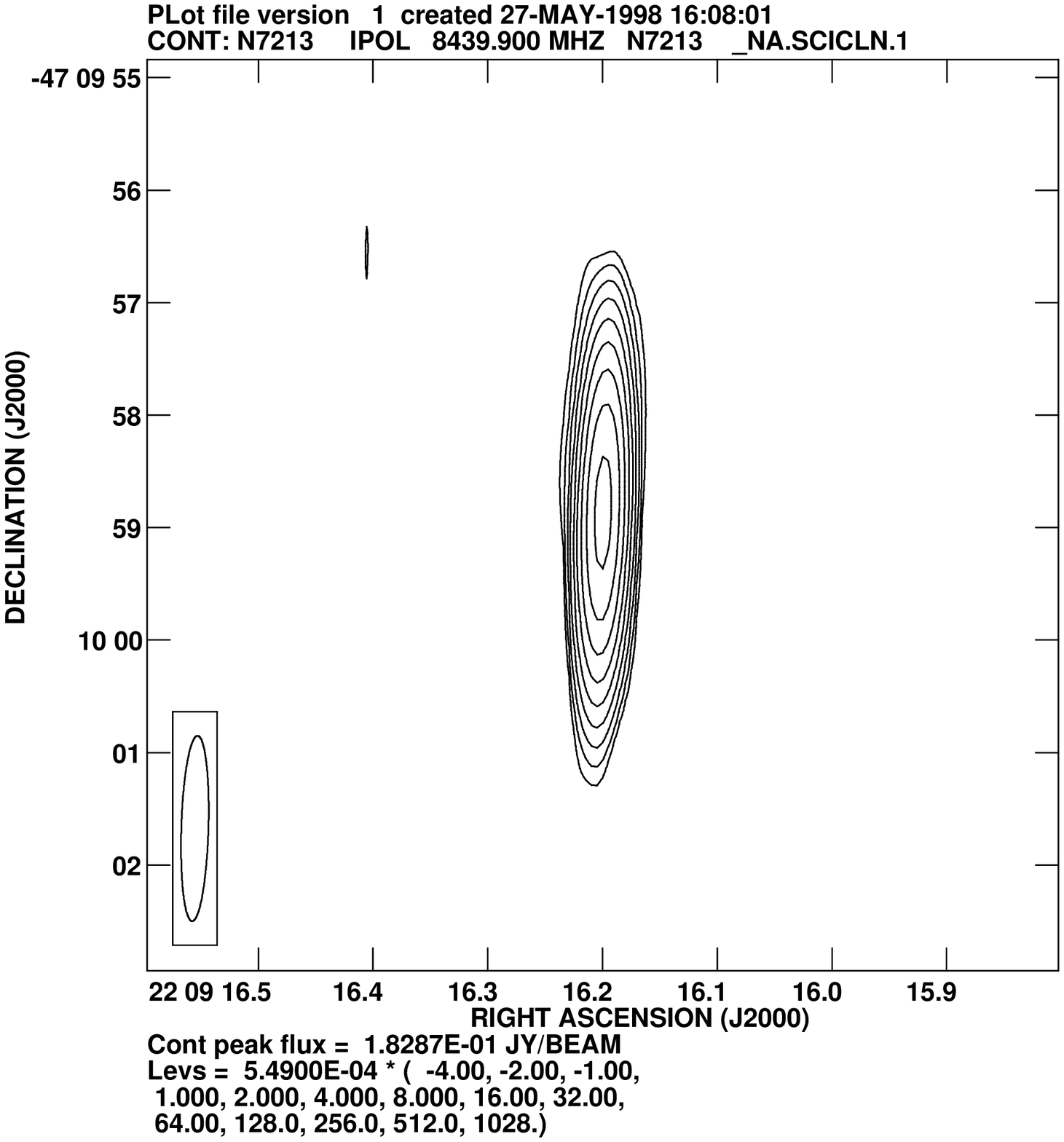}}
      \subfigure[) 3C 445]{
        \includegraphics[width=4.9cm,clip,trim=0 62 0 35]{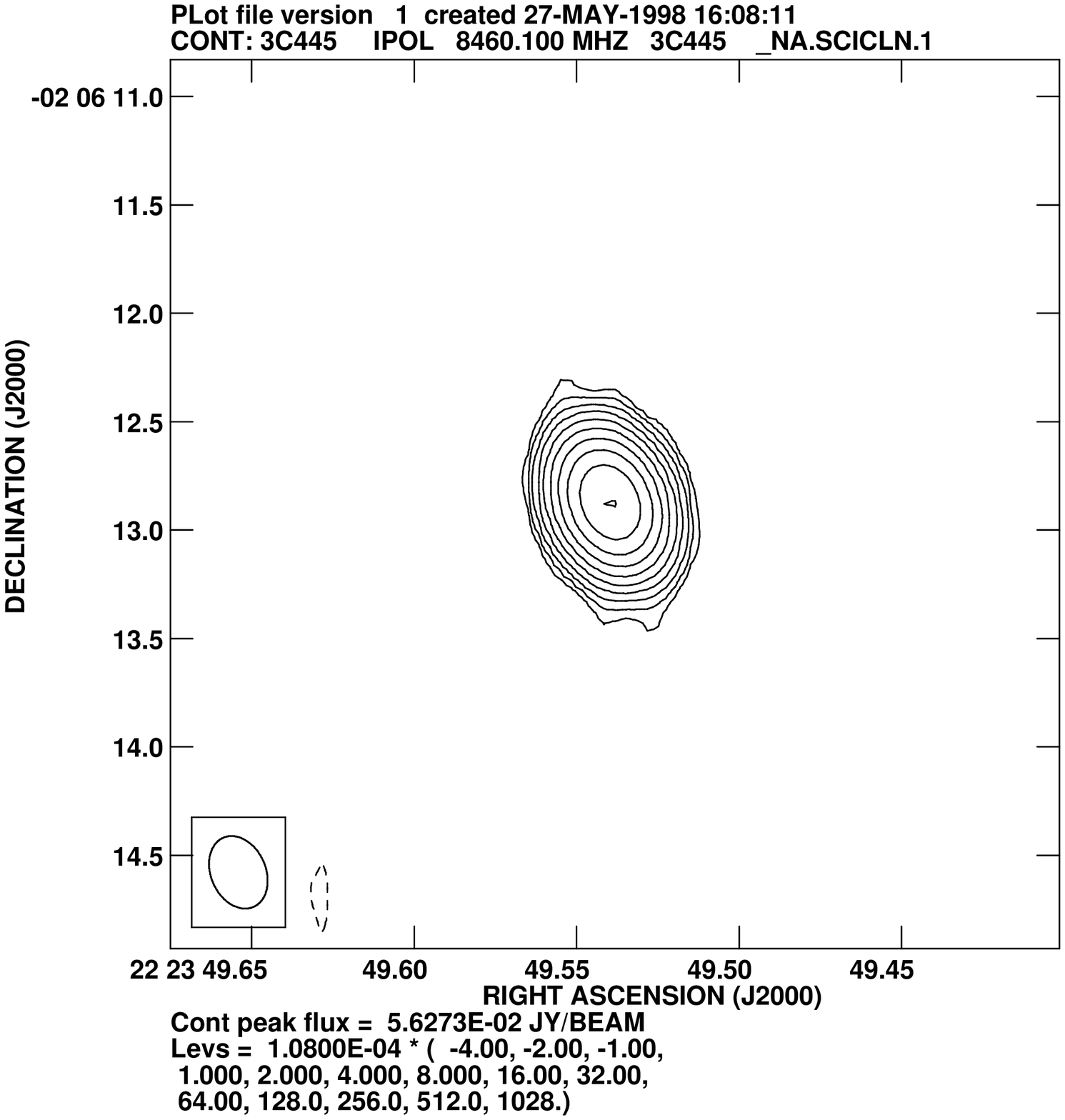}}
      \subfigure[) NGC 7314]{
        \includegraphics[width=4.8cm,clip,trim=0 62 0 35]{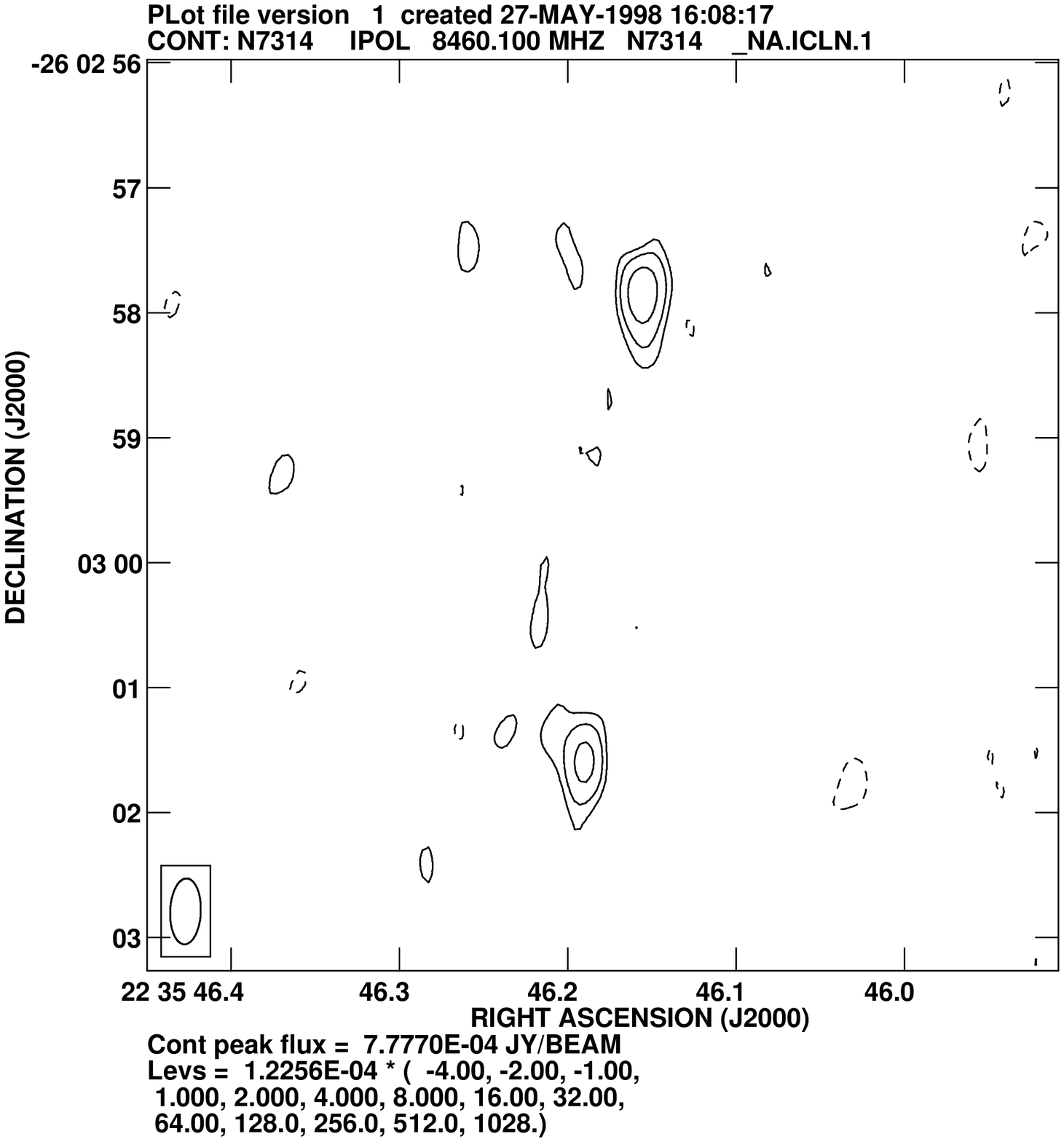}}
\caption{A-configuration 8.4 GHz images ({\it continued}).}
\end{figure*}

\setcounter{figure}{0}
\setcounter{subfigure}{72}

\begin{figure*}
\centering 
      \subfigure[) MCG-3-58-7]{
        \includegraphics[width=4.8cm,clip,trim=0 62 0 35]{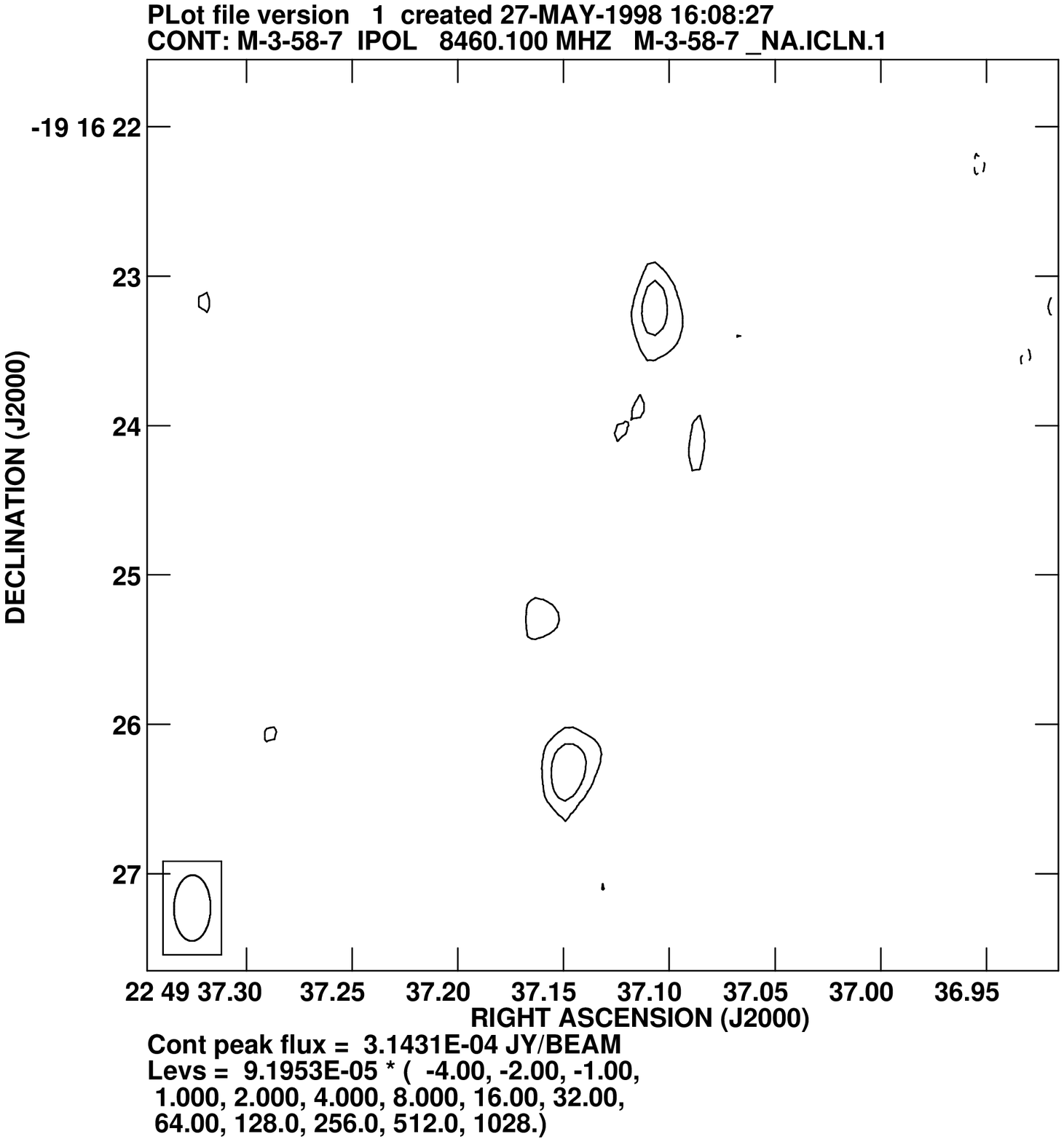}}
      \subfigure[) NGC 7496]{
        \includegraphics[width=4.8cm,clip,trim=0 62 0 35]{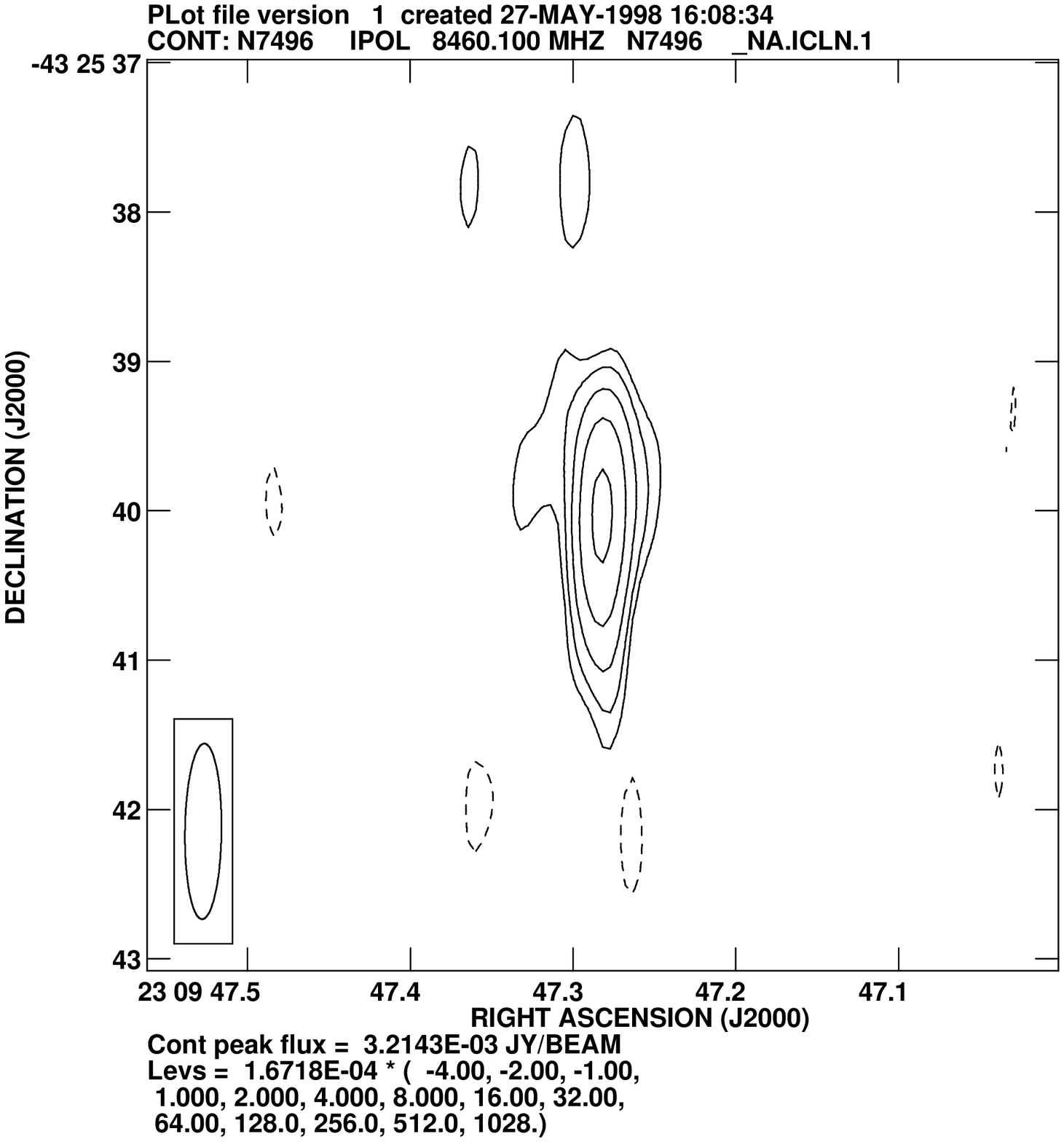}}

      \subfigure[) N7582]{
        \includegraphics[width=4.8cm,clip,trim=0 160 0 134]{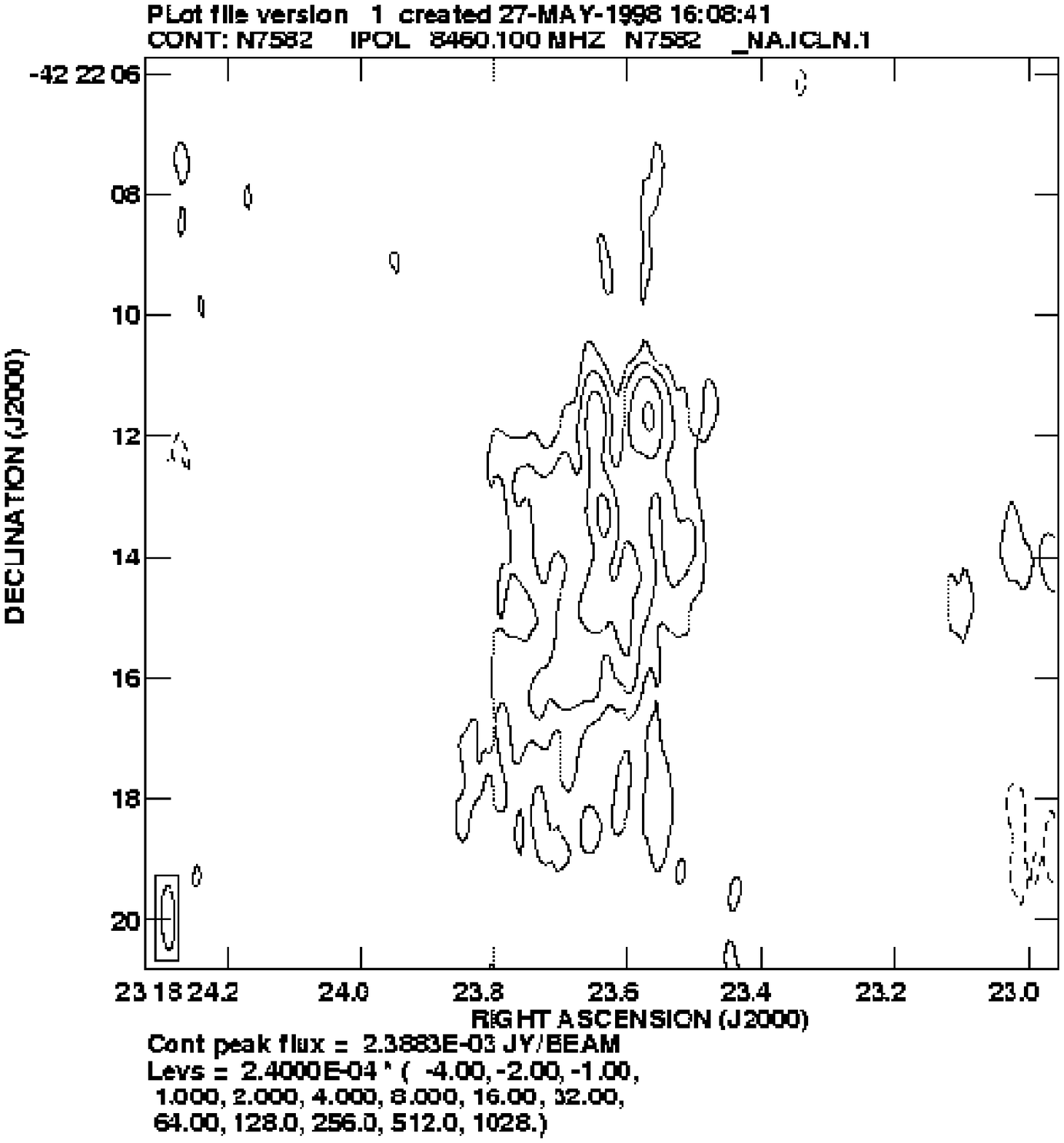}}
      \subfigure[) CG 381]{
        \includegraphics[width=4.9cm,clip,trim=0 62 0 35]{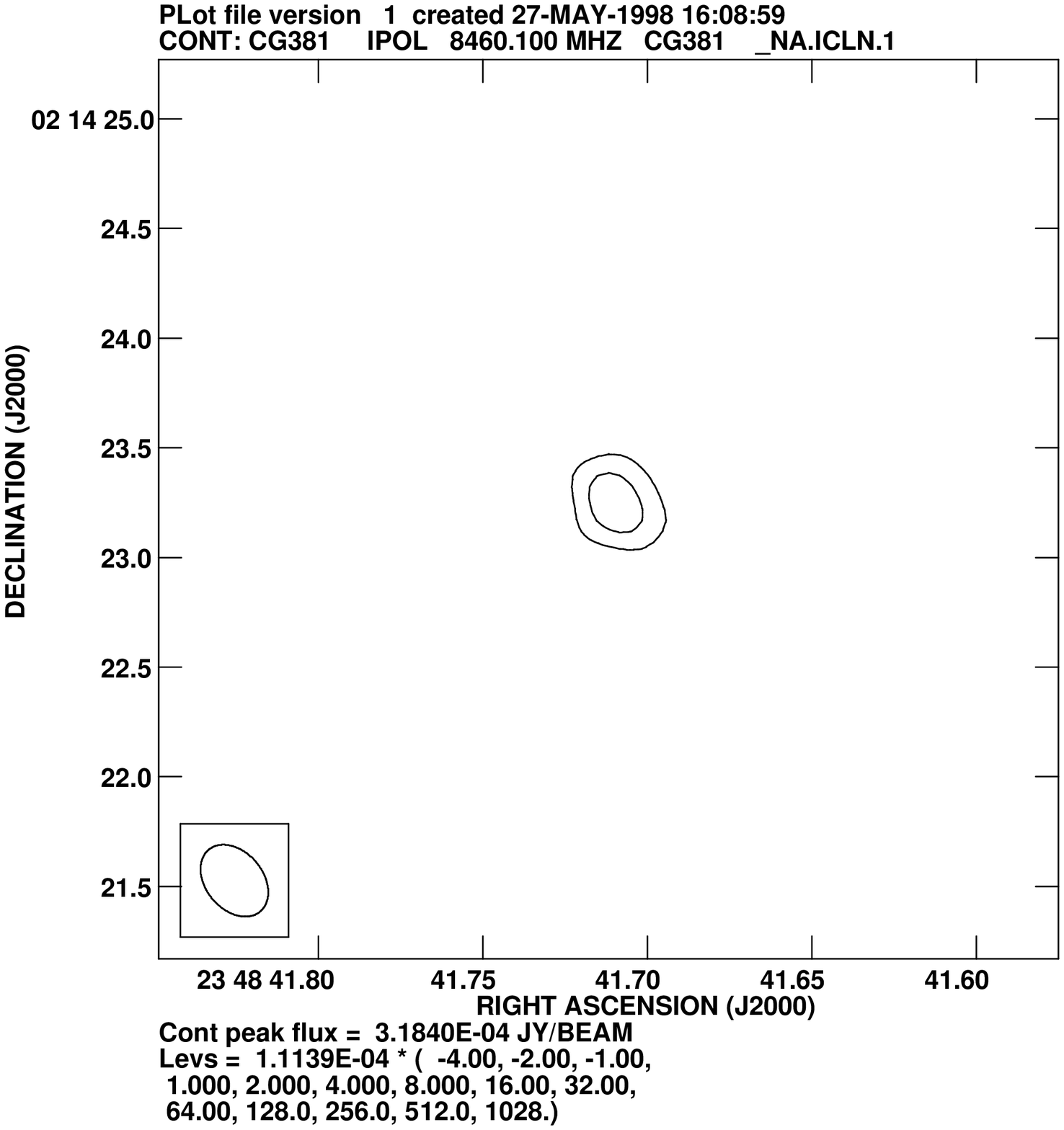}}
\caption{A-configuration 8.4 GHz images ({\it continued}).}
\end{figure*}

\renewcommand{\thesubfigure}{\alph{subfigure}}

\subsection{Description of sources}

A brief description of those sources with unusual radio morphologies is given
below (all positions are given in J2000 co--ordinates). 

{\it NGC 34 (Markarian 938):} 
This infrared--luminous galaxy is in the advanced stage of a merger
with two nuclei separated by approximately 6 kpc.
The weak [O III] emission and strong 
H$_{\alpha}$ emission, which is distributed over the entire galaxy, 
indicate that this galaxy is more properly classified as 
a starburst galaxy rather than a Seyfert galaxy \cite{Mulchaey96}. 

Note that NGC 34 contains a bright radio source more luminous than
two--thirds of the Seyferts observed (P$_{8.4GHz}$=1$\times$10$^{22}$
WHz$^{-1}$) and slightly resolved with a deconvolved
size of 0.4 arcsec (150 pc) and position angle of 140$^{\circ}$. 
High--resolution mid--infrared images show a double source separated
by approximately 1.2 arcsec, with the fainter source at a position angle of
around 180$^{\circ}$ from the brighter source \cite{Miles96}. 
Radio observations by \scite{Condon91} show a faint southern radio 
extension, not found in our map, which is consistent with the infrared
structure.

{\it NGC 526A:}
NGC 526A is strongly interacting with a galaxy to the east. 
The position of the slightly resolved radio source is closer to 
the peak in the optical continuum than to the apex of the
putative emission--line wedge identified by \scite{Mulchaey96};
whereas the emission--line wedge has a position angle of
123$^{\circ}$, that of the radio source is 43$^{\circ}$.

{\it Markarian 1034 (V Zw 233):} 
Markarian 1034 is an interconnected pair of Seyfert 1 galaxies, MCG+05-06-035 
(PGC 0009071) at $\alpha$=02$^{h}$ 23$^{m}$ 18.84$^{s}$, 
$\delta$=+32$^{\circ}$ 11$'$ 18.2$''$ and MCG+05-06-036 (PGC 0009074)
at $\alpha$=02$^{h}$ 23$^{m}$ 21.99$^{s}$,  
$\delta$=+32$^{\circ}$ 11$'$ 49.6$''$; positions are taken from the 
NASA/IPAC Extragalactic Database (NED) as described by \scite{NED}.  
The new observations show MCG +05-06-036. This source has extended `wings'
of radio emission symmetrical about a central bright component.  

{\it NGC 1125:}
The radio source has a clear linear structure with 3 aligned compact
components at a position angle of 120$^{\circ}$. The radio structure is 
not aligned with the slightly resolved  [O III] emission at a position angle of
56$^{\circ}$ \cite{Mulchaey96}.   

{\it NGC 1365:} 
There is no alignment between
the four compact radio components observed and none is co--incident 
with the photographic position of the nucleus which lies at
$\alpha$=03$^{h}$ 33$^{m}$ 35.57$^{s}$, $\delta$=-36$^{\circ}$ 08$'$
22.9$''$ (NED), or either of the two 
infrared sources observed by \scite{Telesco93}, the brightest of which
lies at $\alpha$=03$^{h}$ 33$^{m}$ 36.17$^{s}$, 
$\delta$=-36$^{\circ}$ 08$'$ 25.9$''$ (positional error of 1.5 arcsec). 

{\it IRAS F04385-0828:} 
We detect another weak radio source to the west at a projected
linear separation of 15.4 kpc, it has an integrated flux of 2.05 mJy
and lies at $\alpha$=04$^{h}$ 40$^{m}$ 51.45$^{s}$, 
$\delta$=-08$^{\circ}$ 22$'$ 23.85$''$.  
This weak source is well beyond the
optical radius of the host galaxy, which extends westwards 
approximately 6 kpc, and is unlikely to be related to the Seyfert nucleus.

{\it Markarian 6 (UGC 3547):} 
This source contains a central, well--collimated radio structure. 
On larger scales,
features suggestive of shells or bubbles are seen at varying position angles
(\pcite{Baum93}; \pcite{KukulaMkn6}).

{\it Markarian 79 (UGC 3973):} 
This large radio source shows a linear structure with 3 clearly aligned
components. Lower resolution measurements, at lower frequency,
show that the southern component
is stronger than the northern component 
\cite{Ulvestad+W84a}, whereas in our
image these components are almost equally bright. 

{\it NGC 2639 (UGC 4544):}
The source has a bright core and symmetrical east--west `wings'. 
In our map, the core--to--wings brightness ratio is an order of 
magnitude higher than in the 6 cm map of 
\scite{Ulvestad+W84a}; probably due to core variability.
NGC 2639 displays the rare properties of H$_{2}$O megamaser 
emission \cite{Wilson95b} and VLBI--scale radio emission \cite{Hummel82}.
It has also been classified as a LINER.

{\it NGC 2992 (Arp 245):}
This edge--on, interacting galaxy has unusual `loops'
of diffuse radio emission (\pcite{Wehrle88}; \pcite{Colbert96}).

{\it NGC 4922A/B:}
NGC 4922 is a system of 3 galaxies situated at $\alpha$=13$^{h}$ 01$^{m}$
24.50$^{s}$, $\delta$=+29$^{\circ}$ 18$'$ 29.9$''$ 
(Seyfert 2), $\alpha$=13$^{h}$ 01$^{m}$ 24.67$^{s}$,
$\delta$=+29$^{\circ}$ 18$'$ 33.0$''$ and $\alpha$=13$^{h}$ 01$^{m}$
25.26$^{s}$, $\delta$=+29$^{\circ}$ 18$'$ 49.58$''$ (PGC 044896/FIRST
J130125.2+291849); positions are taken from NED.  
We have detected PGC 044896 (an unresolved 7.8 mJy 
source at $\alpha$=13$^{h}$ 01$^{m}$ 25.26$^{s}$, $\delta$=29$^{\circ}$ 18$'$
49.53$''$, shown in Figure \ref{contours.fig}), but not
the nearby Seyfert nucleus. 

{\it NGC 5135:} 
Despite the bright radio flux of the nucleus
at lower resolutions and frequencies \cite{Ulvestad+W89},
no radio emission is observed at the nucleus in the current observations.
\scite{Wynn-Williams93} have
suggested that, ``most of the radio
emission of this Seyfert galaxy emanates from structures on either side
of the nucleus rather than from the nucleus itself''.
We detect a weak component with a flux density of 2.33
mJy at $\alpha$=13$^{h}$ 25$^{m}$ 44.9$^{s}$, $\delta$=-29$^{\circ}$ 50$'$
16.17$''$, but it is unlikely to be  
related to the active nucleus (see Section \ref{radprop.sec}).

{\it Markarian 273 (UGC 8696, I Zw 071):} 
This double--nucleus source is one of the
most ultra--luminous infrared galaxies known. 
The radio continuum shows two compact radio components separated by
approximately 600 pc. The radio structure has been mapped
previously by \scite{Condon91}, who classify it as a compact
starburst, and \scite{Knapen97}.
 
{\it Markarian 463 (UGC 8850):} 
Three aligned north--south radio components are detected in this well--studied
double--nucleus galaxy. North--south 
radio structures have previously been observed
from VLBI--scales (\pcite{Hummel82}; \pcite{Kukula99}) 
to up to 18 kpc south of the nucleus 
\cite{Mazzarella91}. 
The new observations match the 6 cm radio structure observed by
\scite{Mazzarella91} with the exception of the weak central radio
component which appears to coincide with a bright knot in the aligned
optical jet \cite{Uomoto93}. 
This is the second most radio--luminous Seyfert in the extended 12
$\mu$m sample.

{\it NGC 5506:}
This edge--on galaxy has a compact core surrounded by a diffuse
halo. The `loop' identified by \scite{Wehrle87} is just traceable to
the north--west of the core (see also \pcite{Colbert96}).

{\it UGC 9913 (Arp 220):}  
The two closely separated radio components are the nuclei of this well--known double--nucleus 
galaxy. The
nature of its activity is uncertain. Despite being 
the 7th most radio--luminous radio--quiet source in the extended 12
$\mu$m sample, recent VLBI observations by
\scite{Smith98arp} provide strong evidence that the radio emission
from the north--western component originates in a 
compact nuclear starburst.

{\it IRAS F22017+0319:}
If the weak northernmost component is included, this source is 
a linearly aligned triple radio source.

{\it NGC 7314:}
This variable X--ray source is a north--south radio double.

{\it MCG-03-58-007:}
A north--south radio double.

{\it NGC 7582:}
The source has a diffuse radio structure consistent 
with the lower--resolution observations of \scite{Ulvestad+W84b}.

\subsection{Radio properties}
\label{radprop.sec}

\begin{table*}
\footnotesize
\begin{center}

\begin{tabular}{|l|c|c|r|r|r|r|r|c|c|} \hline 
{\bf Galaxy}  & {\bf Type} & {\bf z} & {\bf RA(J2000)} & {\bf Dec(J2000)} & {\bf S(mJy)} & 
{\bf $\theta$($''$)} & {\bf D(pc)} & {\bf T$_{rad}$} & {\bf PA($^{\circ}$)} \\ \hline 

Mrk335$^{\ast}$        & 1 & 0.02586 & 00 06 19.541 &  20 12 10.63 &     2.1 & $<$   0.1 & $<$     50 &   U 
&      \\ 
 
NGC34(Mrk938)    & sb(2) & 0.01919 & 00 11 06.553 & -12 06 27.71 &    14.5 &       0.4 &        
149 &   S & 140  \\ 
 NGC262(Mrk348)   & 2 & 0.01509 & 00 48 47.144 &  31 57 25.07 &   346.0 & $<$   0.1 & $<$     29 
&   U &     \\ 
IZW1$^{\ast}$         & 1 & 0.06039 & 00 53 34.938 &  12 41 35.96 &     0.9 & $<$   0.1 & $<$    117 &   U 
&     \\ 
 E541-IG12    & 2 & 0.05636 & 01 02 17.381 & -19 40 08.52 &     0.8 & $<$   0.2 & $<$    219 &   
U &     \\ 
 NGC424(TOL0109) & 2 & 0.01169 & 01 11 27.647 & -38 05 00.72 &    11.9 & $<$   0.3 & $<$     68 
&   U &     \\ 
 NGC526A        & 1 & 0.01905 & 01 23 54.382 & -35 03 55.68 &     7.1 &       0.3 &        111 &   S &  
~43 \\ 
 NGC513         & 2 & 0.01949 & 01 24 26.803 &  33 47 58.24 &     1.2 & $<$   0.1 & $<$     38 &   U 
&     \\ 
 F01475-0740  & 2 & 0.01739 & 01 50 02.696 & -07 25 48.53 &   129.8 & $<$   0.2 & $<$     67 
&   U &     \\ 
 Mrk1034       & 1 & 0.03797 & 02 23 21.965 &  32 11 48.90 &    11.6 &       2.1 &       1546 &   L 
&  ~65 \\ 
 MCG-3-7-11     & 1 & 0.03378 & 02 24 40.546 & -19 08 30.50 &     1.9 &       2.1 &       1376 &   L 
&  ~34 \\ 
              &   &         & 02 24 40.587 & -19 08 29.48 & $^{\dagger}$0.2 &&&&\\ 
              & & & & Total&   2.1 &&&&\\
 NGC931(Mrk1040)  & 1 & 0.01639 &              &              & $<$   0.4 &   &   &   &     \\ 
 NGC1056(Mrk1183) & 2 & 0.00520 & 02 42 48.483 &  28 34 25.69 &     0.6 & $<$   0.2 & $<$     20 
&   U &     \\ 
 NGC1068$^{\ast}$        & 2 & 0.00384 & 02 42 40.608 & -00 00 51.52 & $^{\dagger}$30.3 &      10.0 &    745 &   L &  ~65  \\ 
              &   &         & 02 42 40.718 & -00 00 47.58 &   282.4 &&&&\\ 
              &   &         & 02 42 40.724 & -00 00 47.40 &   252.6 &&&&\\ 
              &   &         & 02 42 40.916 & -00 00 43.07 &    70.4 &&&&\\ 
 & & & &   Total &   762.4 &&&&\\
 NGC1097        & L(2) & 0.00430 & 02 46 18.955 & -30 16 28.43 &     3.1 & $<$   0.2 & $<$     17 &   U &     \\ 
 NGC1125        & 2 & 0.03178 & 02 51 40.438 & -16 39 02.33 & $^{\dagger}$3.0 &       0.8 &      493 &   L & 120 \\ 
              &   &         & 02 51 40.458 & -16 39 02.53 &     3.7 &&&&\\ 
              &   &         & 02 51 40.487 & -16 39 02.74 &     5.3 &&&&\\ 
 & & & &Total &   11.7 &&&&\\ 
NGC1143/4$^{\ast}$       & 2 & 0.02880 & 02 55 12.233 & -00 11 00.73 & $^{\dagger}$2.3 &       7.5   &     4189 &   L &  ~62 \\ 

              &   &         & 02 55 12.487 & -00 10 59.03 &     1.6 &&&&\\ 
              &   &         & 02 55 12.560 & -00 10 57.03 &     3.1 &&&&\\ 
              &   &         & 02 55 12.653 & -00 10 57.22 &     0.9 &&&&\\ 
  & & & & Total&   10.5 \\ 
MCG-2-8-39      & 2 & 0.03008 &       &         & $<$   0.3&   &   &   &     \\ 
 NGC1194        & 2 & 0.01339 & 03 03 49.106 & -01 06 13.63 &     0.9 & $<$   0.2 & $<$     52 &   U 
&     \\ 
 NGC1241        & 2 & 0.00720 & 03 11 14.618 & -08 55 18.88 &     6.8 & $<$   0.1 & $<$     14 &   U 
&     \\ 
 NGC1320(Mrk607)  & 2 & 0.00989 & 03 24 48.675 & -03 02 32.38 &     1.0 & $<$   0.2 & $<$     38 
&   U &     \\ 
 NGC1365        & 1 & 0.00550 & 03 33 36.036 & -36 08 28.25 &     2.1 &      13.2 &       1407 &   A 
&     \\ 
              &   &         & 03 33 36.402 & -36 08 18.46 & $^{\dagger}$2.0 &&&&\\ 
              &   &         & 03 33 36.647 & -36 08 28.19 &     1.8 &&&&\\ 
              &   &         & 03 33 36.765 & -36 08 18.49 &     3.3 &&&&\\ 
 & & & & Total&   9.3 \\
 NGC1386        & sb(2) & 0.00310 & 03 36 46.197 & -35 59 57.41 &     6.8 & $<$   0.3 & $<$     18 &   U &      \\ 
 F03362-1642  & 2 & 0.03598 & 03 38 33.557 & -16 32 18.40 &     1.5 & $<$   0.3 & $<$    209 &   U &     \\ 
 F03450+0055  & 1 & 0.03098 & 03 47 40.193 &  01 05 13.97 &     6.8 &       0.3 &        180 &   S &  ~19 \\ 
 Mrk618        & 1 & 0.03468 & 04 36 22.299 & -10 22 34.03 &     2.9 &       0.2 &        134 &   S & ~87  \\ 
 F04385-0828  & 2 & 0.01519 & 04 40 54.964 & -08 22 22.08 &     6.0         & $<$   0.3 & $<$     88 &   U &     \\ 
 NGC1667        & 2 & 0.01529 & 04 48 37.168 & -06 19 11.97 &     1.5         & $<$   0.2 & $<$     59 &   U &     \\ 
 MCG-5-13-17    & 1 & 0.01249 & 05 19 35.794 & -32 39 28.23 & $^{\dagger}$1.0 &       1.3 &        315 &   S &  ~27 \\ 
              &   &         & 05 19 35.815 & -32 39 27.82 &     1.4 &&&&\\ 
 & & & & Total&   3.24 &&&&\\
 F05189-2524  & 2 & 0.04147 & 05 21 01.405 & -25 21 45.30 &     6.9 &       0.4 &        322 &   S & 177 \\ 
 E253-G3      & 2 & 0.04067 &       &         & $<$   2.4 &&&&\\ 
 F05563-3820  & 1 & 0.03438 & 05 58 01.725 & -38 20 04.11 &     3.7 &      24.3 &      16200 &   A &     \\ 
              &   &         & 05 58 01.814 & -38 19 58.37 &     1.6 &&&&\\ 
              &   &         & 05 58 02.030 & -38 20 04.34 &     6.9 &&&&\\ 
              &   &         & 05 58 02.273 & -38 20 01.59 &     1.6 &&&&\\ 
              &   &         & 05 58 02.288 & -38 19 49.04 &     4.0 &&&&\\ 
              &   &         & 05 58 02.384 & -38 20 02.29 &     2.1 &&&&\\ 
              &   &         & 05 58 02.676 & -38 20 12.86 & $^{\dagger}$2.6 &&&&\\ 
 & & & & Total&   22.4 &&&&\\
 Mrk6          & 2(1) & 0.01849 & 06 52 12.331 &  74 25 37.05 & $^{\dagger}$27.9 &       2.4 &        860 &   L & 170 \\ 
              &   &         & 06 52 12.327 &  74 25 38.13 &    10.9 &&&&\\ 
 & & & & Total&   38.8 &&&&\\
 Mrk9          & 1 & 0.00630 & 07 36 57.016 &  58 46 13.48 & $^{\dagger}$0.6  &       0.9 &        110 &   L & 105 \\ 
              &   &         & 07 36 56.900 &  58 46 13.72 &     0.2 &&&&\\ 
 & & & & & Total  0.8 &&&&\\
\hline
\end{tabular}
\caption[Observational results.]{Observational results.}
\label{results.tab}
\end{center}
\end{table*}

\begin{table*}
\footnotesize
\begin{center}
\begin{tabular}{|l|c|c|r|r|r|r|r|c|c|} \hline 
{\bf Galaxy}  & {\bf Type} & {\bf z} & {\bf RA(J2000)} & {\bf Dec(J2000)} & {\bf S(mJy)} & 
{\bf $\theta$($''$)} & {\bf D(pc)} & {\bf T$_{rad}$} & {\bf PA($^{\circ}$)} \\ \hline
 Mrk79         & 1 & 0.02208 & 07 42 32.809 &  49 48 34.90 &     1.2 &       3.5 &       1499 &   L &  ~12 \\ 
              &   &         & 07 42 32.841 &  49 48 36.64 &     1.1 &&&&\\ 
              &   &         & 07 42 32.790 &  49 48 33.67 & $^{\dagger}$0.8 &&&&\\ 
 & & & & Total&   3.1 &&&&\\
 F07599+6508  & 1 & 0.15000 & 08 04 30.465 &  64 59 52.85 &     5.8 & $<$   0.1 & $<$    291 &   U &     \\ 
 NGC2639        & 1 & 0.01079 & 08 43 38.077 &  50 12 19.99 &   118.0 &       1.6 &        335 &   L & 
109 \\ 
 OJ287        & 1 & 0.30579 & 08 54 48.876 &  20 06 30.59 &  1520.7 &&&&\\ 
 F08572+3915  & 2 & 0.05826 & 09 00 25.379 &  39 03 54.15 &     3.8 & $<$   0.1 & $<$    113 &   U &     \\ 
 Mrk704        & 1 & 0.02928 & 09 18 25.995 &  16 18 19.65 &     0.9 & $<$   0.1 & $<$     57 &   U &     \\ 
 UGC5101        & 1 & 0.03997 & 09 35 51.605 &  61 21 11.74 &    45.1 &       0.2 &        155 &   S &  ~88 \\ 
 NGC2992        & 2(1) & 0.00769 & 09 45 41.945 & -14 19 34.60 &     5.5 &       8.2 &       1223 &   D &     \\ 
 & & & & Total&   40.3 \\
 Mrk1239       & 1 & 0.01989 & 09 52 19.096 & -01 36 43.46 &    10.5 & $<$   0.1 & $<$     39 &   U &     \\ 
 NGC3031(M81)    & L(1) & 0.00110 & 09 55 33.174 &  69 03 55.04 &   221.0 & $<$   0.1 & $<$      2 &   U &     \\ 
 NGC3079$^{\ast}$        & 2 & 0.00371 & 10 01 57.801 &  55 40 47.24 &    93.3 & $<$   0.1 & $<$      7 &   U &     \\ 
 3C234        & Q(1) & 0.18477 & 10 01 44.908 &  28 46 52.28 &    38.8 &&&&\\ 
              &   &         & 10 01 46.098 &  28 46 54.17 &   162.3 &&&&\\ 
              &   &         & 10 01 46.930 &  28 46 40.10 &    11.6 &&&&\\ 
              &   &         & 10 01 47.280 &  28 47 18.67 &     7.7 &&&&\\ 
              &   &         & 10 01 49.522 &  28 47 08.89 &    38.8 &&&&\\ 
              &   &         & 10 01 53.824 &  28 47 38.97 &    63.8 &&&&\\ 
 & & & & Total&   323.0 &&&&\\
NGC3227$^{\ast}$       & 1 & 0.00390 & 10 23 30.574 &  19 51 54.24 &    12.2 &       0.5  &         38 &   S & 173 \\ 
 NGC3511        & 1 & 0.00370 &       &         & $<$   0.3 &&&&\\ 
 NGC3516$^{\ast}$        & 1 & 0.00883 & 11 06 47.466 &  72 34 07.30 &     3.1 & $<$   0.1 & $<$     17 &   U &      \\
 M+0-29-23    & 2 & 0.02488 &       &         & $<$   0.3 &&&&\\ 
 NGC3660        & 2 & 0.01229 &       &         & $<$   0.3 &&&&\\ 
 NGC3982$^{\ast}$        & 2 & 0.00396 & 11 56 28.134 &  55 07 30.95 &     0.8 & $<$   0.2 & $<$     15 &   U &     \\ 
 NGC4051$^{\ast}$        & 1 & 0.00204 & 12 03 09.605 &  44 31 52.73 &     0.6 & $<$   0.1 & $<$      4 &   U &     \\ 
 UGC7064        & 1 & 0.02498 &       &         & $<$   0.3 &&&&\\ 
 NGC4151$^{\ast}$        & 1 & 0.00323 & 12 10 32.424 &  39 24 20.69 &     8.0 &       5.0 &        314 &   L &  ~77 \\ 
              &   &         & 12 10 32.508 &  39 24 20.89 & $^{\dagger}$10.9 &&&&\\ 
              &   &         & 12 10 32.545 &  39 24 21.03 &    17.7 &&&&\\ 
              &   &         & 12 10 32.582 &  39 24 21.09 &    30.5 &&&&\\ 
              &   &         & 12 10 32.658 &  39 24 21.35 &     5.2 &&&&\\ 
 & & & & Total&   72.3 &&&&\\
 Mrk766$^{\ast}$        & 1 & 0.01279 & 12 18 26.517 &  29 48 46.50 &     8.7         &       0.3 &         74 &   S &  ~27\\ 
 NGC4388$^{\ast}$        & 2 & 0.00848 & 12 25 46.735 &  12 39 41.87 & $^{\dagger}$1.4 &       3.3 &        543 &   L &  ~23 \\ 
         &  &  & 12 25 46.780 &  12 39 43.68 &     3.4 &&&&\\ 
 & & & & Total&   9.4 &&&&\\
 NGC4501        & 2 & 0.00769 &       &         & $<$   0.2 &&&&\\ 
 NGC4579        & 1 & 0.00500 & 12 37 43.511 &  11 49 05.42 &    36.5 & $<$   0.4 & $<$     39 &   
U &     \\ 
 NGC4593        & 1 & 0.00829 & 12 39 39.445 & -05 20 39.02 &     1.9 & $<$   0.2 & $<$     32 &   U 
&     \\
 NGC4594        & 1 & 0.00380 & 12 39 59.435 & -11 37 23.11 &    86.6 & $<$   0.3 & $<$     22 &   
U &     \\
 NGC4602        & 1 & 0.00849 &              &              & $<$ 0.2 &           &            &     &     \\ 
 TOL1238-364  & 2 & 0.00120 & 12 40 52.783 & -36 45 20.63 &     3.5 &       2.4 &         56 &   A 
&     \\ 
              &   &         & 12 40 52.841 & -36 45 21.22 & $^{\dagger}$2.3 &   &            &     &     \\ 
 & & & &  Total&   9.6 &&&&\\
 MCG-2-33-34    & 1 & 0.01389 & 12 52 12.479 & -13 24 53.31 &     1.5 & $<$   0.1 & $<$     27 &   
U &     \\ 
Mrk231(UGC8058)$^{\ast}$  & 1 & 0.04096 & 12 56 14.238 &  56 52 25.21 &   234.5 & $<$   0.1 & $<$     79 &   U &     \\ 
 NGC4922A/B     & 2 & 0.02368 &              &             & $<$ 0.3 &
 &            &     &     \\ 
 NGC4941        & 2 & 0.00370 & 13 04 13.096 & -05 33 05.69 &     1.7 &       1.0 &         72 &   S & 127 \\ 
              &   &         & 13 04 13.109 & -05 33 05.91 & $^{\dagger}$2.1 &&&&\\ 
 & & & & Total&   4.8 &&&&\\
 NGC4968        & 2 & 0.00989 & 13 07 05.915 & -23 40 36.83 & $^{\dagger}$2.1 &       1.3 &        249 &   L &  ~25 \\ 
         & 2 & 0.00989 & 13 07 05.928 & -23 40 36.47 &     3.7 &&&&\\ 
 & & & & Total&   5.8 &&&&\\
 NGC5005        & 2 & 0.00400 & 13 10 56.258 &  37 03 32.88 &     8.8 &       0.7 &         54 &   S & 155 \\ 
 NGC5033$^{\ast}$        & 1 & 0.00297 & 13 13 27.469 &  36 35 37.93 &     2.1 & $<$   0.1 & $<$      6 &   U &     \\ 
 MCG-3-34-63    & 2 & 0.01719 & 13 22 24.465 & -16 43 42.45 &    42.2 &       1.3 &        433 &   S &  ~30 \\ 
 NGC5135        & 2 & 0.01369 &        &              & $<$   2.3 &&&&\\ 
 NGC5194(M51)    & 2 & 0.00160 & 13 29 52.704 &  47 11 42.79 &     0.5 & $<$   0.2 & $<$      6 &   U &     \\ 
 MCG-6-30-15    & 1 & 0.00769 &       &         & $<$   0.3 &&&&\\ 
\hline
\end{tabular}
\end{center}
\contcaption{}  
\end{table*}

\begin{table*}
\footnotesize
\begin{center}
\begin{tabular}{|l|c|c|r|r|r|r|r|c|c|} \hline 
{\bf Galaxy}  & {\bf Type} & {\bf z} & {\bf RA(J2000)} & {\bf Dec(J2000)} & {\bf S(mJy)} 
& {\bf $\theta$($''$)} & {\bf D(pc)} & {\bf T$_{rad}$} & {\bf PA($^{\circ}$)}  \\ \hline
 F13349+2438  & 1 & 0.10693 & 13 37 18.722 &  24 23 03.32 &     4.7  &   $<$    0.2 &      $<$  415 &   U &  \\ 
 NGC5256(Mrk266)  & 2 & 0.02748 & 13 38 17.238 &  48 16 43.94 &     1.3 &      12.8 &       6821 &   
A &     \\ 
              &   &         & 13 38 17.241 &  48 16 32.18 &     4.6 &&&&\\ 
              &   &         & 13 38 17.600 &  48 16 35.67 & $^{\dagger}$2.9 &&&&\\ 
              &   &         & 13 38 17.789 &  48 16 41.09 &     5.7 &&&&\\ 
 & & & & Total&   14.5 &&&&\\
 Mrk273(UGC8696)  & 2 & 0.03727 & 13 44 42.127 &  55 53 13.49 & $^{\dagger}$30.5 &       0.8 &        578 &   L & 149 \\ 
              &   &         & 13 44 42.179 &  55 53 12.77 &     2.7 &&&&\\ 
 & & & & Total&   33.5 &&&&\\
 IC4329A       & 1 & 0.01609 & 13 49 19.263 & -30 18 34.09 & $^{\dagger}$10.4 &      23.0 &       7177 &   A &     \\ 
              &   &         & 13 49 19.271 & -30 18 36.70 &     1.8 &&&&\\ 
              &   &         & 13 49 20.427 & -30 18 19.29 &     1.5 &&&&\\ 
              &   &         & 13 49 20.686 & -30 18 40.52 &     1.9 &&&&\\ 
 & & & & Total&   15.5 &&&&\\
 NGC5347        & 2 & 0.00779 & 13 53 17.803 &  33 29 26.77 &     0.8 & $<$   0.3 & $<$     45 &   U &     \\ 
 Mrk463        & 2 & 0.05047 & 13 56  2.886 &  18 22 18.68 & $^{\dagger}$43.6 &       1.2 &       1174 &   L & 177 \\ 
              &   &         & 13 56 2.890 &  18 22 17.46 &     3.6 &&&&\\ 
              &   &         & 13 56 2.895 &  18 22 17.85 &     0.4 &&&&\\ 
 & & & & Total&   47.2 &&&&\\
 NGC5506        & 2 & 0.00580 & 14 13 14.880 & -03 12 27.68 &    67.6 &       3.2 &        360 & D+U &     \\ 
 NGC5548$^{\ast}$        & 1 & 0.01728 & 14 17 59.541 &  25 08 12.65 &     2.2 & $<$   0.1 & $<$     34 &   U &     \\ 
 Mrk817        & 1 & 0.03148 & 14 36 22.084 &  58 47 39.38 &     2.8 & $<$   0.2 & $<$    122 &   U &     \\ 
 F15091-2107  & 1 & 0.04457 & 15 11 59.803 & -21 19 01.52 &    7.8 &    $<$   0.2 &  $<$   173 &   S & ~~2 \\ 
 NGC5929$^{\ast}$        & 2 & 0.00835 & 15 26 06.109 &  41 40 14.05 & $^{\dagger}$9.1 &       1.3 &        210 &   L &  ~60 \\ 
              &   &         & 15 26 06.167 &  41 40 14.42 &     1.1 &&&&\\ 
              &   &         & 15 26 06.208 &  41 40 14.68 &     6.5 &&&&\\ 
 & & & & Total&   16.8 &&&&\\
 NGC5953        & 2 & 0.00660 & 15 34 32.382 &  15 11 37.44 &     1.1 & $<$   0.3 & $<$     38 &   U &     \\ 
 UGC9913(Arp220) & sb(2) & 0.01819 & 15 34 57.267 &  23 30 12.13 &    77.2 &       1.0 &        353 &   L & 100 \\ 
              &   &         & 15 34 57.336 &  23 30 11.96 & $^{\dagger}$63.0 &&&& \\ 
 & & & & Total&   140.3 &&&&\\
 MCG-2-40-4     & 2 & 0.02438 & 15 48 24.962 & -13 45 27.22 & $^{\dagger}$2.4 &       0.8 &        378 &   L & 121 \\ 
              &   &         & 15 48 25.006 & -13 45 27.61 &     0.6 &&&&\\ 
 & & & & Total&   3.0 &&&&\\
 F15480-0344  & 2 & 0.03038 & 15 50 41.498 & -03 53 18.06 &    12.4 & $<$   0.1 & $<$     59 &   U &     \\ 
 NGC6890        & 2 & 0.00809 & 20 18 16.467 & -44 48 36.50 &     0.6 &      22.7 &       3563 &   A &     \\ 
              &   &         & 20 18 17.346 & -44 48 17.30 &     0.4 &&&&\\ 
              &   &         & 20 18 18.092 & -44 48 21.78 & $^{\dagger}$0.5 &&&&\\ 
 & & & & Total&   1.5 &&&&\\
 Mrk509        & 1 & 0.03458 & 20 44 9.744 & -10 43 24.73 &     2.2  & $<$   0.1 & $<$     67 &   U &     \\ 
 UGC11680(Mrk897) & 2 & 0.02638 & 21 07 45.858 &  03 52 40.45 &     3.5 & $<$   0.1 & $<$     51 &   U &     \\ 
 NGC7130(IC5135)  & 2 & 0.01619 & 21 48 19.522 & -34 57 04.79 &    18.1 &       0.6 &        188 &   S & 165 \\ 
 NGC7172        & 2 & 0.00859 & 22 02 1.888 & -31 52 10.47 &     4.7  &       0.4 &         67 &   S &  ~18 \\ 
 F22017+0319  & 2 & 0.06595 & 22 04 19.141 &  03 33 52.17 &     0.2 &       2.1 &       2686 &   L 
& 163 \\ 
              &   &         & 22 04 19.163 &  03 33 51.22 &     0.6 &&&&\\ 
              &   &         & 22 04 19.184 &  03 33 50.12 & $^{\dagger}$1.2 &&&&\\ 
 & & & & Total&   2.0 &&&&\\
 NGC7213        & 1 & 0.00590 & 22 09 16.200 & -47 09 58.86 &   183.8 & $<$   0.6 & $<$     69 &   U &     \\ 
 3C445        & 1 & 0.05616 & 22 23 49.540 & -02 06 12.87 &    58.1 &&&&\\ 
 NGC7314        & 1 & 0.00480 & 22 35 46.155 & -26 02 57.87 &     1.0 &       3.8 &        353 &   L & 173 \\ 
              &   &         & 22 35 46.191 & -26 03 01.60 & $^{\dagger}$0.7 &&&&\\ 
 & & & & Total&   1.7 &&&&\\
 MCG-3-58-7     & 2 & 0.03158 & 22 49 37.106 & -19 16 23.23 &     0.5 &       3.2&       1960 &   L & 169 \\ 
              &   &         & 22 49 37.148 & -19 16 26.34 & $^{\dagger}$0.4 &&&&\\ 
 & & & & Total&   0.9 &&&&\\
 NGC7469$^{\ast}$        & 1 & 0.01598 & 23 03 15.616 &  08 52 26.12 &    16.0 &       0.2 &         62 &   S & 107 \\ 
 NGC7496        & 2 & 0.00550 & 23 09 47.282 & -43 25 40.05 &     3.8 & $<$   0.4 & $<$     43 &   U 
&     \\ 
 NGC7582        & 2 & 0.00530 & 23 18 23.638 & -42 22 13.35 &    51.8 &      10.2 &       1046 &   D 
&     \\ 
 NGC7590        & 2 & 0.00500 &       &         & $<$   0.2 &&&&\\ 
 NGC7603(Mrk530)$^{\ast}$  & 1 & 0.02957 & 23 18 56.653 &  00 14 37.96 &     3.3 & $<$   0.2 & $<$    115 &   U &     \\ 
 NGC7674(Mrk533)$^{\ast}$  & 2 & 0.02899 & 23 27 56.680 &  08 46 44.33 & $^{\dagger}$12.8 &       0.8 &        450 &   L & 117 \\ 
              &   &         & 23 27 56.712 &  08 46 44.13 &    27.0 &&&&\\ 
 & & & & Total&   39.8 &&&&\\
 CGCG381-051  & 2 & 0.03048 & 23 48 41.709 &  02 14 23.25 &     0.6 & $<$   0.1 & $<$     59 &   U &     \\ 
\hline 
\end{tabular}
\end{center}
\contcaption{} 
\end{table*}

Radio parameters for all observed sources from the
extended 12 $\mu$m AGN sample are given in Table \ref{results.tab}; 
radio parameters derived by \scite{Kukula95} for 19 sources which 
belong to the CfA sample have also been included. 
The table is organized as follows;
{\it Column 1}: Galaxy name. An asterisk ($\ast$) is used to
indicate those sources whose radio parameters are taken from \scite{Kukula95}. 
{\it Column 2}: Seyfert type. For this paper 
objects are classified simply as type 1 or
type 2 following \scite{RMS93}
except where alternative classifications have been
proposed by \scite{Dopita98} (9 sources) and \scite{Mulchaey96} (NGC
34). Reclassified sources are labeled using the following the
abbreviations; Q for Quasar, L for LINER and sb for Starburst. 
For these sources, the 
classification given by Rush {\it et al} is given in parenthesis. 
{\it Column 3}: Redshift, as taken from \scite{RMS93}.
{\it Columns 4 and 5}: Right ascension, RA (h, m, s),
and declination, Dec (deg, arcmin, arcsec), of each radio
component in J2000 co--ordinates. 
Positions were determined by Gaussian fitting. 
For point--like sources the accuracy of fit, estimated by comparing
the modelled and measured flux densities, was found to be around 5\%.
{\it Column 6}: Integrated flux density of each component,
S (mJy/beam). 
The mean flux density of each component was measured directly from the
map. 
Care was taken to ensure that the effects of non--zero background levels were
taken into account.
The uncertainty on each flux measurement may be estimated by
combining the calibration error (\pcite{Weiler86} estimated a value of
around 4\%) and the map error for each source (around 4\% for strong
sources and 14\% for weak sources). 
For those sources which were not detected an
upper limit on their radio flux of 5--$\sigma$ has been assumed.
In multiple component sources the component nearest the photographic
position (taken from NED) is labeled with a dagger symbol ($\dagger$). 
Ninety--three percent of the sources had at least one radio component
within 2--$\sigma$ of the available photographic position; of the
others, four show close alignment between the radio position and the galactic
centre as judged from Palomar Sky Survey images 
(MCG-3-7-11, NGC 5194 = M51, NGC 5033 and NGC
5005) and 2  
have been excluded from further analysis (NGC 4922A/B and NGC 5135).
The sky density of sources above the average 
3--$\sigma$ detection threshold of 159
$\mu$Jy  at 8.4 GHz has been estimated  by \scite{Windhorst93} and
implies a 2\% probability of detecting an unrelated  
source within the 51 $\times$51 arcsec$^{2}$ field--of--view.
{\it Column 7}: Maximum angular size of radio structure in
arcseconds, $\theta$. For unresolved sources
an upper limit to the angular size was taken as one third of
the major axis of the beam.
For slightly resolved sources the angular size
was taken as the length of the major
axis of the nominal deconvolution\footnote{The nominal deconvolution
is obtained when the Gaussian fit to a component is deconvolved from the 
CLEAN beam} for single Gaussian fits, or the 
maximum separation of the peaks for multiple Gaussian fits.
For all other sources the maximum size measured was
either the maximum peak separation (for those sources 
with point--like components) or the maximum size at the lowest contour
(for those sources with diffuse components). 
{\it Column 8}: Maximum linear size in parsecs, D, assuming H$_{\circ}$ = 75
km$\,$s$^{-1}$Mpc$^{-1}$.
{\it Column 9}: Type of radio structure, T$_{rad}$, according to the 
notation used by \scite{Ulvestad+W84a}: U for single
unresolved sources, S for single slightly--resolved sources, A for 
sources with ambiguous structures, D for sources with diffuse structures
and L for sources with possible linear
structures (sources with two components, sources with three or more aligned
components or sources with extended linear components). 
Gaussian fitting was used to
distinguish between types U and S, type S sources had a signal to noise
ratio greater than 20 and a nominal deconvolution size greater than
one third of the beam at FWHM.
{\it Column 10}: Position angle of radio structure, PA, 
measured North 
to East from 0$^{\circ}$ to 180$^{\circ}$. For partially
resolved sources (type S) the position angle of the nominal deconvolution was
used. The position angle of clearly resolved linear sources
(type L) was
measured directly from the map. 

\section{THE IDENTIFICATION OF RADIO-LOUD SOURCES}
\label{radio-loud.sec}

\begin{figure} 
\centerline{
       \includegraphics[angle=0,width=8.5cm,clip,trim=0 0 0 216]{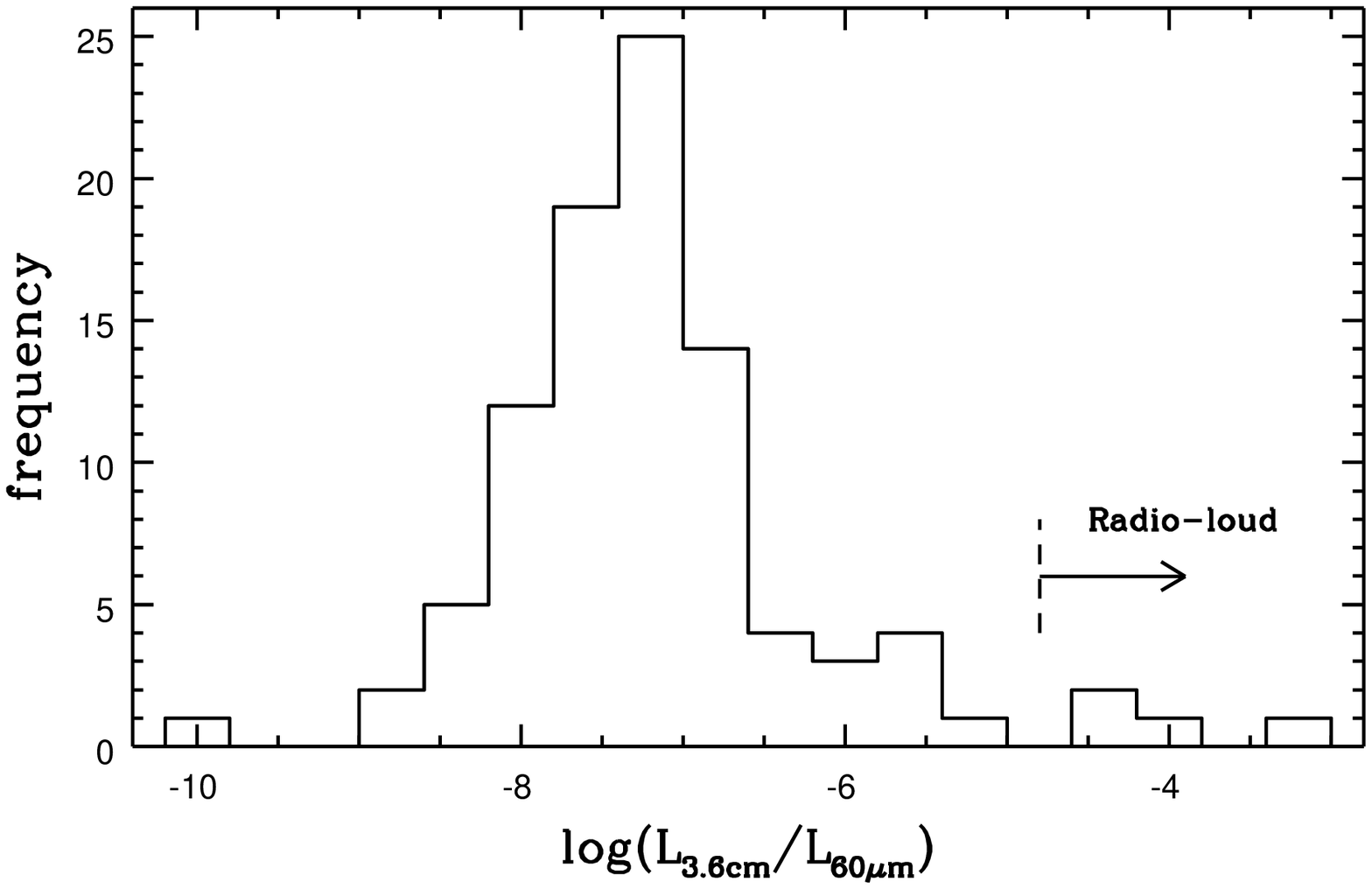}} 
\caption{A histogram showing the criterion used to exclude 5
radio--loud sources from the Seyfert sample. 
The frequency of sources per logarithmic
radio to far--infrared luminosity ratio interval
$\Delta$log(L$_{3.6cm}$/L$_{60\mu m}$) = 0.4 is plotted. 
A value of log(L$_{3.6cm}$/L$_{60\mu m}$) of around -4.8 may be used 
to distinguish the two classes, above this limit sources are
considered radio--loud.}
\label{rad-60um_hist.fig}
\end{figure}

The distinction between `radio--loud' and `radio--quiet' AGN 
(\pcite{Kellerman89}; \pcite{Miller90}) is widely accepted and 
usually thought to result from truly distinguishable physical processes
(e.g. \pcite{Wilson95a}).
In this section we describe the identification of 
five radio--loud objects in the extended 12 $\mu$m AGN
sample (OJ 287, 3C 120, 3C 234, 3C 273 and 3C 445);
these sources will be excluded from further statistical analysis and 
henceforth the remaining sources will be referred to 
as the extended 12 $\mu$m Seyfert sample.

We have chosen to use the radio to far--infrared luminosity ratio as
the main diagnostic of `radio--loudness'.
This has been done by using IRAS FSC 60 $\mu$m
luminosities and our newly--measured 8.4 GHz (3.6 cm) radio
luminosities; single--dish radio  
observations from \pcite{Wright91} were used for the two unobserved
radio--loud objects, 3C 120 and 3C 273, and 3C 445 for which the
majority of the flux is outside the field--of--view of the new observations.
The radio to far--infrared luminosity ratio is probably a 
more useful indicator of radio--loudness than the radio luminosity
alone given that the bolometric luminosities of the sources in the
sample are likely to span several orders of magnitude. 
Figure \ref{rad-60um_hist.fig} is a histogram showing the frequency of
sources in the extended 12 $\mu$m AGN sample per logarithmic radio to
far--infrared luminosity ratio interval
$\Delta$log(L$_{3.6cm}$/L$_{60\mu m}$) = 0.4; the 
L$_{3.6cm}$/L$_{60\mu m}$ ratio for each detected source 
is given in Table \ref{radioloud.tab}. 
The 5 excluded sources have the highest radio to far--infrared luminosity 
ratios of the sample; 
all have log(L$_{3.6cm}$/L$_{60\mu m}$) ratios greater than -4.6, whereas
log(L$_{3.6cm}$/L$_{60\mu m}$) ratios are less than -5.1 for the Seyferts. These values are in 
agreement with the radio to far--infrared
luminosity ratios used by \scite{RushM+E96} to 
identify 3 of the same radio--loud sources 
in the original 12 $\mu$m AGN sample (OJ287, 3C 120 and 3C 273).
The mean ratio between the 8.4 GHz A--configuration flux to the 5 GHz
D--configuration flux for those sources observed by \scite{RushM+E96}   
is around 0.3; this ratio is consistent with a mean radio spectral
index of $\alpha$ = -0.7 \cite{RushM+E96} and a A--configuration to
D--configuration flux ratio of around 0.5.  
Using this flux ratio, 
the radio to far--infrared luminosity ratios used by 
\scite{RushM+E96} translate to -4.7 $<$ log(L$_{3.6cm}$/L$_{60\mu m}$)
$<$ -2.3 for radio--loud sources and -6.3 $<$ log(L$_{3.6cm}$/L$_{60\mu m}$)
$<$ -4.8 for radio--quiet sources.

Note that there is no clear evidence for a bimodal distribution of
the radio to far--infrared luminosity ratio, possibly
because of the infrared flux-limit used to define the sample.
As well as having the highest radio to far--infrared luminosity
ratios, the radio--loud sources we have identified are also the 5 most
radio luminous sources in the sample, being the only  
sources more luminous than L$_{3.6cm}$ $>$ 10$^{24}$ WHz. 
They are all well--known objects with powerful jets;
three show super--luminal motions in their jet
components (OJ287, 3C120 and 3C273, as cited by \pcite{Ghosh95}) and
the other two are classical FR--II radio galaxies with radio structures 
hundreds of kiloparsecs in size (see \pcite{Leahy86} 
and \pcite{Leahy97} for images of 3C234 and 3C445 respectively).
A strong reason for excluding these sources is that 
they are broad--line objects which, when grouped
with the Seyfert 1 subsample, would systematically affect
comparisons between the two Seyfert types. 

\begin{table}
\scriptsize
\begin{center}
\begin{tabular}{|l|c||l|c|} \hline 
\bf{Source} & \bf{log(L$_{rad}$/L$_{ir}$)} & \bf{Source} &
\bf{log(L$_{rad}$/L$_{ir}$)}\\ \hline
3C 273        &  -3.10 & Mrk 1034       &  -7.29 \\  
3C 120        &  -4.15 & Mrk 509        &  -7.32 \\    
OJ 287        &  -4.31 & NGC 1241       &  -7.33 \\  
3C 234        &  -4.39 & F22017+0319    &  -7.35 \\  
3C 445        &  -4.54 & Mrk 273        &  -7.35 \\  
Mrk 348       &  -5.14 & NGC 3516       &  -7.35 \\  
F01475-0740   &  -5.45 & F03362-1642    &  -7.36 \\  
NGC 7213      &  -5.69 & NGC 3227       &  -7.36 \\  
F05563-3820   &  -5.75 & MGC-3-7-11     &  -7.40 \\  
NGC 2639      &  -5.76 & Arp 220        &  -7.41 \\  
Mrk 6         &  -6.03 & Mrk 817        &  -7.43 \\  
Mrk 231       &  -6.10 & MGC-2-33-34    &  -7.45 \\  
Mrk 463       &  -6.19 & UGC 11680      &  -7.45 \\  
NGC 4594      &  -6.21 & NGC 1386       &  -7.47 \\  
NGC 4151      &  -6.41 & Mrk 9          &  -7.48 \\  
F15480-0344   &  -6.47 & Mrk 618        &  -7.49 \\  
NGC 5506      &  -6.51 & NGC 7130       &  -7.49 \\  
F03450+0055   &  -6.63 & TOL1238-364    &  -7.50 \\  
IC 4329A      &  -6.67 & NGC 1194        &  -7.52 \\ 
NGC 7674      &  -6.67 & NGC 7582        &  -7.53 \\ 
MGC-3-34-63   &  -6.69 & NGC 4388        &  -7.57 \\ 
Mrk 1239      &  -6.72 & NGC 34          &  -7.58 \\ 
NGC 4579      &  -6.74 & E541-IG12       &  -7.59 \\ 
Mrk 335       &  -6.75 & NGC 7172        &  -7.61 \\ 
NGC 424       &  -6.75 & MGC-2-40-4      &  -7.66 \\ 
F13349+2438   &  -6.78 & NGC 7469        &  -7.77 \\ 
F15091-2107   &  -6.83 & NGC 4593        &  -7.79 \\ 
NGC 3031      &  -6.83 & NGC 5347        &  -7.80 \\ 
UGC 5101      &  -6.93 & F05189-2524     &  -7.81 \\ 
NGC 1068      &  -6.94 & F08572+3915     &  -7.81 \\ 
NGC 2992      &  -6.95 & NGC 1320        &  -7.85 \\ 
F07599+6508   &  -7.00 & NGC 5005        &  -7.93 \\ 
NGC 1125      &  -7.02 & I ZW 1          &  -7.94 \\ 
NGC 526A      &  -7.04 & NGC 7496        &  -7.95 \\ 
NGC 513       &  -7.07 & NGC 6890        &  -7.96 \\ 
Mrk 704       &  -7.11 & CGCG381-051     &  -7.99 \\ 
NGC 4941      &  -7.11 & MGC-3-58-7      &  -8.00 \\ 
NGC 7603      &  -7.11 & NGC 7314        &  -8.01 \\ 
MGC-5-13-17   &  -7.12 & NGC 1667        &  -8.13 \\ 
NGC 4968      &  -7.15 & NGC 5033        &  -8.44 \\ 
Mrk 766       &  -7.17 & NGC 3982        &  -8.46 \\ 
NGC 5548      &  -7.20 & NGC 1365        &  -8.48 \\ 
F04385-0828   &  -7.21 & NGC 1056        &  -8.50 \\ 
Mrk 79        &  -7.22 & NGC 5953        &  -8.55 \\ 
NGC 5256      &  -7.22 & NGC 1097        &  -8.71 \\
NGC 1143/4    &  -7.23 & NGC 4051        &  -8.77 \\
NGC 3079      &  -7.26 & NGC 5194        &  -9.86 \\       
NGC 5929      &  -7.28 &                 &        \\
\hline
\end{tabular}
\caption[Detected sources ranked according to decreasing L$_{3.6 cm}$/L$_{60
\mu m}$ ratio]{
Detected sources ranked according to decreasing L$_{3.6 cm}$/L$_{60 \mu m}$
(L$_{rad}$/L$_{ir}$) luminosity ratio. 
}  
\label{radioloud.tab}
\end{center}
\end{table}

\section{SUMMARY}

The maps presented in this paper reveal for the first time the
sub--arcsecond radio structures of Seyferts contained in the extended
12 $\mu$m AGN sample. They provide a large and 
homogeneously--selected database for investigating the
generic properties of compact radio cores in Seyfert nuclei.  

Seventy--five of the 87 sources observed were detected; 
36 contain single unresolved radio sources, 
13 contain single slightly--resolved radio sources, 
9 contain radio sources with diffuse or ambiguous
structures, 8 contain radio sources with two distinguishable 
components and 9 contain radio sources with three or more 
linearly--aligned components or extended linear structures.
Subsequent papers will discuss the statistical properties of the
sample in detail, paying particular attention to comparisons of the radio powers
and radio morphologies of the two Seyfert types.

\section{ACKNOWLEDGMENTS}

AHCT would like to acknowledge the receipt of a studentship from
the Particle Physics and Astronomy Research Council and a visit funded
by the STScI visitor program. Part of this research was 
supported by the European Commission, TMR Programme, Research Network Contract
ERBFMRXCT96-0034 ``CERES''. 
We have made use of NASA's Astrophysics Data
System Abstract Service, the NASA/IPAC Extragalactic database (NED),
which is operated by the Jet Propulsion Laboratory.

\bibliography{12mic}
\bibliographystyle{mnras}

\end{document}